\documentclass[letterpaper,11pt,usenames,dvipsnames,cmyk]{article}

\usepackage[totalwidth=480pt, totalheight=680pt]{geometry}
\usepackage{amsmath}
\usepackage{amsfonts}
\usepackage{dsfont}
\usepackage{bbm}
\usepackage[colorlinks=true, linktoc=page, citecolor=blue, urlcolor=blue]{hyperref}
\usepackage[retainorgcmds]{IEEEtrantools}
\usepackage{verbatim}
\usepackage{graphics}
\usepackage{tikz}
\usepackage{tikz-feynman}
\usepackage{slashed}
\usepackage{float}
\usepackage{bbold}
\usepackage{scalerel}
\usepackage{multirow}
\usepackage[bulletsep]{collref}
\usepackage{enumitem}

\newcommand{\OO}{\mathcal{O}}
\newcommand{\x}{\textrm{x}}
\newcommand{\y}{\textrm{y}}
\newcommand{\uu}{\mathfrak{u}}
\newcommand{\kk}{\text{k}}
\newcommand{\z}{{\textrm{z}}}
\newcommand{\N}{\mathcal{N}}
\newcommand{\X}{\mathbb{X}}
\newcommand{\Y}{\mathbb{Y}}
\newcommand{\W}{\mathcal{W}}
\newcommand{\K}{\mathcal{K}}
\newcommand{\G}{\mathcal{G}}
\newcommand{\LL}{\mathcal{L}}
\newcommand{\C}{\mathcal{C}}
\newcommand{\B}{\mathcal{B}}
\newcommand{\CC}{\mathfrak{C}}
\newcommand{\I}{\mathcal{I}}
\newcommand{\J}{\mathcal{J}}
\newcommand{\F}{\mathfrak{F}}

\newcommand{\CP}{{\mathbb{CP}}}
\newcommand{\A}{{\hat{A}}}
\newcommand{\tr}{{\text{tr}}}
\newcommand{\Z}{\mathds{Z}}
\newcommand{\eq}{\stackrel{!}{=}}
\newcommand{\U}{\mathbbm{U}}
\newcommand{\T}{\mathcal{T}}
\newcommand{\st}{\text{st}}
\newcommand{\Q}{\mathfrak{Q}}
\newcommand{\Pexp}{\overrightarrow{\rm P}\exp}
\newcommand{\M}{\mathcal{M}}

\newcommand{\hodge}{{\star}}
\newcommand{\quotes}[1]{``#1''}

\makeatletter \@addtoreset{equation}{section} \makeatother

\hypersetup{pdfstartview={XYZ null null 1.25}}
\begin{document}
\begin{flushright}
APCTP Pre2025-001
\end{flushright}
\title{{\color{Blue}\textbf{String theory methods for defect CFTs}}\\[12pt]} \date{}
\author{\textbf{Georgios Linardopoulos}\footnote{E-mail: \href{mailto:georgios.linardopoulos@apctp.org}{georgios.linardopoulos@apctp.org}.}\\[12pt]
Asia Pacific Center for Theoretical Physics (APCTP) \\ Hogil Kim Memorial Building, \#501 POSTECH \\ 77 Cheongam-Ro Nam-gu, Pohang Gyeongsangbuk-do 37673, Korea. \\[12pt]}
{\let\newpage\relax\maketitle}
\begin{abstract}
\normalsize{\noindent We discuss string theory methods for the study of strongly coupled holographic defect conformal field theories (CFTs) which are dual to probe-brane systems. First, we examine whether the string theory duals of such defect CFTs are classically integrable. This amounts to identifying all string boundary conditions on probe branes which preserve the integrability of the underlying Green-Schwarz sigma model. Second, we compute holographic defect correlation functions with semiclassical strings. After reviewing the computation of three and higher-point functions in strongly coupled AdS/CFT, and one-point functions in its probe-brane descendants, we outline the computation of two and higher-point correlation functions in strongly coupled defect CFTs.\\}
\end{abstract}
\newpage\tableofcontents\normalsize
\newpage\section[Introduction and motivation]{Introduction and motivation \label{Section:IntroductionMotivation}}
\noindent Understanding the dynamics of gauge theories at strong coupling is one of the greatest challenges in theoretical physics. Owing to the seminal work of Wilson \cite{Wilson74a}, it has been known for a long time that strongly coupled Yang-Mills theory can be reformulated as an effective theory of color flux tubes which form between quark-antiquark pairs and are responsible for quark confinement. This mechanism is inevitably reminiscent of relativistic string theory. Yet another fascinating connection between gauge and string theory was uncovered in 1973 by 't Hooft \cite{Hooft74} who noticed that the perturbative behavior of Yang-Mills theory in the planar (or large-$N_c$) limit bears a striking resemblance to the topological expansion of string theory. \\
\indent The first direct proof of concept for these ideas was provided by holography. The AdS/CFT correspondence \cite{Maldacena97} suggests that certain highly supersymmetric gauge theories with exact conformal symmetry can be identified with superstring theory in anti-de Sitter space. At weak gauge theory coupling, Feynman perturbation theory can be used to calculate the basic observables of the theory such as the spectrum, correlation functions, etc.\ \cite{MAGOO99}. At strong (gauge theory) coupling, string theory becomes weakly coupled and therefore suitable for calculations in the nonperturbative regime of the gauge theory. However, because the gauge and the string theory coupling constants are inversely proportional, the two perturbative regimes are practically disconnected from each other and almost impossible to compare except in special cases (such as for example the BMN sector \cite{BMN02}). \\
\indent The weak/strong coupling dilemma of AdS/CFT can be circumvented by integrability which was first discovered in the spectrum of planar $\N = 4$ super Yang-Mills (SYM) theory \cite{MinahanZarembo03, BeisertKristjansenStaudacher04, Beisert03}. Integrability solves the spectral problem of (planar) AdS/CFT and confirms its validity from weak to strong coupling. One similarly hopes that integrability could lead to the solution of the full planar theory, that is the computation of all of its observables (correlators, scattering amplitudes, Wilson loops, etc.). If this program (which is currently underway, see \cite{Beisertetal10}) succeeds, it will provide unequivocal evidence for the validity of the particle/string duality in the planar limit. \\
\indent The price to pay for entering the nonperturbative regime of gauge theories with holography is the high level of symmetry (supersymmetry, exact conformal invariance and planar integrability) that is present in holographic dualities. This makes them highly unrealistic and arguably very far removed from reality. The real-world gauge theories that we would like to study at strong coupling (such as QCD, the theory of strong interactions) are neither finite, nor supersymmetric, integrable, or holographic. It is therefore very natural to wonder whether we could still construct other holographic examples which are more realistic than the original AdS/CFT correspondence but remain solvable at strong coupling. Because holography is clearly affected by the weak/strong coupling dilemma, we are also eager to keep integrability come what may. \\
\indent There exist many ways to obtain new integrability-preserving holographic dualities by deforming known ones.\footnote{See e.g.\ the nice review article \cite{Zoubos10} by K.\ Zoubos.} In the present work we focus on just one of these ways, namely we will consider holographic dualities which are obtained by inserting a probe brane on the string theory side of the AdS/CFT correspondence. The gauge theory side of the new holographic dualities will be a defect conformal field theory (dCFT), whence the name AdS/defect CFT correspondence (or AdS/{\color{red}d}CFT for short) which we will use to describe the dualities. As we will see, AdS/{\color{red}d}CFT dualities may or may not preserve the integrability of the original AdS/CFT duality they came from. \\
\indent We begin by introducing classical string integrability in holographic dualities with and without probe branes in the following section \ref{Section:StringIntegrabilityHolographicDefects}. Before revisiting classical string integrability in AdS/CFT (subsection \ref{SubSection:StringIntegrabilityCFT}), we briefly review the gauge theory sides of AdS$_5$/CFT$_4$ and AdS$_4$/CFT$_3$ duality, that is $\N = 4$ SYM and ABJM theory (subsection \ref{SubSection:SolvableHolographicCFTs}). Subsection \ref{SubSection:StringIntegrabilitydCFT} is about classical string integrability in AdS/{\color{red}d}CFT. The basic probe-brane deformations of AdS$_5$/CFT$_4$ and AdS$_4$/CFT$_3$ are examined in three subsequent paragraphs, namely \S\ref{SubSubSection:D3D5intersection}--\S\ref{SubSubSection:D2D4intersection}. In \S\ref{SubSubSection:D3D5intersection} we show that the deformation of AdS$_5$/CFT$_4$ with a probe D5-brane (D3-D5 intersection) is classically integrable, i.e.\ string boundary conditions on the probe D5-brane give rise to classically integrable string dynamics. In \S\ref{SubSubSection:D3D7intersection} we examine two possible deformations of AdS$_5$/CFT$_4$ duality with a probe D7-brane (D3-D7 intersection), one with $SU(2)\times SU(2)$ global symmetry and another with $SO(5)$ global symmetry. In \S\ref{SubSubSection:D2D4intersection} we show that the deformation of AdS$_4$/CFT$_3$ duality with a probe D4-brane (D2-D4 intersection) is also classically integrable. We conclude section \ref{Section:StringIntegrabilityHolographicDefects} with an outlook and a summary of open problems in subsection \ref{SubSection:IntegrabilityOutlook}. \\
\indent In section \ref{Section:HolographicDefectCorrelators} we discuss the computation of correlation functions in strongly coupled AdS/{\color{red}d}CFT dualities with semiclassical strings. In \S\ref{SubSection:ConformalCorrelationFunctions} we provide a broad overview of conformal field theories with (\S\ref{SubSubSection:ConformalFieldTheories}) and without (\S\ref{SubSubSection:DefectConformalFieldTheories}) codimension-1 defects, highlighting correlation functions, operator product expansions and bootstrapping. In \S\ref{SubSection:HolographicCorrelatorsCFTs} we briefly review correlation functions in weakly and strongly coupled holographic CFTs (subsections \S\ref{SubSection:HolographicCorrelatorsCFTsWeak} and \S\ref{SubSection:HolographicCorrelatorsCFTsStrong}). We provide details for the computation of three and higher point functions at strong coupling with semiclassical strings. In subsection \ref{SubSection:HolographicCorrelatorsdCFTs} we take up holographic defect correlators at weak and strong coupling (\S\ref{SubSection:HolographicCorrelatorsdCFTsWeak} and \S\ref{SubSection:HolographicCorrelatorsdCFTsStrong} respectively). We demonstrate how to compute strongly coupled correlation functions with open semiclassical strings which end on probe branes, by also reviewing the corresponding literature. Subsection \ref{SubSection:IntegrabilityOutlook} contains a discussion and an outlook.
\section[String integrability of holographic defects]{String integrability of holographic defects \label{Section:StringIntegrabilityHolographicDefects}}
\noindent In the present section we discuss classical string integrability in holographic theories with and without probe branes. Let us first outline the notion of integrability. A dynamical system is Liouville integrable if for its every degree of freedom there exists an independent conserved quantity (charge). By solving the set of algebraic equations that is specified by the conserved charges, all the degrees of freedom can theoretically be determined in a closed form. In this sense, Liouville-integrable systems are also solvable. Liouville integrability can naturally be extended to classical field theories with infinite degrees of freedom. Quantum integrability can also be defined in theories which support scattering as the absence of particle production/annihilation and the factorization of multi-particle scattering processes into chains of two-body events.\footnote{See e.g.\ the book \cite{BabelonBernardTalon03} and the recent lecture notes \cite{Driezen21, Retore21} for more.} \\
\indent String theory models which are relevant for holography correspond to classical field theories in 1+1 dimensions. More often than not, the emergence of classical integrability on the string theory side of a holographic duality signifies the presence of a quantum integrable spin chain on the dual gauge theory side. Theories with such built-in quantum integrable structures are most likely solvable in the planar limit (in the sense defined in \S\ref{Section:IntroductionMotivation} above) for all values of their coupling constants. Indeed, the spectral problem has been solved in certain holographic dualities by non-perturbative integrability methods such as the thermodynamic Bethe ansatz (TBA), the Y/T/Hirota system,\footnote{For more, see e.g.\ the review \cite{Beisertetal10}.} and more recently by the quantum spectral curve (QSC) method. \\
\indent Inserting a probe brane on the string theory side of an integrable holographic duality breaks many of the theory's (super) symmetries and possibly string integrability. Apart from closed string states which propagate far from the brane, the defect theory now includes open strings which have their endpoints on the brane. Closed string states are dual to single-trace operators in the undeformed theory. These are described by closed spin chains. Both closed strings and their dual spin chains are integrable as we have already explained. On the other hand, open string states are dual to operators which include defect fields. These are described by open spin chains \cite{DeWolfeFreedmanOoguri01, DeWolfeMann04}. Based on what we have said above, demonstrating classical integrability for open string states in the brane-deformed theory is a strong indication that the dual defect theory has an underlying (quantum) integrable structure. This quantum integrable structure could lead to the full solution of the defect theory (i.e.\ the calculation of all of its observables at any coupling), provided non-perturbative integrability methods (similar to the TBA/Y-system/QSC) become available. \\
\indent The existence of an underlying integrability structure within a brane-deformed holographic duality can also be examined from the gauge theory point of view by means of a set of integrable quench criteria. These are based on an interesting argument by Ghoshal and Zamolodchikov \cite{GhoshalZamolodchikov93} who viewed initial conditions for the creation/annihilation of particle pairs (\quotes{quench channel}) in 2-dimensional integrable quantum field theories (IQFTs) as Wick-rotated boundary conditions for the scattering of particles (\quotes{defect channel}). According to the quench criteria of \cite{PiroliPozsgayVernier17}, a boundary (initial) state in an integrable lattice model/spin chain is characterized as integrable if it is annihilated by all the parity-odd conserved charges in the integrable hierarchy of the spin chain. Equivalently, overlaps of the boundary state with Bethe eigenstates of the integrable spin chain are expressible in terms of closed-form determinant formulas which involve the corresponding Bethe roots. Moreover, the overlaps can only be nontrivial when the Bethe roots are fully balanced, i.e.\ when they come in pairs of equal magnitudes and opposite signs.
\subsection[Solvable holographic CFTs]{Solvable holographic CFTs \label{SubSection:SolvableHolographicCFTs}}
\noindent We are basically going to focus on holographic defects which are derived from two main \quotes{solvable} holographic dualities. The first one involves 4-dimensional $\N = 4$ SYM theory, and the other one involves 3-dimensional $\N = 6$ super Chern-Simons (CS) matter theory. This theory is also known as the Aharony-Bergman-Jafferis-Maldacena (ABJM) theory. By \quotes{solvable} we mean that the (planar) spectrum of the gauge theory (CFT) is described by an integrable spin chain, and this spin chain can be used to fully solve the CFT (i.e.\ determine all its observables at any coupling strength). Before concentrating on the string theory side of the corresponding holographic dualities, let us briefly review the gauge theory sides.
\paragraph{$\N = 4$ SYM theory}$\N = 4$ SYM is a 4-dimensional maximally supersymmetric gauge theory which is defined in terms of three sets of fields (4d gluon, six real scalars, 4d Majorana-Weyl fermions) and the following Lagrangian density \cite{BrinkSchwarzScherk76, GliozziScherkOlive76}:
\begin{IEEEeqnarray}{ll}
\mathcal{L}_{\N = 4} = &\frac{2}{g_{\text{\scalebox{.8}{YM}}}^2} \cdot \text{tr}\bigg\{-\frac{1}{4} F_{\mu\nu} F^{\mu\nu} - \frac{1}{2} \left(D_{\mu}\varphi_i\right)^2 + i\,\bar{\psi}_{\alpha}\slashed{D}\psi_{\alpha} + \frac{1}{4}\left[\varphi_i,\varphi_j\right]^2 + \sum_{i = 1}^{3}G^i_{\alpha\beta}\bar{\psi}_{\alpha}\left[\varphi_i,\psi_{\beta}\right] + \nonumber \\
& + \sum_{i = 4}^{6}G^i_{\alpha\beta} \bar{\psi}_{\alpha} \gamma_5 \left[\varphi_i,\psi_{\beta}\right]\bigg\}, \quad \mu,\nu = 0,1,2,3, \quad i,j = 1,\ldots,6, \quad \alpha,\beta = 1,\ldots,4.\footnote{In the Weyl/chiral basis, the 4d gamma matrices are given by \eqref{GammaMatrices1} and the $4\times 4$ matrices $G^i$ by \eqref{G-Matrices1}--\eqref{G-Matrices2}.} \qquad \label{LagrangianSYM}
\end{IEEEeqnarray}
According to the AdS$_5$/CFT$_4$ correspondence \cite{Maldacena97}, $\N = 4$ SYM is holographically dual to type IIB superstring theory on AdS$_5\times\text{S}^5$:
\begin{IEEEeqnarray}{c}
\begin{array}{c} \N = 4, \ U\left(N_c\right) \ \text{SYM theory in 4d} \end{array} \ \Leftrightarrow \ \begin{array}{c} \text{Type IIB string theory on AdS}_5 \times \text{S}^5 \text{ with } N_c \text{ units} \\ \text{of self-dual 5-form RR flux through AdS}_5 \text{ \& S}^5. \end{array} \qquad \label{AdS5CFT4duality}
\end{IEEEeqnarray}
$\N = 4$ SYM is one of the few finite quantum field theories in 4 dimensions. Its beta function vanishes to all orders in perturbation theory so that the theory possesses exact superconformal symmetry $PSU(2,2|4)$. The spectrum of $\N = 4$ SYM is described by an integrable spin chain in the planar limit $N_c \rightarrow \infty$, $\lambda \equiv g_{\text{\scalebox{.8}{YM}}}^2 N_c = \text{const.}$ \cite{MinahanZarembo03, BeisertKristjansenStaudacher04, Beisert03} ($\lambda$ is the 't Hooft coupling). As we have mentioned, the spectral problem of planar $\N = 4$ SYM is solved by integrability and the quantum spectral curve method \cite{GromovKazakovLeurentVolin13}. Solving the full planar theory by computing all of its observables (correlators, scattering amplitudes, Wilson loops, etc.) is currently in progress \cite{Beisertetal10}. Let us also mention that half-BPS boundary conditions (BCs) in $\N = 4$ SYM theory have been studied by Gaiotto and Witten in \cite{GaiottoWitten08a}.
\paragraph{ABJM theory} ABJM theory is a 3-dimensional superconformal gauge theory with product gauge group $U\left(N_c\right) \times \hat{U}\left(N_c\right)$. The theory has 3 sets of fields (two 3d gluons, four complex scalars, four 3d Weyl fermions) and Lagrangian density \cite{BennaKlebanovKloseSmedback08}:
\begin{IEEEeqnarray}{ll}
\mathcal{L}_{\text{ABJM}} = \frac{k}{4\pi} \cdot \bigg[&\epsilon^{\mu\nu\rho} \tr\bigg\{A_{\mu}\partial_{\nu}A_{\rho} + \frac{2i}{3}A_{\mu}A_{\nu}A_{\rho} - \A_{\mu}\partial_{\nu}\A_{\rho} - \frac{2i}{3}\A_{\mu}\A_{\nu}\A_{\rho}\bigg\} - \tr\bigg\{\left(D_{\mu}Y_{B}\right)^{\dag}D^{\mu}Y_{B} + \nonumber \\
& + i \psi^{\dag}_{B}\slashed{D}\psi_{B}\bigg\} - V_{\text{ferm}} - V_{\text{bos}}\bigg], \quad \mu,\nu,\rho = 0,1,2, \quad B = 1,\ldots,4, \qquad \label{LagrangianABJM}
\end{IEEEeqnarray}
where $k\in\Z$ is the CS level. The potential contains mixed quartic and sextic bosonic terms,
\begin{IEEEeqnarray}{ll}
V_{\text{ferm}} &= \frac{i}{2} \tr\bigg\{Y_A^{\dag} Y_A \psi^{\dag}_B \psi_B - Y_A Y_A^{\dag} \psi_B \psi^{\dag}_B + 2Y_A Y_B^{\dag} \psi_A \psi_B^{\dag} - 2Y_A^{\dag} Y_B \psi^{\dag}_A \psi_B - \epsilon^{ABCD} Y_A^{\dag} \psi_B Y_C^{\dag} \psi_D + \nonumber \\
&\hspace{1cm} + \epsilon^{ABCD} Y_A \psi_B^{\dag} Y_C \psi_D^{\dag}\bigg\} \qquad \\
V_{\text{bos}} &= -\frac{1}{12} \tr\bigg\{Y_{A}Y_{A}^{\dag}Y_{B}Y_{B}^{\dag}Y_{C}Y_{C}^{\dag} + Y_{A}^{\dag}Y_{A}Y_{B}^{\dag}Y_{B}Y_{C}^{\dag}Y_{C} + 4Y_{A}Y_{B}^{\dag}Y_{C}Y_{A}^{\dag}Y_{B}Y_{C}^{\dag} - 6Y_{A}Y_{B}^{\dag}Y_{B}Y_{A}^{\dag}Y_{C}Y_{C}^{\dag}\bigg\}. \qquad
\end{IEEEeqnarray}
According to the AdS$_4$/CFT$_3$ correspondence \cite{ABJM08}, ABJM theory is holographically dual to M-theory on AdS$_4\times\text{S}^7/\Z_k$ (M/ABJM correspondence). In the limit $k^5 \gg N_c$, M-theory on AdS$_4\times\text{S}^7/\Z_k$ reduces to type IIA superstring theory on AdS$_4\times\CP^3$ (IIA/ABJM correspondence), so that
\begin{IEEEeqnarray}{c}
\begin{array}{c} \N = 6, \ U\left(N_c\right)_k \times U\left(N_c\right)_{-k} \\ \text{ super CS matter theory in 4d} \\ \text{with CS levels } k \text{ \& } -k, \ k^5 \gg N_c \end{array} \ \Leftrightarrow \ \begin{array}{c} \text{Type IIA string theory on AdS}_4 \times \CP^3 \\ \text{ with } N_c \text{ units of 4-form RR flux through AdS}_4 \\ \text{\& } k \text{ units of 2-form RR flux through }\CP^1 \subset \CP^3. \end{array}\qquad \label{AdS4CFT3duality}
\end{IEEEeqnarray}
ABJM theory has exact superconformal symmetry $Osp(2,2|6)$. In the planar limit, i.e.\ for $N_c,k \rightarrow \infty$, $\lambda \equiv g_{\text{\scalebox{.8}{CS}}}^2 N_c = \text{const.}$, $g_{\text{\scalebox{.8}{CS}}}^2 \equiv 1/k$, the IIA/ABJM spectrum is described by an integrable spin chain \cite{MinahanZarembo09}. The theory's spectral problem has also been solved by the quantum spectral curve method \cite{CavagliaFioravantiGromovTateo14}.
\subsection[String integrability in AdS/CFT]{String integrability in AdS/CFT \label{SubSection:StringIntegrabilityCFT}}
\noindent In theory, there exist two possible formulations of superstring theory in the AdS/CFT backgrounds \eqref{AdS5CFT4duality} and \eqref{AdS4CFT3duality}.\footnote{The corresponding supergravity solutions can be found in appendix \ref{Appendix:Conventions}.} In the Ramond-Neveu-Schwarz (RNS) formalism, supersymmetry is manifest on the string worldsheet, while in the Green-Schwarz (GS) formalism, supersymmetry is manifest in target space. In practice however, both of these formulations are too complicated to be applied to the backgrounds of the AdS/CFT correspondence. \\
\indent An alternative approach was developed by Metsaev and Tseytlin (MT) \cite{MetsaevTseytlin98} following the work of Henneaux and Mezincescu \cite{HenneauxMezincescu85}, who had shown that the Green-Schwarz superstring action in flat space can be reproduced by a Wess-Zumino-Witten (WZW) sigma model on the supercoset, $\left\{\text{10d super Poincar\'{e} group}\right\}/SO(9,1)$. Based on the observation that the AdS$_5\times\text{S}^5$ target space can be written as a product of two cosets which span the bosonic section of a supercoset
\begin{IEEEeqnarray}{c}
\text{AdS}_5\times\text{S}^5 = \frac{SO(4,2)}{SO(4,1)} \times \frac{SO(6)}{SO(5)} \subseteq \frac{PSU(2,2|4)}{SO(4,1)\times SO(5)}. \qquad \label{AdS5S5supercoset}
\end{IEEEeqnarray}
Metsaev and Tseytlin \cite{MetsaevTseytlin98} formulated type IIB superstring theory on AdS$_5\times\text{S}^5$ as a WZW sigma model. By the same token, type IIA superstring theory on AdS$_4\times\CP^3$ has been formulated as a WZW sigma model on the coset superspace \cite{ArutyunovFrolov08, Stefanski08}:
\begin{IEEEeqnarray}{c}
\text{AdS}_4\times\CP^3 = \frac{SO(3,2)}{SO(3,1)} \times \frac{SO(6)}{U(3)} \subseteq \frac{Osp(2,2|6)}{SO(3,1)\times U(3)}, \qquad \label{AdS4CP3supercoset}
\end{IEEEeqnarray}
although the equivalence of the sigma model on the supercoset space \eqref{AdS4CP3supercoset} to the full GS superstring action on AdS$_4\times\CP^3$ is much more subtle and in fact only partial in this case \cite{GomisSorokinWulff08, SorokinWulff11}. \\
\indent The two AdS/CFT supercosets \eqref{AdS5S5supercoset}--\eqref{AdS4CP3supercoset} are semi-symmetric superspaces. A supercoset $G/H_0$ is a semi-symmetric space when the Lie algebra of the denominator group $H_0 \subset G$ is $\Z_4$ invariant \cite{Serganova83}. Semi-symmetric spaces are known to give rise to classically integrable nonlinear sigma models.\footnote{The same is true for symmetric spaces, that is when the Lie algebra of the denominator group $H_0 \subset G$ is $\Z_2$ invariant \cite{EichenherrForger79}. See e.g.\ the set of lectures \cite{Zarembo17} for more details and definitions.} The nonlinear sigma model that describes Green-Schwarz superstrings on the semi-symmetric spaces \eqref{AdS5S5supercoset}--\eqref{AdS4CP3supercoset} reads:
\begin{IEEEeqnarray}{c}
S = -\frac{T_2}{2} \int \ell^2\text{str}\left[J^{(2)}\wedge \hodge J^{(2)} + J^{(1)} \wedge J^{(3)}\right], \qquad T_2 \equiv \frac{1}{2\pi\alpha'}, \label{MetsaevTseytlinAction}
\end{IEEEeqnarray}
where $T_2$ is the string tension and $\ell$ is the AdS radius. The $\Z_4$ components of the current $J^{(n)}$ will be defined below. Parameter matching between the two sides of the AdS/CFT dualities \eqref{AdS5CFT4duality}, \eqref{AdS4CFT3duality} yields the following identifications:
\begin{IEEEeqnarray}{l}
{\color{red}\text{AdS}_5\times\text{S}^5}: \ \lambda = \frac{\ell^4}{\alpha'^2} \ \Leftrightarrow \ T_2 = \frac{\sqrt{\lambda}}{2\pi\ell^2} \label{ParameterMetchingAdS5xS5} \\[6pt]
{\color{red}\text{AdS}_4\times\CP^3}: \ \lambda = \frac{\ell^4}{2\pi^2\alpha'^2} \ \Leftrightarrow \ T_2 = \frac{1}{\ell^2}\sqrt{\frac{\lambda}{2}}. \qquad \quad \label{ParameterMetchingAdS4xCP3}
\end{IEEEeqnarray}
By definition, semi-symmetric superspaces are endowed with a $\Z_4$ grading. In other words, every element of a semi-symmetric superspace affords a $\Z_4$ decomposition. The (moving-frame) current $J$ is decomposed into four components $J^{(n)}$, three of which ($J^{(1,2,3)}$) showed up in the action \eqref{MetsaevTseytlinAction}:
\begin{IEEEeqnarray}{c}
J \equiv \mathfrak{g}^{-1}d\mathfrak{g} = J^{(0)} + J^{(1)} + J^{(2)} + J^{(3)}, \qquad \Omega\big[J^{(n)}\big] = i^n J^{(n)}, \label{MovingFrameCurrent}
\end{IEEEeqnarray}
where $\mathfrak{g}$ is an element of the corresponding superspace. The $\Z_4$ automorphism $\Omega$ is defined as
\begin{IEEEeqnarray}{ll}
\Omega\left(M\right) = -\K M^{\text{st}}\K^{-1}, \label{Z4automorphism}
\end{IEEEeqnarray}
where the matrix $\K$ is given in our two cases of interest (namely AdS$_5\times\text{S}^5$ and AdS$_4\times\CP^3$) by
\begin{IEEEeqnarray}{ll}
{\color{red}\text{AdS}_5\times\text{S}^5}: \ \K = \left[\begin{array}{cc} K_4 & 0 \\ 0 & K_4 \end{array}\right], \quad K_4 =\gamma_{13}, \quad M \in \mathfrak{psu}(2,2|4) \qquad \label{Z4automorphismAdS5xS5} \\[6pt]
{\color{red}\text{AdS}_4\times\CP^3}: \ \K = \left[\begin{array}{cc} K_4 & 0 \\ 0 & -K_6\end{array}\right], \quad K_4 = \gamma_{12}, \quad K_6 \equiv \textrm{I}_3 \otimes \left(i\sigma_{2}\right), \quad M \in \mathfrak{osp}(2,2|6). \label{Z4automorphismAdS4xCP3} \qquad
\end{IEEEeqnarray}
The definition of the superalgebras $\mathfrak{psu}(2,2|4)$ and $\mathfrak{osp}(2,2|6)$ (elements of which are the matrices $M$) can be found in many places, see e.g.\ \cite{ArutyunovFrolov09b, ArutyunovFrolov08}. The 5d gamma matrices (appendix \ref{Appendix:5dGammaMatrices}) which are relevant for AdS$_5\times\text{S}^5$ are given in \eqref{GammaMatrices4}--\eqref{GammaMatrices6}, while the 4d gamma matrices (appendix \ref{Appendix:4dGammaMatrices}) which are relevant for AdS$_4\times\CP^3$ are given by \eqref{GammaMatrices2}--\eqref{GammaMatrices3}. \\
\indent Differentiating the definition \eqref{MovingFrameCurrent} of the moving-frame current $J$, it can be shown that its curvature vanishes. Moreover, the equations of motion that follow from the superstring action \eqref{MetsaevTseytlinAction} imply that the Lax connection $L$ is also flat:
\begin{IEEEeqnarray}{c}
dJ + J \wedge J = 0, \qquad dL + L \wedge L = 0, \label{FlatnessConditionsMovingFrame}
\end{IEEEeqnarray}
where $L$ is given by
\begin{IEEEeqnarray}{c}
L\left(\y\right) = J^{(0)} + \frac{{\y}^2+1}{{\y}^2-1}\,J^{(2)} - \frac{2{\y}}{{\y}^2-1} \hodge J^{(2)} + \sqrt{\frac{{\y}+1}{{\y}-1}} \, J^{(1)} + \sqrt{\frac{\y-1}{\y+1}} \, J^{(3)}, \label{LaxConnectionMovingFrame}
\end{IEEEeqnarray}
and $\y$ is the spectral parameter. We may also define the fixed-frame (or left) current $j$
\begin{IEEEeqnarray}{c}
j \equiv \mathfrak{g} J \mathfrak{g}^{-1} = d\mathfrak{g}\,\mathfrak{g}^{-1} = j^{(0)} + j^{(1)} + j^{(2)} + j^{(3)}, \qquad j^{(n)} \equiv \mathfrak{g} J^{(n)} \mathfrak{g}^{-1}. \label{FixedFrameCurrent}
\end{IEEEeqnarray}
In this frame, the (fixed-frame) Lax connection $a$ reads
\begin{IEEEeqnarray}{c}
a\left(\y\right) = \frac{2}{\y^2 - 1} \left(j^{(2)} - \y \hodge j^{(2)}\right) + \left({\z} - 1\right) \, j^{(1)} + \left(\frac{1}{{\z}} - 1\right) \, j^{(3)}, \qquad \z \equiv \sqrt{\frac{\y+1}{\y-1}}. \qquad \label{LaxConnectionFixedFrame}
\end{IEEEeqnarray}
This form (rather than the moving-frame connection \eqref{LaxConnectionMovingFrame}) was originally used by Bena, Polchinski and Roiban \cite{BenaPolchinskiRoiban03} to show that the MT sigma model \eqref{MetsaevTseytlinAction}, \eqref{ParameterMetchingAdS5xS5}, \eqref{Z4automorphismAdS5xS5} is classically integrable on AdS$_5\times\text{S}^5$. In terms of the fixed-frame current $j$ and the Lax connection $a$, the flatness conditions \eqref{FlatnessConditionsMovingFrame} take the form:
\begin{IEEEeqnarray}{c}
dj - j \wedge j = 0, \qquad da + a \wedge a = 0. \label{FlatnessConditionsFixedFrame}
\end{IEEEeqnarray}
\paragraph{Closed strings} The mere existence of a flat Lax connection directly implies classical Liouville integrability for the string sigma model \eqref{MetsaevTseytlinAction}. To generate the infinite set of conserved charges, we define the (single-row) monodromy matrix
\begin{equation}
\M(\sigma_1,\sigma_2,\tau;\y) \equiv \Pexp\left(\int_{\sigma_1}^{\sigma_2} ds \,a_\sigma (s,\tau;\y)\right), \label{MonodromyMatrix1}
\end{equation}
as the holonomy (or Wilson line) of the Lax connection. Taking the derivative with respect to worldsheet time $\tau$ and using the flatness of the Lax connection \eqref{FlatnessConditionsFixedFrame} we are led to
\begin{equation}
\partial_{\tau}\M(\sigma_1,\sigma_2,\tau;\y) = \M(\sigma_1,\sigma_2,\tau;\y) \, a_\tau(\sigma_2,\tau;\y) - a_\tau(\sigma_1,\tau;\y) \, \M(\sigma_1,\sigma_2,\tau;\y). \label{MonodromyMatrix2}
\end{equation}
It readily follows that the corresponding transfer matrix $\text{str}\M$ is conserved upon imposing periodic boundary conditions, i.e.\ in the case of closed strings,
\begin{equation}
\partial_{\tau}\text{str}\M(0,2\pi,\tau;\y) = 0.
\end{equation}
More charges are obtained by forming powers of the monodromy matrix $\text{str}\M^n$ for $n = 1,2,\ldots$ These are also conserved. The same conclusion can also be reached by considering the (super) determinant of the monodromy matrix in \eqref{MonodromyMatrix2}. The upshot is that the conserved charges are fully encoded in the spectrum of the infinite-dimensional monodromy matrix $\M$. In other words, the motion integrals of the string sigma model \eqref{MetsaevTseytlinAction} are given by the eigenvalues of the monodromy matrix $\M$. By Taylor-expanding $\M$ in inverse powers of the spectral parameter $x \rightarrow \infty$, we may obtain analytic expressions for the conserved charges:
\begin{IEEEeqnarray}{ll}
\M(\y) &= \mathbbm{1} - \frac{2}{\y}\int_0^{2\pi} ds \left[j^{(2)}_{\tau} + \frac{j^{(3)}_{\sigma} - j^{(1)}_{\sigma}}{2}\right] + \frac{2}{\y^2} \Bigg\{\int_0^{2\pi} ds \bigg[j^{(2)}_{\sigma} + \frac{j^{(1)}_{\sigma} + j^{(3)}_{\sigma}}{4}\bigg] + \nonumber \\[6pt]
&\hspace{0.4cm} + \int_0^{2\pi}\int_0^{s} ds ds' \left[2j^{(2)'}_{\tau} j^{(2)}_{\tau} + \ldots\right]\Bigg\} - \ldots \equiv \\[6pt]
& \equiv \exp\left[2\sum_{r = 0}^{2\pi} \left(-\frac{1}{\y}\right)^{r+1} \Q_{r}\right] = \mathbbm{1} - \frac{2}{\y}\,\Q_{0} + \frac{2}{\y^2}\left(\Q_{1} + \Q_{0}^2\right) - \ldots \qquad
\end{IEEEeqnarray}
As it turns out, the conserved charges of the Green-Schwarz sigma model \eqref{MetsaevTseytlinAction} Poisson-commute \cite{MikhailovSchaferNameki07b, Magro08}, so that classical Liouville integrability also holds in the strong sense for this system. Because the sigma model on the supercoset \eqref{AdS4CP3supercoset} is not fully equivalent to the Green-Schwarz superstring on AdS$_4\times\CP^3$ as we mentioned above, further evidence for the classical integrability of type IIA superstring theory on AdS$_4\times\CP^3$ has been provided in \cite{SorokinWulff10}. \\
\indent Showing that the string sigma models on AdS$_5\times\text{S}^5$ and AdS$_4\times\CP^3$ are classically integrable is an important step forward but certainly not the end of the story. Closed string states in the AdS/CFT dualities \eqref{AdS5CFT4duality} and \eqref{AdS4CFT3duality} are dual to single-trace operators of the gauge theory. Because single-trace operators are described by quantum integrable spin chains in the planar limit, the dual string sigma model \eqref{MetsaevTseytlinAction} is also expected to describe quantum integrable string states on both supercosets \eqref{AdS5S5supercoset} and \eqref{AdS4CP3supercoset}. This is still unproven however. To the best of our knowledge, there have been two main approaches to the issue of quantum integrability of the MT sigma model \eqref{MetsaevTseytlinAction}, \eqref{ParameterMetchingAdS5xS5}, \eqref{Z4automorphismAdS5xS5}. First, the absence of particle production/annihilation and the factorization of multi-particle scattering has been shown up to next-to-leading order in the strong coupling expansion of the AdS$_5\times\text{S}^5$ sigma model in \cite{HentschelPlefkaSundin07, GiangrecoMarottaPulettiKloseSax07}. Secondly, it has been conjectured in \cite{Berkovits04a} (based on the pure spinor formalism) that the conserved charges of the AdS$_5\times\text{S}^5$ string sigma model exist to all orders in $\alpha'$. This property has been confirmed up to one loop in \cite{MikhailovSchaferNameki07a}.
\paragraph{Open strings} In contrast to closed strings which are associated with periodic boundary conditions, open strings have their endpoints on certain dynamical objects of string theory which are called D-branes. Open strings can be associated to various types of open boundary conditions such as Dirichlet, Neumann or mixed Neumann-Dirichlet. Commonly, the D-brane cuts the string worldsheet at a constant-$\sigma$ section (say at $\sigma = 0$). \\
\indent Opening out a classically integrable closed string by placing its endpoints on a D-brane does not necessarily preserve its integrability. To examine open string (or D-brane) integrability for the string sigma model \eqref{MetsaevTseytlinAction} we need to modify our previous formalism in order to accommodate the boundary at $\sigma = 0$. Let us define the double-row monodromy matrix \cite{Sklyanin87a},
\begin{equation}
\T(\tau;\y) = \M^{\text{st}}(\tau;-\y) \cdot \U(\tau;\y) \cdot \M(\tau;\y), \label{DoubleRowMonodromyMatrix}
\end{equation}
by \quotes{folding} two (single-row) monodromy matrices $\M$ via a reflection matrix $\U$. The reflection matrix $\U$ generally depends on the spectral parameter $\y$ and is \quotes{dynamical}, i.e.\ it depends indirectly on the worldsheet time $\tau$ through its explicit dependence on the string embedding coordinates $\X_M(\tau)$. Note that the folding of the two holonomies $\M$ and the reflection matrix $\U$ occurs at the same point of the string worldsheet, i.e.\ $\left\{\tau,\sigma = 0\right\}$. Attaching holonomies at different worldsheet points is also possible and relevant to the presence of crosscaps instead of boundaries/defects in target space \cite{Gombor22b, Ekman22, Gombor22c}. \\
\indent The two single-row monodromy matrices $\M$ which enter the definition \eqref{DoubleRowMonodromyMatrix} contain the infinite set of conserved charges of the classically integrable closed string, as we have explained above. The presence of a reflection matrix $\U$ in between them however, eliminates a great deal of the conserved charges which will be absent from the spectrum of the double-row monodromy matrix $\T$. Taking the derivative as before,
\begin{equation}
\dot{\T}(\y) = \M^{\text{st}}(-\y) \cdot \left(\dot{\U}(\y) - a^{\text{st}}_\tau(-\y)\cdot\U(\y) - \U(\y)\cdot a_\tau(\y)\right) \cdot \M(\y),
\end{equation}
we conclude that the corresponding transfer matrix will be conserved if
\begin{equation}
\dot{\U}(\y) \eq a^{\text{st}}_\tau(-\y)\cdot\U(\y) + \U(\y)\cdot a_\tau(\y), \label{DbraneIntegrabilityCondition1}
\end{equation}
where $a(\y)$ stands for the flat fixed-frame Lax connection \eqref{LaxConnectionFixedFrame}. The exclamation mark over the equals sign ($\eq$) will henceforth mark an equation which is valid on the boundary (e.g.\ a boundary condition). In terms of the fixed-frame current $j$, the D-brane integrability condition becomes:
\begin{IEEEeqnarray}{ll}
\dot{\U} &\eq \frac{2}{\y^2 - 1} \cdot \left\{j^{(2)\,\st}_{\tau} \, \U + \U \, j^{(2)}_{\tau}\right\} + \frac{2\y}{\y^2 - 1} \cdot \left\{j^{(2)\,\st}_{\sigma} \, \U - \U \, j^{(2)}_{\sigma}\right\} + \nonumber \\[6pt]
& + \left({\z} - 1\right) \cdot \left\{\U \, j^{(1)}_{\tau} + j^{(3)\,\st}_{\tau} \, \U\right\} + \left(\frac{1}{{\z}} - 1\right) \cdot \left\{\U \, j^{(3)}_{\tau} + j^{(1)\,\st}_{\tau} \, \U\right\}. \qquad \label{DbraneIntegrabilityCondition2}
\end{IEEEeqnarray}
There are two interesting special cases of the integrability condition \eqref{DbraneIntegrabilityCondition2}. The first one was considered by Dekel and Oz \cite{DekelOz11b} and occurs when the reflection matrix $\U$ is a constant matrix that does not depend on the spectral parameter $\y$ or the worldsheet time $\tau$:
\begin{IEEEeqnarray}{l}
j^{(2)\,\st}_{\tau} \, \U + \U \, j^{(2)}_{\tau} \eq j^{(2)\,\st}_{\sigma} \, \U - \U \, j^{(2)}_{\sigma} \eq 0, \quad
\U \, j^{(1)}_{\tau} + j^{(3)\,\st}_{\tau} \, \U \eq \U \, j^{(3)}_{\tau} + j^{(1)\,\st}_{\tau} \, \U \eq 0. \qquad \label{DbraneIntegrabilityCondition3}
\end{IEEEeqnarray}
The second interesting special case of \eqref{DbraneIntegrabilityCondition2} is the bosonic one. In this case, the two fermionic components $j^{(1,3)}$ of the fixed-frame current \eqref{FixedFrameCurrent} vanish, while supertransposes get replaced by simple transposes,
\begin{IEEEeqnarray}{c}
\dot{\U} \eq \frac{2}{\y^2 - 1} \cdot \left\{j^{(2)\,\text{t}}_{\tau} \, \U + \U \, j^{(2)}_{\tau}\right\} + \frac{2\y}{\y^2 - 1} \cdot \left\{j^{(2)\,\text{t}}_{\sigma} \, \U - \U \, j^{(2)}_{\sigma}\right\}. \qquad \label{DbraneIntegrabilityCondition4}
\end{IEEEeqnarray}
Moreover, the reflection matrix $\U$ is block diagonal with its top-left block belonging in AdS and its lower-right block belonging in either S$^5$ or $\CP^3$:
\begin{equation}
\U_{\text{AdS}\times\text{S}} = \K\cdot\left[\begin{array}{cc} \U_{\text{AdS}} & 0 \\ 0 & \U_{\text{S}}\end{array}\right], \qquad \U_{\text{AdS}\times\CP} = \left[\begin{array}{cc} \U_{\text{AdS}} & 0 \\ 0 & \U_{\CP}\end{array}\right]. \label{ReflectionMatrices}
\end{equation}
A given set of open string BCs of the Green-Schwarz sigma model \eqref{MetsaevTseytlinAction} is integrable, if a reflection matrix $\U$ can be found such that it satisfies the integrability condition \eqref{DbraneIntegrabilityCondition2} upon imposing the BCs.
\subsection[\texorpdfstring{String integrability in AdS/{\color{red}d}CFT}{String integrability in AdS/dCFT}]{String integrability in AdS/{\color{red}d}CFT \label{SubSection:StringIntegrabilitydCFT}}
\noindent Holographic dCFTs were first realized in the context of the AdS$_5$/CFT$_4$ correspondence by Karch and Randall \cite{KarchRandall01a, KarchRandall01b}, in an effort to provide specific examples of gravity localization on AdS branes. The work of Karch and Randall suggested that probe D-branes on the string theory side of holographic dualities (such as \eqref{AdS5CFT4duality} and \eqref{AdS4CFT3duality}) correspond to defects on the dual gauge theory (CFT) side. The codimensionality of the CFT defect is given by the codimensionality of the AdS component of the probe D-brane. The defect can host either new degrees of freedom (which couple to those in the ambient space of the CFT) or BCs for the CFT fields. According to the AdS/{\color{red}d}CFT correspondence, the two deformed theories (i.e.\ string theory with a probe brane and the dCFT) are holographic duals. Two of the most studied cases are the D5 \cite{DeWolfeFreedmanOoguri01, ErdmengerGuralnikKirsch02, DeWolfeMann04} and the D7 \cite{DavisKrausShah08, Rey08, MyersWapler08} defect branes of type IIB string theory on AdS$_5\times\text{S}^5$. These are holographically dual to (codimension-1) domain wall versions of $\N = 4$ SYM theory \eqref{LagrangianSYM} which are respectively known as the D3-D5 and the D3-D7 dCFT. \\
\indent One-point functions of scalar local operators are among the most important observables in a defect CFT. From them (and the CFT data) one can in principle determine all scalar correlation functions of the theory. Details about correlation functions and the bootstrap method in CFTs and codimension-1 dCFTs can be found in \S\ref{SubSection:ConformalCorrelationFunctions} below. Apart from global symmetries, defects usually also break gauge symmetries. In such cases, the fields of the CFT acquire vacuum expectation values (vevs) so that one-point functions of gauge invariant operators can be seen as Higgs condensates. The vevs are given by classical solutions of the theory's equations of motion (aka fuzzy funnel solutions). For the D3-D5 \cite{NagasakiTanidaYamaguchi11, NagasakiYamaguchi12} and the D3-D7 \cite{KristjansenSemenoffYoung12b} dCFT, the computed values of one-point functions agree with the string theory calculations, thus confirming the validity of the AdS/{\color{red}d}CFT duality in these cases. The computation of defect correlation functions in string theory is reviewed in \S\ref{SubSection:HolographicCorrelatorsdCFTs} below. \\
\indent Integrability methods were introduced in the AdS/{\color{red}d}CFT correspondence by de Leeuw, Kristjansen and Zarembo \cite{deLeeuwKristjansenZarembo15}. Starting from a \quotes{solvable} holographic gauge theory (such as $\N =4$ SYM), de Leeuw, Kristjansen and Zarembo considered gauge invariant operators that correspond to Bethe (or highest-weight) states of the integrable spin chain which describes the dilatation operator of the CFT. It is generally safe to assume that the dilatation operator will not be altered by the presence of a defect, although this property has so far only been shown at one-loop order in the D3-D5 case \cite{DeWolfeMann04, IpsenVardinghus19}. Defect CFT one-point functions are then given by overlaps of (normalized) Bethe states and matrix product states (MPSs) which implement the Higgs mechanism for the fields. At tree-level, one-point functions in the D3-D5 and the $SO(5)$ symmetric version of the D3-D7 dCFT are given (as functions of the Bethe roots) by closed-form determinant formulas \cite{Buhl-MortensenLeeuwKristjansenZarembo15, deLeeuwKristjansenMori16, deLeeuwKristjansenLinardopoulos16, deLeeuwKristjansenLinardopoulos18a, deLeeuwGomborKristjansenLinardopoulosPozsgay19, KristjansenMullerZarembo20a}. In perturbation theory, one-loop one-point functions have been computed for the D3-D5 and the D3-D7 dCFT (either $SO(5)$ or $SU(2)\times SU(2)$ symmetric) \cite{Buhl-MortensenLeeuwIpsenKristjansenWilhelm16a, Buhl-MortensenLeeuwIpsenKristjansenWilhelm16c, GimenezGrauKristjansenVolkWilhelm18, GimenezGrauKristjansenVolkWilhelm19}, while asymptotic generalizations to all-loop orders have been proposed for the D3-D5 system in \cite{Buhl-MortensenLeeuwIpsenKristjansenWilhelm17a, GomborBajnok20a, GomborBajnok20b, KristjansenMullerZarembo20b, KristjansenMullerZarembo21}. Two-point functions in the D3-D5 dCFT have been studied in \cite{deLeeuwIpsenKristjansenVardinghusWilhelm17, Widen17, deLeeuwKristjansenLinardopoulosVolk23, BaermanChalabiKristjansen24}. See also the reviews \cite{deLeeuwIpsenKristjansenWilhelm17, deLeeuw19, Linardopoulos20, KristjansenZarembo24a} and the theses \cite{BuhlMortensen17, Vardinghus19, Widen19, Gombor20, Volk21}. \\
\indent Another holographic dCFT which has attracted a lot of attention is dual to type IIA string theory on AdS$_4\times\mathbb{CP}^3$ in the presence of a probe D4-brane. This system is holographically dual to a domain wall version of ABJM theory \eqref{LagrangianABJM} that is known as the D2-D4 dCFT. Tree-level one-point functions of Bethe states have been studied for the ABJM defect in \cite{Gombor21, KristjansenVuZarembo21, GomborKristjansen22}, finding closed-form determinant formulas for the overlaps of Bethe states with matrix product states, as well as agreement with the corresponding calculations at strong coupling. The spectrum of quantum fluctuations for the ABJM defect was studied recently in \cite{KristjansenQianSu24}. \\
\indent The existence of closed-form expressions for one-point functions of Bethe states is a prime indication that the holographic dCFT is \quotes{solvable}. We have mentioned in the introduction of \S\ref{Section:StringIntegrabilityHolographicDefects} that this constitutes one of the three integrable quench criteria which were proposed by Piroli, Pozsgay and Vernier \cite{PiroliPozsgayVernier17} and can be applied to dCFTs thanks to the Ghoshal-Zamolodchikov argument \cite{GhoshalZamolodchikov93}. Another quench criterion requires the annihilation of the matrix product state (MPS) by the parity-odd conserved charges of the integrable spin chain that describes the dilatation operator:
\begin{IEEEeqnarray}{c}
\mathbb{Q}_{2s+1}\left|\text{MPS}\right\rangle = 0, \qquad s = 1,2,\ldots \qquad \label{QuenchIntegrabilityCriterion}
\end{IEEEeqnarray}
Equivalently, one-point functions of Bethe states can only be nonzero when all the Bethe roots are fully balanced, that is they appear in pairs of opposite signs. So far, the D5, the $SO(5)$ symmetric D7, and the D4 defect branes have been found to satisfy all the integrable quench criteria \cite{PiroliPozsgayVernier17, deLeeuwKristjansenLinardopoulos18a, deLeeuwGomborKristjansenLinardopoulosPozsgay19, GomborKristjansen22}. On the flipside, defect branes such as the NS5-brane, the $\beta$-deformed D5-brane and the $SU(2)\times SU(2)$ symmetric D7-brane, are most probably non-integrable \cite{Rapcak15, Widen18, deLeeuwKristjansenVardinghus19}. \\
\indent At strong coupling, the dual string theory becomes weakly coupled and contains open string states which are emitted from the probe D-brane. Open strings are holographically dual to gauge invariant operators which contain defect fields and are described by open spin chains. In the D3-D5 case the open spin chain has been shown to be integrable at one loop order \cite{DeWolfeMann04, IpsenVardinghus19}. However, if integrability persists to all orders of perturbation theory, it should manifest itself at strong coupling as classical open string integrability. On the other hand, open strings have their endpoints on D-branes which impose boundary conditions on their dynamics and may break or preserve their integrability. \\
\indent D-brane integrability (i.e.\ the integrability of open strings attached to D-branes) can be examined by means of the double-row monodromy matrix formalism which we described in \S\ref{SubSection:StringIntegrabilitydCFT} above. Indeed, substantial evidence which supports classical integrability for the D5 and the D4 brane has been found \cite{DekelOz11b, LinardopoulosZarembo21, Linardopoulos22}.\footnote{See also \cite{MannVazquez06, CorreaYoung08, CorreaRegelskisYoung11, MacKayRegelskis11b} for earlier work on the subject.} This evidence will be reviewed below. For simplicity we will only consider the bosonic case \eqref{DbraneIntegrabilityCondition4}. In the absence of fermions, the GS sigma model action \eqref{MetsaevTseytlinAction} reduces to the string Polyakov action,
\begin{IEEEeqnarray}{c}
S = -\frac{T_2}{2}\int d\tau d\sigma \sqrt{-\gamma}\gamma^{ab} \partial_a \X^{M} \partial_b \X^{N}g_{MN} + \int d\tau \left[A_{\textrm{i}}\dot{\X}_{\textrm{i}}\right]_{\sigma=0}, \label{PolyakovAction1}
\end{IEEEeqnarray}
where the boundary term at $\sigma = 0$ accounts for the presence of an Abelian flux field $A_{\textrm{i}}$ on the worldvolume of the D-brane. $\X_M$ are the target-space coordinates aka the embedding coordinates of the string worldsheet in either AdS$_5\times\text{S}^5$ or AdS$_4\times\CP^3$. These can be either longitudinal ($\X_{\textrm{i}}$) or transverse ($\X_{\textrm{a}}$) with respect to the D-brane. \\
\indent Because the D-brane cuts the string worldsheet at a constant-$\sigma$ section ($\sigma = 0$), the variations of the initial ($\tau_1$) and the final ($\tau_2$) string states vanish:
\begin{IEEEeqnarray}{c}
\delta \X_M\left(\tau_1,\sigma = 0\right) \eq \delta \X_M\left(\tau_2,\sigma = 0\right) \eq 0.
\end{IEEEeqnarray}
By varying the bosonic string action \eqref{PolyakovAction1}, we derive the following string boundary conditions on the D-brane:
\begin{IEEEeqnarray}{rl}
\acute{\X}_{\textrm{i}} - 2\pi\alpha' \, F_{\textrm{ij}} \, \dot{\X}_{\textrm{j}} \eq 0 \qquad &\text{(longitudinal: Neumann-Dirichlet)} \label{StringBoundaryConditions1} \\[6pt]
\dot{\X}_{\textrm{a}} \eq 0 \qquad &\text{(transverse: Dirichlet)}, \label{StringBoundaryConditions2}
\end{IEEEeqnarray}
which are mixed Neumann-Dirichlet for the longitudinal and Dirichlet for the transverse coordinates of the string.
\subsubsection[The D3-D5 intersection]{The D3-D5 intersection \label{SubSubSection:D3D5intersection}}
\noindent The D3-D5 intersection is made up from $N_c$ coincident D3-branes and $M$ coincident D5-branes. We will only be interested in the case of a single probe D5-brane, i.e.\ we set $M=1$. The relative orientation of the branes in flat space is shown in table \ref{Table:D3D5system} below.
\renewcommand{\arraystretch}{1.1}\setlength{\tabcolsep}{5pt}
\begin{table}[H]\begin{center}\begin{tabular}{|c||c|c|c|c|c|c|c|c|c|c|}
\hline
& $x_0$ & $x_1$ & $x_2$ & $x_3$ & $x_4$ & $x_5$ & $x_6$ & $x_7$ & $x_8$ & $x_9$ \\ \hline
\text{D3} & $\bullet$ & $\bullet$ & $\bullet$ & $\bullet$ &&&&&& \\ \hline
\text{D5} & $\bullet$ & $\bullet$ & $\bullet$ & & $\bullet$ & $\bullet$ & $\bullet$&&& \\ \hline
\end{tabular}\caption{Brane orientation for the D3-D5 intersection in flat space.\label{Table:D3D5system}}\end{center}\end{table}
\vspace{-.5cm}\noindent The D3-branes source the fields of type IIB supergravity. The graviton field develops a singularity at the location of the D3-branes. Very close to the horizon, the spacetime becomes AdS$_5\times\text{S}^5$ hosting type IIB string theory, i.e.\ it is the system on the rhs of the AdS$_5$/CFT$_4$ duality \eqref{AdS5CFT4duality}. \\
\indent Because the D5 is a probe brane, it does not backreact with the geometry of D3-branes. It follows that the near-horizon region, which now contains type IIB string theory on AdS$_5\times\text{S}^5$ (rhs of \eqref{AdS5CFT4duality}) as well as the probe D5-brane, should remain holographic. According to the AdS$_5$/{\color{red}d}CFT$_4$ correspondence \cite{KarchRandall01a, KarchRandall01b}, the dual field theory is a defect superconformal field theory (SCFT) which consists of $\N = 4$ SYM theory \eqref{LagrangianSYM} on the lhs of \eqref{AdS5CFT4duality}, coupled to a $2+1$ dimensional SCFT which lives on a codimension-1 defect. The total action of the dCFT takes the form \cite{DeWolfeFreedmanOoguri01}:\footnote{Supersymmetry of the D3-D5 system has been studied in \cite{DomokosRoyston22}.}
\begin{IEEEeqnarray}{ll}
S = S_{\N = 4} + S_{2+1}. \label{DefectAction}
\end{IEEEeqnarray}
Defect correlation functions can be computed at weak coupling with the aid of domain walls \cite{NagasakiTanidaYamaguchi11, deLeeuwKristjansenZarembo15}, see the reviews \cite{deLeeuwIpsenKristjansenWilhelm17, deLeeuw19, Linardopoulos20, KristjansenZarembo24a} for more. See also \S\ref{SubSection:HolographicCorrelatorsdCFTsWeak}. At strong coupling, defect correlators can be computed with semiclassical strings \cite{NagasakiYamaguchi12, KristjansenSemenoffYoung12b, GeorgiouLinardopoulosZoakos23}. Details will be provided in \S\ref{SubSection:HolographicCorrelatorsdCFTsStrong} below. Non-perturbative methods based on supersymmetric localization have also been developed \cite{RobinsonUhlemann17, Robinson17, Wang20a, KomatsuWang20, BeccariaCaboBizet23}. \\
\indent The embedding of the probe D5-brane in AdS$_5\times\text{S}^5$ can be found by solving the equations of motion which arise from the action of the D5-brane (spelled out in \eqref{D5braneAction}). The D5-brane wraps an AdS$_4\times\text{S}^2$ geometry which is parametrized by \cite{KarchRandall01b}:
\begin{IEEEeqnarray}{c}
x_3 = \kappa \cdot z, \qquad \kappa \equiv \frac{\pi \kk}{\sqrt{\lambda}} \equiv \tan\alpha, \qquad \psi = 0, \label{D5braneEmbedding1}
\end{IEEEeqnarray}
where the metric of AdS$_5$ is given by \eqref{MetricAdS5xS5} (the Minkowski signature is relevant here), and the metric of the unit 5-sphere in \eqref{MetricS5so3so3}. Setting $\psi = 0$ in \eqref{MetricS5so3so3}--\eqref{CartesianCoordinatesS5c}, we obtain the D5-brane embedding in the Cartesian coordinate system:
\begin{IEEEeqnarray}{c}
x_4^2 + x_5^2 + x_6^2 = \ell^2, \qquad x_7 = x_8 = x_9 = 0, \label{D5braneEmbedding2}
\end{IEEEeqnarray}
by also including the 5-sphere radius $\ell$. The D5-brane preserves the $SO(3,2)\times SO(3)\times SO(3)$ subgroup of $SO(4,2)\times SO(6)$. The corresponding supersymmetry is halved, i.e.\ the supergroup $PSU(2,2|4)$ becomes $Osp(4|4)$ and the D-brane is half-BPS \cite{SkenderisTaylor02a}. The D5-brane embedding \eqref{D5braneEmbedding1} is stable \cite{KarchRandall01b}; although there is a tachyonic mode in the fluctuations of the angle $\psi$ in \eqref{MetricS5so3so3}, it does not violate the Breitenlohner-Freedman (BF) bound \cite{BreitenlohnerFreedman82a}. \\
\indent The stability of the system is enhanced by the presence of $\kk$ units of magnetic flux through S$^2$,
\begin{IEEEeqnarray}{c}
\int_{\text{S}^2}\frac{F}{2\pi} = \kk, \qquad F = dA = \frac{\kk}{2} \cdot d\cos\theta\wedge d\chi \qquad \text{(first Chern class)}, \label{FieldStrengthD5a}
\end{IEEEeqnarray}
so that the gauge potential which corresponds to the field strength \eqref{FieldStrengthD5a} has the form:
\begin{IEEEeqnarray}{c}
A = \frac{\kk}{2} \left(c + \cos\theta\right)d\chi. \label{PotentialD5}
\end{IEEEeqnarray}
Changing variables, we may express the field strength $F$ in \eqref{FieldStrengthD5a} in Cartesian coordinates as
\begin{IEEEeqnarray}{c}
F = dA = -\frac{\kk}{2}\,\left(x_4 \, dx_5 \wedge dx_6 + x_5 \, dx_6 \wedge dx_4 + x_6 \, dx_4 \wedge dx_5\right). \qquad \label{FieldStrengthD5b}
\end{IEEEeqnarray}
In component form, the field strength reads
\begin{IEEEeqnarray}{c}
F_{ij} = -\frac{\kk}{2}\,\epsilon_{ijl}x_l = \partial_i A_j - \partial_j A_i, \qquad i,j,l = 4,5,6. \label{FieldStrengthD5c}
\end{IEEEeqnarray}
The flux \eqref{FieldStrengthD5a}--\eqref{FieldStrengthD5b} forces exactly $\kk$ of the D3-branes to end on one side of the D5-brane. In the nonzero flux case ($\kk\neq 0$), the dual field theory has no boundary degrees of freedom \cite{GaiottoWitten08a, IpsenVardinghus19}. The codimension-1 defect only serves as a host of the BCs for the (ambient) fields of $\N = 4$ SYM \eqref{LagrangianSYM}.
\paragraph{String boundary conditions on D5} The longitudinal and transverse string coordinates can be directly identified from the D5-brane parametrization \eqref{D5braneEmbedding1}--\eqref{D5braneEmbedding2}. Obviously,
\begin{IEEEeqnarray}{lrl}
\text{longitudinal: } &x_0,x_1,x_2, \ x_3\sin\alpha + z\cos\alpha, \quad &x_4,x_5,x_6 \\
\text{transverse: } &x_3\cos\alpha - z\sin\alpha, \quad & x_7,x_8,x_9
\end{IEEEeqnarray}
and the bosonic string boundary conditions \eqref{StringBoundaryConditions1}--\eqref{StringBoundaryConditions2} take the following form in AdS$_5$:
\begin{IEEEeqnarray}{l}
\acute{x}_{0,1,2} \eq \acute{z} + \kappa \, \acute{x}_3 \eq 0 \quad \text{(Neumann)} \qquad \& \qquad \dot{x}_3 - \kappa \, \dot{z} \eq 0 \quad \text{(Dirichlet)}. \label{D5braneStringBCsAdS}
\end{IEEEeqnarray}
Taking into account the field strength \eqref{FieldStrengthD5c}, the string boundary conditions \eqref{StringBoundaryConditions1}--\eqref{StringBoundaryConditions2} take the following form on S$^5$:
\begin{IEEEeqnarray}{lll}
\acute{x}_4 \eq \kappa \left(x_5 \, \dot{x}_6 - x_6 \, \dot{x}_5\right) \quad & \text{(Neumann-Dirichlet)} \quad \& \quad & x_4\dot{x}_4 + x_5\dot{x}_5 + x_6\dot{x}_6 \eq 0 \quad \text{(Dirichlet)} \qquad \label{D5braneStringBCsS1a} \\
\acute{x}_5 \eq \kappa \left(x_6 \, \dot{x}_4 - x_4 \, \dot{x}_6\right) && \dot{x}_7 \eq \dot{x}_8 \eq \dot{x}_9 \eq 0 \label{D5braneStringBCsS1b} \\
\acute{x}_6 \eq \kappa \left(x_4 \, \dot{x}_5 - x_5 \, \dot{x}_4\right). && \label{D5braneStringBCsS1c}
\end{IEEEeqnarray}
Setting $\textbf{\textit{x}}_\parallel = \left(x_4,x_5,x_6\right)$ and $\textbf{\textit{x}}_\perp = \left(x_7,x_8,x_9\right)$, we may write these BCs as follows:
\begin{IEEEeqnarray}{l}
\acute{\textbf{\textit{x}}}_{\parallel} - \kappa \left(\textbf{\textit{x}}_{\parallel} \times \dot{\textbf{\textit{x}}}_{\parallel}\right) \eq 0 \quad \text{(Neumann-Dirichlet)} \quad \& \quad \textbf{\textit{x}}_{\parallel} \cdot \dot{\textbf{\textit{x}}}_{\parallel} = \dot{\textbf{\textit{x}}}_{\perp} \eq 0 \quad \text{(Dirichlet)}. \label{D5braneStringBCsS2}
\end{IEEEeqnarray}
\indent We will see below that the coset representation of the 5-sphere is expressed by means of a slightly modified set of coordinates. We thus need to transform the S$^5$ boundary conditions \eqref{D5braneStringBCsS1a}--\eqref{D5braneStringBCsS2} to the modified coordinate system that we will use to parametrize S$^5$. Let us first recall the standard sine parametrization of S$^5$:
\begin{IEEEeqnarray}{ll}
x_a = \ell \, m_{(a-3)}\sin\theta, \qquad x_9 = \ell \cos\theta, \qquad a = 4,\ldots,8, \label{SineParametrization1}
\end{IEEEeqnarray}
where $\theta \in \left[0,\pi\right]$ and, abbreviating $\cos\theta_a$ to $c_a$ and $\sin\theta_a$ to $s_a$,
\begin{IEEEeqnarray}{ll}
m_{1} = c_1, \quad m_{2} = s_1 c_2, \quad m_{3} = s_1 s_2 c_3, \quad m_{4} = s_1 s_2 s_3 c_4, \quad m_{5} = s_1 s_2 s_3 s_4, \quad \sum_{a=1}^{5} m_{a}^2 = 1, \qquad \label{SineParametrization2}
\end{IEEEeqnarray}
for $\theta_{1,2,3} \in \left[0,\pi\right]$ and $\theta_4 \in \left[0,2\pi\right)$. The corresponding line element becomes
\begin{IEEEeqnarray}{l}
ds^2 = \sum_{a = 4}^9 dx_a^2 = \ell^2\left(d\theta^2 + \sin^2\theta\sum_{a = 1}^{5}dm_a^2\right), \label{MetricS5so6a}
\end{IEEEeqnarray}
where $dm_a^2$ is just the line element of a unit 4-sphere parametrized again with sines:
\begin{IEEEeqnarray}{l}
\sum_{a = 1}^{5}dm_a^2 = d\theta_1^2 + s_1^2 d\theta_2^2 + s_1^2s_2^2 d\theta_3^2 + s_1^2s_2^2s_3^2 d\theta_4^2. \qquad \label{MetricS5so6b}
\end{IEEEeqnarray}
We now define a new set of S$^5$ coordinates $n_1, \ldots, n_6$ as follows:
\begin{eqnarray}
n_a = m_a \, \sin\frac{\theta}{2}, \qquad n_6 = \cos\frac{\theta}{2}, \qquad a = 1,\ldots,5, \label{CosetCoordinatesS5}
\end{eqnarray}
so that the parametrization of the S$^2$ part of the D5-brane \eqref{D5braneEmbedding2} becomes:
\begin{IEEEeqnarray}{ll}
n_1^2 + n_2^2 + n_3^2 = \frac{1}{2}, \qquad n_4 = n_5 = 0, \qquad n_6 = \frac{1}{\sqrt{2}}. \label{D5braneEmbedding3}
\end{IEEEeqnarray}
Therefore $n_1, n_2, n_3$ are the longitudinal and $n_4,n_5,n_6$ are the transverse string coordinates of the S$^2$ boundary \eqref{D5braneEmbedding3}, obeying Neumann-Dirichlet and Dirichlet boundary conditions respectively. The BCs \eqref{D5braneStringBCsS1a}--\eqref{D5braneStringBCsS1c} can be written as
\begin{IEEEeqnarray}{ll}
n_1\acute{n}_6 + n_6\acute{n}_1 \eq 2\ell \kappa n_6^2 \left(n_2 \, \dot{n}_3 - n_3 \, \dot{n}_2\right) \qquad \& \qquad & n_1\dot{n}_1 + n_2\dot{n}_2 + n_3\dot{n}_3 \eq 0 \qquad \label{D5braneStringBCsS3a} \\
n_2\acute{n}_6 + n_6\acute{n}_2 \eq 2\ell \kappa n_6^2 \left(n_3 \, \dot{n}_1 - n_1 \, \dot{n}_3\right) & \dot{n}_4 \eq \dot{n}_5 \eq \dot{n}_6 \eq 0 \label{D5braneStringBCsS3b} \\
n_3\acute{n}_6 + n_6\acute{n}_3 \eq 2\ell \kappa n_6^2 \left(n_1 \, \dot{n}_2 - n_2 \, \dot{n}_1\right). & \label{D5braneStringBCsS3c}
\end{IEEEeqnarray}
\paragraph{Integrable D5-brane in AdS$_5$} To examine whether open string dynamics in AdS$_5$ which obeys the D5-brane BCs \eqref{D5braneStringBCsAdS} is integrable, we express the coset representative of AdS$_5$ in terms of the standard 4d conformal generators \eqref{GeneratorsAdS5}:
\begin{equation}
\mathfrak{g}_{\text{AdS}} = e^{P_\mu x^\mu }z^{D}, \label{CosetParametrizationAdS5}
\end{equation}
where $\mu = 0,\ldots,3$. From the coset representative of AdS$_5$ we may obtain the moving-frame current \eqref{MovingFrameCurrent} and from the $\Z_4$ automorphism \eqref{Z4automorphism}--\eqref{Z4automorphismAdS5xS5} the corresponding components $J^{(n)}_{\text{AdS}}$. The bosonic $Z_4$ component $j^{(2)}_{\text{AdS}}$ of the fixed-frame current (entering the fixed-frame Lax connection \eqref{LaxConnectionFixedFrame}) then follows from \eqref{FixedFrameCurrent}:
\begin{equation}
j^{(2)}_{\text{AdS}} = \frac{1}{2z^2}\left[2\left(zdz + x dx\right)\left(D - x P\right) + \left(z^2 + x^2\right)P dx + K dx + L_{\mu\nu}x^\mu dx^\nu\right]. \label{FixedFrameCurrentAdS5}
\end{equation}
Notice that the trace of the square of the fixed-frame current component \eqref{FixedFrameCurrentAdS5} returns the target space metric, that is the line element of AdS$_5$ (in Minkowski signature) in \eqref{MetricAdS5xS5}:
\begin{equation}
\tr\left[\left(j^{(2)}_{\text{AdS}}\right)^2\right] = \tr\left[\left(J^{(2)}_{\text{AdS}}\right)^2\right] = \frac{1}{z^2} \left(-dx_0^2 + dx_1^2 + dx_2^2 + dx_3^2 + dz^2\right),
\end{equation}
which is consistent with the form of the bosonic part of the MT action \eqref{MetsaevTseytlinAction}, \eqref{ParameterMetchingAdS5xS5}, \eqref{Z4automorphismAdS5xS5} and the Polyakov action \eqref{PolyakovAction1}. This is a useful crosscheck of the AdS$_5$ coset parametrization \eqref{CosetParametrizationAdS5}. \\
\indent To show that the set \eqref{D5braneStringBCsAdS} of bosonic string boundary conditions on the probe D5-brane \eqref{D5braneEmbedding1} is classically integrable in AdS$_5$, we need to specify a reflection matrix $\U$ which satisfies the integrability condition \eqref{DbraneIntegrabilityCondition4} upon imposing the BCs. Indeed, the dynamical reflection matrix \cite{LinardopoulosZarembo21}
\begin{equation}
\U_{\text{AdS}} = \gamma_3 + \frac{2\kappa}{\y^2 + 1}\cdot\frac{x^{\mu} \gamma_{\mu} - \Pi_+ - (x^2 + z^2)\Pi_-}{z}, \qquad \Pi_{\pm} \equiv \frac{1}{2}\left(1 \pm \gamma_4\right), \qquad \label{ReflectionMatrixAdS5}
\end{equation}
depends on the spectral parameter $\y$ and the flux $\kappa$, and satisfies the integrability condition \eqref{DbraneIntegrabilityCondition4}. In the zero-flux case $\kappa = 0$ and we trivially recover the constant, non-dynamical and independent of the spectral parameter $\y$ reflection matrix of Dekel-Oz \cite{DekelOz11b}.
\paragraph{Integrable D5-brane in S$^5$} Similarly, to examine whether open string dynamics on S$^5$ (obeying the D5-brane BCs \eqref{D5braneStringBCsS1a}--\eqref{D5braneStringBCsS2}, or equivalently \eqref{D5braneStringBCsS3a}--\eqref{D5braneStringBCsS3c}) is integrable, we need to write down the coset representation of the unit 5-sphere. The coset parametrization is given by:
\begin{eqnarray}
\mathfrak{g}_{\text{S}} = n_6 + i\gamma_{a}n_{a}, \label{CosetParametrizationS5}
\end{eqnarray}
where the $n_a$ coordinates were defined in \eqref{CosetCoordinatesS5} and the analytic expressions of the 5d gamma matrices can be found in \eqref{GammaMatrices4}--\eqref{GammaMatrices6} of appendix \ref{Appendix:5dGammaMatrices}. As before, we can calculate the moving-frame current \eqref{MovingFrameCurrent} and use the $\Z_4$ automorphism \eqref{Z4automorphism}--\eqref{Z4automorphismAdS5xS5} to find its $\Z_4$ components $J^{(n)}_{\text{S}}$. The (bosonic) $Z_4$ component $j^{(2)}_{\text{S}}$ of the fixed-frame current follows directly from \eqref{FixedFrameCurrent}; it reads:
\begin{eqnarray}
j^{(2)}_{\text{S}} = i(2n_6^2 - 1) n_6 dn_{a} \gamma_{a} - i(2n_6^2 + 1) dn_6 n_{a} \gamma_{a} - 2n_6^2 n_{a}dn_{b} \gamma_{ab}.
\end{eqnarray}
Taking the supertrace (which is minus the trace in this case) of the square of $j^{(2)}_{\text{S}}$, we recover the target space metric \eqref{MetricS5so6a}--\eqref{MetricS5so6b} (for unit radius $\ell$) as we should:
\begin{equation}
-\tr\left[\left(j^{(2)}_{\text{S}}\right)^2\right] = -\tr\left[\left(J^{(2)}_{\text{S}}\right)^2\right] = d\theta^2 + \sin^2\theta\sum_{a = 1}^{5}dm_a^2.
\end{equation}
The dynamical reflection matrix $\U$, which satisfies the integrability condition \eqref{DbraneIntegrabilityCondition4} upon imposing the bosonic string BCs \eqref{D5braneStringBCsS3a}--\eqref{D5braneStringBCsS3c} on the S$^5$ part of the D5-brane, is given by \cite{LinardopoulosZarembo21}:
\begin{eqnarray}
\U_{\text{S}} = \gamma_{45} + \frac{2\kappa \ell \y}{\y^2 + 1} \cdot \frac{n_{i}\gamma_{i}}{n_6}, \qquad i = 1,2,3. \label{ReflectionMatrixS5}
\end{eqnarray}
This reflection matrix obviously depends on the spectral parameter $\y$ and the flux number $\kappa$. In the zero-flux case $\kappa = 0$, we once more recover the constant, non-dynamical and independent of the spectral parameter $\y$ reflection matrix of Dekel-Oz \cite{DekelOz11b}.
\subsubsection[The D3-D7 intersection]{The D3-D7 intersection \label{SubSubSection:D3D7intersection}}
\noindent Another interesting format of the AdS$_5$/{\color{red}d}CFT$_4$ correspondence can be obtained by deforming the AdS$_5$/CFT$_4$ correspondence \eqref{AdS5CFT4duality} with D7 instead of D5-branes. The relative orientation of the $N_c$ coincident D3 and the $M$ coincident D7-branes in flat space is shown in table \ref{Table:D3D7system} right below. As before, we only consider the case of a single probe D7-brane, i.e.\ we take $M=1$.
\renewcommand{\arraystretch}{1.1}\setlength{\tabcolsep}{5pt}
\begin{table}[H]\begin{center}\begin{tabular}{|c||c|c|c|c|c|c|c|c|c|c|}
\hline
& $x_0$ & $x_1$ & $x_2$ & $x_3$ & $x_4$ & $x_5$ & $x_6$ & $x_7$ & $x_8$ & $x_9$ \\ \hline
\text{D3} & $\bullet$ & $\bullet$ & $\bullet$ & $\bullet$ &&&&&& \\ \hline
\text{D7} & $\bullet$ & $\bullet$ & $\bullet$ & & $\bullet$ & $\bullet$ & $\bullet$ & $\bullet$ & $\bullet$ & \\ \hline
\end{tabular}\caption{Brane orientation for the D3-D7 intersection in flat space.\label{Table:D3D7system}}\end{center}\end{table}
\vspace{-.5cm}\noindent There are two possible embeddings of probe D7-branes in AdS$_5\times\text{S}^5$, dual to codimension-1 dCFTs with $\N = 4$ SYM. The D7-brane embeddings can be found by solving the equations of motion that arise from the D7-brane action \eqref{D7braneAction}. In the first scenario, the D7-brane wraps an AdS$_4\times\text{S}^2\times\text{S}^2$ geometry that is parametrized by
\begin{IEEEeqnarray}{c}
x_3 = \Lambda \cdot z, \qquad \Lambda \equiv \frac{\kappa_1 \kappa_2}{\sqrt{\left(\kappa_1^2+\cos^4\psi_0\right) \left(\kappa_2^2+\sin^4\psi_0\right) - \kappa_1^2\kappa_2^2}}, \qquad \kappa_{\mathfrak{i}} \equiv \frac{\pi \kk_{\mathfrak{i}}}{\sqrt{\lambda}}, \qquad \psi = \psi_0, \qquad \label{D7braneEmbedding1}
\end{IEEEeqnarray}
where the (Minkowski) metric of AdS$_5$ is given by \eqref{MetricAdS5xS5} and the relevant parametrization of the unit 5-sphere is \eqref{MetricS5so3so3}. There are also $\kk_{\mathfrak{i}}$ ($\mathfrak{i}=1,2$) units of magnetic flux through each S$^2$:
\begin{IEEEeqnarray}{c}
\int_{\text{S}^2}\frac{F_{\mathfrak{i}}}{2\pi} = \kk_{\mathfrak{i}}, \qquad F_1 = \frac{k_1}{2} \cdot d\cos\theta\wedge d\chi, \qquad F_2 = \frac{\kk_2}{2} \cdot d\cos\vartheta\wedge d\varrho, \qquad \label{FieldStrengthD7a}
\end{IEEEeqnarray}
which forces exactly $\kk = \kk_1 \kk_2$ (out of the $N_c$ total) D3-branes to end on one side of the D7-brane. The constant value $\psi_0$ of the angle $\psi$ in \eqref{D7braneEmbedding1} satisfies an identity which follows from the D7-brane equations of motion:
\begin{IEEEeqnarray}{c}
\left(\kappa_1^2+\cos^4\psi_0\right)\sin^2\psi_0 = \left(\kappa_2^2+\sin^4\psi_0\right)\cos^2\psi_0. \label{D7braneConstraint}
\end{IEEEeqnarray}
The D7-brane embedding in the Cartesian coordinate system can be found by setting $\psi = \psi_0$ in \eqref{MetricS5so3so3}--\eqref{CartesianCoordinatesS5c}, so that by also including the 5-sphere radius $\ell$ we are led to
\begin{IEEEeqnarray}{c}
x_4^2 + x_5^2 + x_6^2 = \ell^2 \cos^2\psi_0, \qquad x_7^2 + x_8^2 + x_9^2 = \ell^2 \sin^2\psi_0. \label{D7braneEmbedding2}
\end{IEEEeqnarray}
In terms of the coordinate system \eqref{CosetCoordinatesS5}, the D7-brane embedding in S$^5$ gets written as
\begin{IEEEeqnarray}{c}
4n_6^2\left(n_1^2 + n_2^2 + n_3^2\right) = \cos^2\psi_0, \qquad 4n_6^2\left(n_4^2 + n_5^2 + n_6^2\right) = 4n_6^2 - \cos^2\psi_0.
\end{IEEEeqnarray}
It is also straightforward to transform the field strengths in \eqref{FieldStrengthD7a} to Cartesian coordinates,
\begin{IEEEeqnarray}{c}
F_{ij}^{(\mathfrak{i})} = -\frac{\kk_{\mathfrak{i}}}{2}\,\epsilon_{ijl}x_l, \qquad i,j,l = 4,5,6 \quad \text{or} \quad i,j,l = 7,8,9, \quad \mathfrak{i} = 1,2.
\end{IEEEeqnarray}
\indent Just like the D5-brane, the D7-brane preserves the $SO(3,2)\times SO(3)\times SO(3)$ subgroup of $SO(4,2)\times SO(6)$. Unlike the D3-D5 brane system however, the D3-D7 system is not supersymmetric, i.e.\ it breaks all the supersymmetries of the original $PSU(2,2|4)$ supergroup. The D7-brane embedding \eqref{D7braneEmbedding1} is also unstable in the zero-flux case (i.e.\ for $\kk_1 = \kk_2 = 0$) \cite{BergmanJokelaLifschytzLippert10}. The tachyonic mode in the fluctuations of the angle $\psi$ violates the BF bound. However, it was shown in \cite{BergmanJokelaLifschytzLippert10} that the $SO(3)\times SO(3)$ symmetric D3-D7 probe-brane system can be stabilized when the fluxes $\kk_{\mathfrak{i}}$ become larger than a certain threshold. \\
\indent Let us now briefly examine the second scenario for embedding a probe D7-brane in AdS$_5\times\text{S}^5$. In this scenario, the D7-brane wraps an AdS$_4\times\text{S}^4$ geometry which is parametrized by
\begin{IEEEeqnarray}{c}
x_3 = \Lambda \cdot z, \qquad \Lambda \equiv \frac{Q}{\sqrt{1+2Q}}, \qquad Q \equiv \frac{\pi^2}{\lambda}\cdot(n+1)(n+3), \qquad \theta = 0, \label{D7braneEmbedding3}
\end{IEEEeqnarray}
where $n = 0,1,2\ldots$, and the relevant parametrization of the 5-sphere is given by \eqref{MetricS5so5}. The geometry is supported by $d_G$ units of instanton flux through the S$^4$ (second Chern class):
\begin{IEEEeqnarray}{c}
\int_{\text{S}^4}\frac{1}{8\pi^2}\cdot\tr F\wedge F = d_G, \qquad d_G \equiv \frac{1}{6}\cdot(n+1)(n+2)(n+3), \label{FieldStrengthD7b}
\end{IEEEeqnarray}
where the $SU(n+2)$ homogeneous instanton is obtained by stereographically projecting the Belavin-Polyakov-Schwartz-Tyupkin (BPST) instanton of $SU(2)$ Yang-Mills theory \cite{BelavinPolyakovSchwartzTyupkin75} on S$^4$ and replacing the Pauli matrix generators with the generators of the $n+2$ dimensional irreducible representation of $SU(2)$ \cite{ConstableMyersTafjord01a},
\begin{IEEEeqnarray}{c}
A_a = \frac{2\eta_{iab} x_b t_i}{x^2 + 1}, \quad A_8 = 0, \qquad \eta_{iab}\equiv \varepsilon_{iab7} + \delta_{ia}\delta_{7b} - \delta_{ib}\delta_{7a}, \qquad a,b = 4, \ldots, 7. \qquad \label{HomogeneousInstanton}
\end{IEEEeqnarray}
In \eqref{HomogeneousInstanton}, $\eta_{iab}$ is the 't Hooft symbol, while the matrices $t_i$ furnish a $n+2$ dimensional irreducible representation of $\mathfrak{so}\left(3\right) \simeq \mathfrak{su}\left(2\right)$ (see e.g.\ \cite{deLeeuwKristjansenZarembo15} for the precise form of $t_i$),
\begin{IEEEeqnarray}{c}
\left[t_i, t_j\right] = i \epsilon_{ijk}t_k, \qquad i,j,k = 4,5,6.
\end{IEEEeqnarray}
The instanton field strength $F_{\mu\nu}$ can be directly computed from \eqref{HomogeneousInstanton}. It obeys \eqref{FieldStrengthD7b} and reads:
\begin{IEEEeqnarray}{c}
F_{ab} = -\frac{4 \eta_{iab} t_i}{\left(1 + x_c x_c\right)^2}.
\end{IEEEeqnarray}
Once more, the flux \eqref{FieldStrengthD7b} forces exactly $d_G$ out of the $N_c$ total D3-branes to terminate on one side of the D7-brane. The subgroup of $SO(4,2)\times SO(6)$ that is preserved by the D7-brane is $SO(3,2)\times SO(5)$ and supersymmetry is completely broken. There is also a tachyonic instability in the fluctuations of the angle $\theta$ which violates the BF bound \cite{DavisKrausShah08, Rey08}, unless $Q>7/2$ in \eqref{D7braneEmbedding3} \cite{MyersWapler08}.
\paragraph{String boundary conditions on D7} Evidently, the analysis of string boundary conditions on the AdS$_4\subset\text{AdS}_5$ component of the D7-brane is pretty much the same with the one that was carried out for the D5-brane in \S\ref{SubSubSection:D3D5intersection}. The only difference is that $\kappa$ in \eqref{D5braneEmbedding1} has been replaced by $\Lambda$ in either \eqref{D7braneEmbedding1} or \eqref{D7braneEmbedding3}. Everything else, including the bosonic string BCs in AdS$_5$ \eqref{D5braneStringBCsAdS}, as well as the dynamical reflection matrix $\U_{\text{AdS}}$ in \eqref{ReflectionMatrixAdS5} is exactly the same. Because the reflection matrix \eqref{ReflectionMatrixAdS5} satisfies the D-brane integrability condition \eqref{DbraneIntegrabilityCondition4}, we conclude that the AdS$_4$ component of the D7-brane is classically integrable. \\
\indent Concerning the analysis of string boundary conditions on the compact components of the D7-brane (either S$^2\times\text{S}^2$ or S$^4$), let us first remark that the BCs \eqref{StringBoundaryConditions1}--\eqref{StringBoundaryConditions2} were derived in \S\ref{SubSection:StringIntegrabilitydCFT} only in the case of classical string worldsheets that are coupled to Abelian flux fields $A_{\textrm{i}}$ in \eqref{PolyakovAction1}. To the best of our knowledge, it is currently not known how to couple a classical string worldsheet and a non-Abelian flux field such as \eqref{HomogeneousInstanton}. Coupling the string Polyakov action to the non-Abelian flux field \eqref{HomogeneousInstanton}, deriving the corresponding (bosonic) string BCs on the S$^4 \subset \text{S}^5$ component of the D7-brane, and showing their classical integrability by finding a (dynamical) reflection matrix $\U$ which satisfies the integrability condition \eqref{DbraneIntegrabilityCondition4}, is a very interesting open problem. \\
\indent There is an extra twist however which makes this problem even more interesting. As we have already mentioned, it has been shown in \cite{deLeeuwKristjansenLinardopoulos18a} that the D3-D7 dCFT that is holographically dual to the $SO(5)$ symmetric D3-D7 probe-brane system satisfies the integrable quench criterion \eqref{QuenchIntegrabilityCriterion}. Moreover, the overlaps of one-loop Bethe states with the $SO(5)$ symmetric MPS of the D3-D7 system are described by a closed-form determinant formula \cite{deLeeuwGomborKristjansenLinardopoulosPozsgay19}. By all means this system is expected to be classically integrable from weak to strong coupling! In an interesting turn of events, it was claimed in \cite{GomborBajnok20a} that the integrability of the $SO(5)$ symmetric D3-D7 dCFT breaks down beyond one-loop order. Treating this problem at strong coupling with the methods that we have developed could possibly shed light on this puzzle. \\
\indent Concerning the dCFT that is dual to the $SO(3)\times SO(3)$ symmetric D3-D7 intersection \eqref{D7braneEmbedding1}, it was shown in \cite{deLeeuwKristjansenVardinghus19} that the corresponding domain wall is not integrable. We therefore do not expect classical integrability to show up at strong coupling. Let us first write down the corresponding boundary conditions. From the parametrization \eqref{D7braneEmbedding1} and the expressions \eqref{FieldStrengthD7a} for the field strength of the $SO(3)\times SO(3)$ symmetric D3-D7 system, we conclude that all six coordinates $x_a$ ($a = 4,\ldots,9$) are longitudinal string coordinates. Therefore, the bosonic string BCs \eqref{StringBoundaryConditions1}--\eqref{StringBoundaryConditions2} are all Neumann-Dirichlet:
\begin{IEEEeqnarray}{lll}
\acute{x}_4 \eq \kappa_1 \left(x_5 \, \dot{x}_6 - x_6 \, \dot{x}_5\right) && \acute{x}_7 \eq \kappa_2 \left(x_8 \, \dot{x}_9 - x_9 \, \dot{x}_8\right) \qquad \label{D7braneStringBCsS1a} \\
\acute{x}_5 \eq \kappa_1 \left(x_6 \, \dot{x}_4 - x_4 \, \dot{x}_6\right) & \qquad \& \qquad & \acute{x}_8 \eq \kappa_2 \left(x_9 \, \dot{x}_7 - x_7 \, \dot{x}_8\right)\label{D7braneStringBCsS1b} \\
\acute{x}_6 \eq \kappa_1 \left(x_4 \, \dot{x}_5 - x_5 \, \dot{x}_4\right) && \acute{x}_9 \eq \kappa_2 \left(x_7 \, \dot{x}_8 - x_8 \, \dot{x}_7\right), \label{D7braneStringBCsS1c}
\end{IEEEeqnarray}
while the parametrization \eqref{D7braneEmbedding2} implies the additional set of Dirichlet BCs
\begin{IEEEeqnarray}{lll}
x_4\dot{x}_4 + x_5\dot{x}_5 + x_6\dot{x}_6 \eq x_7\dot{x}_7 + x_8\dot{x}_8 + x_9\dot{x}_9 \eq 0. \label{D7braneStringBCsS2}
\end{IEEEeqnarray}
For $\textbf{\textit{x}}_\parallel = \left(x_4,x_5,x_6\right)$ or $\textbf{\textit{x}}_\parallel = \left(x_7,x_8,x_9\right)$, we may write the BCs \eqref{D7braneStringBCsS1a}--\eqref{D7braneStringBCsS2} in a compact form as
\begin{IEEEeqnarray}{c}
\acute{\textbf{\textit{x}}}_{\parallel} - \kappa_i \left(\textbf{\textit{x}}_{\parallel} \times \dot{\textbf{\textit{x}}}_{\parallel}\right) \eq 0 \quad \text{(Neumann-Dirichlet)} \quad \& \quad \textbf{\textit{x}}_{\parallel} \cdot \dot{\textbf{\textit{x}}}_{\parallel} \eq 0 \quad \text{(Dirichlet)}, \label{D7braneStringBCsS3}
\end{IEEEeqnarray}
where $i=1$ (i.e.\ $\kappa_1$ in \eqref{D7braneStringBCsS3}) relates to the former (i.e.\ when $\textbf{\textit{x}}_\parallel = \left(x_4,x_5,x_6\right)$) and $i=2$ (i.e.\ $\kappa_2$ in \eqref{D7braneStringBCsS3}) relates to the latter case (i.e.\ when $\textbf{\textit{x}}_\parallel = \left(x_7,x_8,x_9\right)$). Transforming the BCs \eqref{D7braneStringBCsS1a}--\eqref{D7braneStringBCsS2} to the coset coordinate system \eqref{CosetCoordinatesS5} is straightforward, obtaining for the Neumann-Dirichlet set \eqref{D7braneStringBCsS1a}--\eqref{D7braneStringBCsS1c},
\begin{IEEEeqnarray}{l}
n_1\acute{n}_6 + n_6\acute{n}_1 \eq 2\ell \kappa_1 n_6^2 \left(n_2 \, \dot{n}_3 - n_3 \, \dot{n}_2\right) \qquad \label{D7braneStringBCsS4a} \\[6pt]
n_2\acute{n}_6 + n_6\acute{n}_2 \eq 2\ell \kappa_1 n_6^2 \left(n_3 \, \dot{n}_1 - n_1 \, \dot{n}_3\right) \qquad \label{D7braneStringBCsS4b} \\[6pt]
n_3\acute{n}_6 + n_6\acute{n}_3 \eq 2\ell \kappa_1 n_6^2 \left(n_1 \, \dot{n}_2 - n_2 \, \dot{n}_1\right) \qquad \label{D7braneStringBCsS4c} \\[6pt]
n_4\acute{n}_6 + n_6\acute{n}_4 \eq \ell \kappa_2 \left[\left(2n_6^2+1\right)n_5 \, \dot{n}_6 - \left(2n_6^2-1\right) n_6 \, \dot{n}_5\right] \qquad \label{D7braneStringBCsS4d} \\[6pt]
n_5\acute{n}_6 + n_6\acute{n}_5 \eq \ell \kappa_2 \left[\left(2n_6^2-1\right)n_6 \, \dot{n}_4 - \left(2n_6^2+1\right) n_4 \, \dot{n}_6\right] \qquad \label{D7braneStringBCsS4e} \\[6pt]
n_6\acute{n}_6 \eq \ell \kappa_2 n_6^2 \left(n_4 \, \dot{n}_5 - n_5 \, \dot{n}_4\right), \qquad \label{D7braneStringBCsS4f}
\end{IEEEeqnarray}
while the Dirichlet BCs \eqref{D7braneStringBCsS2} become
\begin{IEEEeqnarray}{l}
n_6^2\left(n_1\dot{n}_1 + n_2\dot{n}_2 + n_3\dot{n}_3\right) + \left(n_1^2 + n_2^2 + n_3^2\right)n_6\dot{n}_6 \eq 0 \qquad \label{D7braneStringBCsS5a} \\
n_6^2\left(n_4\dot{n}_4 + n_5\dot{n}_5\right) + \left(n_4^2 + n_5^2 + 2n_6^2 - 1\right)n_6\dot{n}_6 \eq 0. \qquad \label{D7braneStringBCsS5b}
\end{IEEEeqnarray}
\subsubsection[The D2-D4 intersection]{The D2-D4 intersection \label{SubSubSection:D2D4intersection}}
\noindent The AdS$_4$/CFT$_3$ duality \eqref{AdS4CFT3duality} can be deformed by a probe D4-brane \cite{FujitaLiRyuTakayanagi09}. The D4-brane is inserted on the rhs of the duality which contains type IIA string theory on AdS$_4\times\CP^3$. This theory arises in the near-horizon limit of $N_c$ coincident D2-branes. The D2-D4 intersection then gives rise to the AdS$_4$/{\color{red}d}CFT$_3$ correspondence which claims that type IIA superstring theory on AdS$_4\times\CP^3$ with a probe D4-brane is holographically dual to a codimension-1 domain wall version of ABJM theory. \\
\indent The embedding of the probe D4-brane inside AdS$_4\times\CP^3$ can be found by solving the equations of motion that arise from the DBI action \eqref{D4braneAction}. The D4-brane wraps an AdS$_3\times\CP^1$ geometry which is parametrized by \cite{ChandrasekharPanda09}
\begin{IEEEeqnarray}{c}
x_2 = Q \cdot z, \qquad \chi = 0, \qquad \theta_2,\phi_2,\psi = \text{const.}, \label{D4braneEmbedding1}
\end{IEEEeqnarray}
where the metric of AdS$_4\times\CP^3$ is given by \eqref{MetricAdS4xCP3}--\eqref{MetricCP3}. The relative orientation of D2 and D4 branes is shown in the following table \ref{Table:D2D4system}.
\renewcommand{\arraystretch}{1.1}\setlength{\tabcolsep}{5pt}
\begin{table}[H]\begin{center}\begin{tabular}{|c||c|c|c|c|c|c|c|c|c|c|}
\hline
& $x_0$ & $x_1$ & $x_2$ & $z$ & $\chi$ & $\theta_1$ & $\phi_1$ & $\theta_2$ & $\phi_2$ & $\psi$ \\ \hline
\text{D2} & $\bullet$ & $\bullet$ & $\bullet$ &&&&&&& \\ \hline
\text{D4} & $\bullet$ && $\bullet$ & $\bullet$ && $\bullet$ & $\bullet$ &&& \\ \hline
\end{tabular}\caption{Brane orientation in the D2-D4 intersection.\label{Table:D2D4system}}\end{center}\end{table}
\vspace{-.5cm}\noindent By setting $\chi = 0$ (i.e.\ the parametrization \eqref{D4braneEmbedding1} of the D4-brane inside $\CP^3$) into the $\CP^3$ metric \eqref{MetricCP3}, we come up with the line element of $\CP^1$, which is just that of a two-sphere:
\begin{IEEEeqnarray}{l}
ds^2_{\CP^1} = \ell^2\left(d\theta_1^2 + \sin^2\theta_1 \, d\phi_1^2\right) = \sum_{i = 4}^6 dx_{i} \, dx_{i}.
\end{IEEEeqnarray}
Therefore, the embedding of the D4-brane inside $\CP^3$ can also be expressed in terms of the Cartesian coordinate set $(x_4,x_5,x_6)$ as
\begin{IEEEeqnarray}{l}
x_4^2 + x_5^2 + x_6^2 = \ell^2. \qquad \label{D4braneEmbedding2}
\end{IEEEeqnarray}
The global bosonic symmetry that is preserved by the D4-brane is given by the $SO(2,2) \times SO(3) \times SO(3) \times U(1)$ subgroup of $SO(3,2)\times SO(6)$. The D4-brane is half-BPS, i.e.\ it preserves exactly half of the supersymmetries of the ABJM global supersymmetry group $Osp(2,2|6)$. We can also add $Q$ units of magnetic flux on $\CP^1 \sim \text{S}^2$ in the standard way,
\begin{IEEEeqnarray}{c}
\int_{\CP^1}\frac{F}{2\pi} = Q, \qquad F = \ell^2 Q \cdot d\cos\theta_1\wedge d\phi_1, \label{FieldStrengthD4a}
\end{IEEEeqnarray}
and then transform the field strength to the Cartesian coordinates \eqref{D4braneEmbedding2} in order to obtain
\begin{IEEEeqnarray}{c}
F_{ij} = -\frac{Q}{\ell}\,\epsilon_{ijk}x_k, \qquad i,j,k = 4,5,6. \label{FieldStrengthD4b}
\end{IEEEeqnarray}
Because of the flux \eqref{FieldStrengthD4a}, exactly $q \equiv \sqrt{2\lambda}\,Q$ out of the $N_c$ initial D2-branes are forced to dissolve on one side of the D4-brane.
\paragraph{String boundary conditions on D4} The set of longitudinal and transverse string coordinates relative to the D4-brane follows directly from the D-brane parametrization \eqref{D4braneEmbedding1}--\eqref{D4braneEmbedding2}. Defining the inclination angle $\alpha$ by $Q \equiv \tan\alpha$ we obtain the two sets of coordinates, namely
\begin{IEEEeqnarray}{lrl}
\text{longitudinal: } &x_0,x_1, \ x_2\sin\alpha + z\cos\alpha, \quad & \theta_1, \phi_1 \\
\text{transverse: } &x_2\cos\alpha - z\sin\alpha, \quad & \chi, \theta_2, \phi_2, \psi.
\end{IEEEeqnarray}
The boundary conditions on the AdS$_3 \subset \text{AdS}_5$ part of the probe D4-brane are almost a carbon copy of the ones for the D5-brane which were specified in \eqref{D5braneStringBCsAdS}:
\begin{IEEEeqnarray}{c}
\acute{x}_{0,1} \eq \acute{z} + Q \, \acute{x}_2 \eq 0 \quad \text{(Neumann)} \qquad \& \qquad \dot{x}_2 - Q \, \dot{z} \eq 0 \quad \text{(Dirichlet)}. \label{D4braneStringBCsAdS}
\end{IEEEeqnarray}
On the other hand, the BCs on the $\CP^1 \sim \text{S}^2 \subset \CP^3$ part of the D4-brane follow from \eqref{StringBoundaryConditions1}--\eqref{StringBoundaryConditions2}
\begin{IEEEeqnarray}{ll}
\acute{\theta}_1 + \tilde{Q}\sin\theta_1\dot{\phi}_1 \eq \sin\theta_1\acute{\phi}_1 - \tilde{Q} \, \dot{\theta}_1 \eq 0, \qquad & \text{(Neumann-Dirichlet)} \label{D4braneStringBCsCP1a} \\
\dot{\chi} \eq \dot{\theta}_2 \eq \dot{\phi}_2 \eq \dot{\psi} \eq 0, \qquad & \text{(Dirichlet)} \label{D4braneStringBCsCP1b}
\end{IEEEeqnarray}
where we have defined $\tilde{Q} \equiv q/\lambda = \sqrt{2/\lambda}\,Q$. The Neumann-Dirichlet boundary condition \eqref{D4braneStringBCsCP1a} can also be expressed via the Cartesian coordinate set \eqref{D4braneEmbedding2},\eqref{FieldStrengthD4b} as follows
\begin{IEEEeqnarray}{ll}
\ell\acute{x}_i \eq \tilde{Q} \, \epsilon_{ijk}x_j\dot{x}_k. \label{D4braneStringBCsCP2}
\end{IEEEeqnarray}
\paragraph{Integrable D4-brane in AdS$_5$} The coset representative of AdS$_4$ is almost identical to the one for AdS$_5$ in \eqref{CosetParametrizationAdS5}. It can be expressed in terms of the standard 3d conformal generators \eqref{GeneratorsAdS4},
\begin{equation}
\mathfrak{g}_{\text{AdS}} = {\rm e}\,^{P_\mu x^\mu }z^{D}, \label{CosetParametrizationAdS4}
\end{equation}
where $\mu = 0,1,2$. From the coset representative \eqref{CosetParametrizationAdS4} we may obtain the moving-frame current \eqref{MovingFrameCurrent} and then, by using the automorphism \eqref{Z4automorphism}, \eqref{Z4automorphismAdS4xCP3}, all the $\Z_4$ components $J^{(n)}_{\text{AdS}}$ ($n = 0,\ldots,3$). The bosonic $Z_4$ component of the fixed-frame current $j^{(2)}_{\text{AdS}}$ follows from \eqref{FixedFrameCurrent}:
\begin{equation}
j^{(2)}_{\text{AdS}} = \frac{1}{2z^2}\left[2\left(zdz + x dx\right)\left(D - x P\right) + \left(z^2 + x^2\right)P dx + K dx + L_{\mu\nu}x^\mu dx^\nu\right] \label{FixedFrameCurrentAdS4}
\end{equation}
and is naturally identical to the one for AdS$_5$ in \eqref{FixedFrameCurrentAdS5}. As expected, the trace of the square of the current reproduces the target space metric in \eqref{MetricAdS4xCP3}, i.e.\ the line element of AdS$_4$:
\begin{equation}
\tr\left[\left(j^{(2)}_{\text{AdS}}\right)^2\right] = \tr\left[\left(J^{(2)}_{\text{AdS}}\right)^2\right] = \frac{1}{z^2} \left(-dx_0^2 + dx_1^2 + dx_2^2 + dz^2\right).
\end{equation}
\indent The reflection matrix $\U$, which satisfies the D-brane integrability condition \eqref{DbraneIntegrabilityCondition4} in AdS$_4$ upon imposing the bosonic string BCs \eqref{D4braneStringBCsAdS}, is almost the same with the one in \eqref{ReflectionMatrixAdS5} for the probe D5-brane in AdS$_5$. It is again dynamical while also depending on the spectral parameter $\y$ and the flux number $Q$ \cite{Linardopoulos22}:
\begin{IEEEeqnarray}{c}
\U_{\text{AdS}} = \gamma_{13} \cdot \Bigg[\gamma_3 + \frac{2Q}{\y^2 + 1}\cdot\frac{x^{\mu}\gamma_{\mu} - \Pi_{+} - \left(z^2 + x^2\right)\Pi_{-}}{z}\Bigg]. \qquad \label{ReflectionMatrixAdS4}
\end{IEEEeqnarray}
We therefore conclude that the set \eqref{D4braneStringBCsAdS} of bosonic string boundary conditions on the probe D4-brane \eqref{D4braneEmbedding1} in AdS$_4$ is classically integrable. For $Q = 0$ the reflection matrix \eqref{ReflectionMatrixAdS4} reduces to the Dekel-Oz result \cite{DekelOz11b}, i.e.\ it is constant, non-dynamical and independent of the spectral parameter $\y$.
\paragraph{Integrable D4-brane in $\CP^3$} The coset parametrization of $\CP^3$ can be expressed in terms of the generators of $\mathfrak{so}\left(6\right)$. As explained in appendix \ref{Appendix:TR-Matrices}, the generators of $\mathfrak{so}\left(6\right)$ are divided into two sets, namely the R-matrices \eqref{Rmatrices1}--\eqref{Rmatrices3} and the T-matrices \eqref{Tmatrices1}--\eqref{Tmatrices2}. The former are the graded-0 generators with respect to the $\mathfrak{u}\left(3\right)$ subalgebra of $\mathfrak{so}\left(6\right)$, while the latter are the graded-2 generators. The coset element of $\CP^3$ reads:
\begin{eqnarray}
\mathfrak{g}_{\CP} = e^{-R_8 \psi} e^{T_3 \phi_1} e^{T_4 \left(\theta_1 + \frac{\pi}{2}\right)} e^{R_3 \phi_2} e^{R_4 \left(\theta_2 + \frac{\pi}{2}\right)} e^{2T_6 \chi}. \label{CosetParametrizationCP3}
\end{eqnarray}
With the coset representative of $\CP^3$ at hand, we may immediately obtain the moving-frame current \eqref{MovingFrameCurrent} and all of its the $\Z_4$ components $J^{(0)}_{\CP}, \ldots J^{(3)}_{\CP}$. The temporal and spatial parts of the bosonic $Z_4$ component of the fixed-frame current $j^{(2)}_{\CP}$ follow from \eqref{FixedFrameCurrent}. Their full expressions are too complicated to be written out, so we only provide their values on the D4-brane \eqref{D4braneEmbedding1}:
\begin{IEEEeqnarray}{ll}
j^{(2)}_\tau \eq \ & T_3 \dot{\phi}_1 \sin^2\theta_1 + T_4 \left(\dot{\theta}_1\cos\phi_1 - \dot{\phi}_1\sin\theta_1\cos\theta_1\sin\phi_1\right) - \frac{1}{4}\left(R_7 + R_9\right)\big(2\dot{\theta}_1\sin\phi_1 + \nonumber \\
& + \dot{\phi}_1\sin2\theta_1\cos\phi_1\big) \\[6pt]
j^{(2)}_\sigma \eq \ & -\tilde{Q}\cdot\frac{d}{d\tau}\left[T_3 \cos\theta_1 + T_4 \sin\theta_1\sin\phi_1 + \frac{1}{2}\left(R_7 + R_9\right) \sin\theta_1\cos\phi_1 - R_8\right] + \nonumber \\
& + f\left(\theta_1,\theta_2,\phi_1,\phi_2,\psi\right)\cdot \chi',
\end{IEEEeqnarray}
where the function $f\left(\theta_1,\theta_2,\phi_1,\phi_2,\psi\right)$ is defined as
\begin{IEEEeqnarray}{ll}
f\left(\theta_1,\theta_2,\phi_1,\phi_2,\psi\right) = &T_1\left(a_-\cos\psi - b_+\sin\psi\right) + T_2\left(a_-\sin\psi + b_+\cos\psi\right) - \nonumber \\
& - T_5\left(c_+\cos\psi + d_-\sin\psi\right) - T_6\left(c_+\sin\psi - d_-\cos\psi\right) - \nonumber \\
& - R_1\left(a_+\cos\psi - b_-\sin\psi\right) + R_2\left(a_+\sin\psi + b_-\cos\psi\right) + \nonumber \\
& + R_5\left(c_-\cos\psi + d_+\sin\psi\right) + R_6\left(c_-\sin\psi - d_+\cos\psi\right), \qquad \label{CoefficientXiPrime1}
\end{IEEEeqnarray}
and, for $\theta_{\pm} = \left(\theta_1 \pm \theta_2\right)/2$, $\phi_{\pm} = \left(\phi_1 \pm \phi_2\right)/2$, we have set
\begin{IEEEeqnarray}{ll}
a_{\pm} = \cos\theta_+\cos\phi_- \pm \sin\theta_-\cos\phi_+, \qquad &b_{\pm} = \sin\theta_+\sin\phi_+ \pm \cos\theta_-\sin\phi_- \qquad \label{CoefficientXiPrime2} \\
c_{\pm} = \sin\theta_-\sin\phi_+ \pm \cos\theta_+\sin\phi_-, \qquad &d_{\pm} = \cos\theta_-\cos\phi_- \pm \sin\theta_+\cos\phi_+. \qquad \label{CoefficientXiPrime3}
\end{IEEEeqnarray}
Note also that the supertrace (i.e.\ minus the trace) of the square of both $n=2$ (bosonic) current components (either moving-frame or fixed) equals the line element of $\CP^3$ in \eqref{MetricCP3}. This is a nice verification of the coset parametrization \eqref{CosetParametrizationCP3} of $\CP^3$. \\
\indent The set \eqref{D4braneStringBCsCP1a}--\eqref{D4braneStringBCsCP1b} of bosonic string boundary conditions on the $\CP^3$ part of the probe D4-brane \eqref{D4braneEmbedding1} is classically integrable. The dynamical reflection matrix $\U$,
\begin{IEEEeqnarray}{ll}
\U_{\CP} = \U_0 + \frac{2\tilde{Q}\,\y}{\y^2 + 1}\cdot S, \quad & S \equiv 2T_3 \cos\theta_1 + 2T_4 \sin\theta_1\sin\phi_1 + \left(R_7 + R_9\right) \sin\theta_1\cos\phi_1 - R_8 \qquad \quad \label{ReflectionMatrixCP3a} \\
& \U_0 \equiv 2\left(T_1^2 - 3T_3^2 + T_5^2\right), \label{ReflectionMatrixCP3b}
\end{IEEEeqnarray}
depends on the spectral parameter $\y$ and the flux $\tilde{Q}$, and satisfies the integrability condition \eqref{DbraneIntegrabilityCondition4} upon imposing the bosonic string BCs \eqref{D4braneStringBCsCP1a}--\eqref{D4braneStringBCsCP1b}. Once more, the zero-flux case $\tilde{Q} = 0$ reduces to the constant, non-dynamical and independent of the spectral parameter $\y$ reflection matrix of Dekel-Oz \cite{DekelOz11b}.
\subsection[Outlook and open problems]{Outlook and open problems \label{SubSection:IntegrabilityOutlook}}
\noindent Summing up, we have shown that the D3-D5 and the D2-D4 probe-brane systems are classically integrable. Based on the formalism that was developed by Sklyanin \cite{Sklyanin87a}, we have managed to construct conserved double-row monodromy matrices \eqref{DoubleRowMonodromyMatrix} for both of these systems. The infinite set of conserved charges that is encoded in the spectrum of double-row monodromy matrices can generally be extracted by Taylor-expanding them around appropriate values of the spectral parameter, e.g.\ around $\y = \infty$. In the two original papers \cite{LinardopoulosZarembo21, Linardopoulos22}, it was shown that the first conserved charge $\tilde{\mathcal{Q}}_0$ in each system's integrable hierarchy is spanned by the generators of the unbroken global symmetries and supersymmetries. This finding is in full agreement with the geometry of the probe branes and the fraction of supersymmetry they preserve. \\
\indent In more practical terms, classical integrability of probe D-branes translates to classical integrability for open string dynamics subject to boundary conditions on the D-branes. However, we have only considered boundary conditions of open bosonic strings. Without the fermions, the expression of the fixed-frame current and its $\Z_4$ components (be they fermionic or bosonic) in \eqref{FixedFrameCurrent}, as well as the integrability condition \eqref{DbraneIntegrabilityCondition4} are much simpler to deal with. Showing classical integrability for the full superstring by including both the fermionic and the bosonic degrees of freedom is an interesting open problem. To get a more concrete idea, let us briefly try to set it up. First, we need to include the fermions in the coset element $\mathfrak{g}$,
\begin{IEEEeqnarray}{l}
\mathfrak{g} = \mathfrak{g}_f\mathfrak{g}_b, \qquad \mathfrak{g}_f \equiv e^{\boldsymbol{\chi}}, \label{CosetParametrization}
\end{IEEEeqnarray}
where $\mathfrak{g}_b$ is the familiar bosonic coset element and the matrix $\boldsymbol{\chi}$ incorporates the fermionic degrees of freedom.\footnote{Note that the choice \eqref{CosetParametrization} is not unique, e.g.\ we could have defined $\mathfrak{g} = \mathfrak{g}_b\mathfrak{g}_f$.} The current of the fermionic coset element $\mathfrak{g}_f$ has a bosonic and a fermionic part:
\begin{IEEEeqnarray}{l}
\mathfrak{g}_f^{-1}d\mathfrak{g}_f \equiv B + F.
\end{IEEEeqnarray}
We may now proceed to obtain the bosonic projection of the fixed-frame current $j^{(2)}$ which enters the fixed-frame Lax connection \eqref{LaxConnectionFixedFrame}. To this end, we must compute the corresponding moving-frame projection $J^{(2)}$ via \eqref{MovingFrameCurrent}; then, by using \eqref{FixedFrameCurrent} we are led to
\begin{IEEEeqnarray}{l}
j^{(2)} = \frac{1}{2}\,\mathfrak{g}_f\left[B + GB^{t}G^{-1} + dG G^{-1}\right]\mathfrak{g}_f^{-1}, \qquad G \equiv \mathfrak{g}_b \K \mathfrak{g}_b^{-1}.
\end{IEEEeqnarray}
Next, we must specify the set of fermionic boundary conditions on the probe D-brane that we are studying (e.g.\ on the D5 or on the D4-brane). A set of superstring boundary conditions would be classically integrable if a reflection matrix $\U$ could be found, so that the full integrability condition \eqref{DbraneIntegrabilityCondition2} is satisfied upon imposing the boundary conditions. \\
\indent All the D-brane systems that we have analyzed descended from integrable holographic dualities which are in principle solvable from weak to strong coupling by integrability methods. We have also seen that the introduction of probe D-branes on the string theory side of an AdS/CFT duality like \eqref{AdS5CFT4duality} or \eqref{AdS4CFT3duality} generally preserves holography but not necessarily integrability. Checking whether integrability is preserved on the (strongly coupled) string theory side of AdS/{\color{red}d}CFT dualities is a useful complement of the various quench integrability criteria which have been developed for weakly coupled dCFTs.
\renewcommand{\arraystretch}{1.5}\setlength{\tabcolsep}{5pt}
\begin{table}[H]\begin{center}\begin{tabular}{|l||c|c|c|c|}
\hline
& D3-D5 & D3-D7 & D3-D7 & D2-D4 \\ \hline
Brane geometry & AdS$_4\times\text{S}^2$ & AdS$_4\times\text{S}^2\times\text{S}^2$ & AdS$_4\times\text{S}^4$ & AdS$_3\times\CP^1$ \\ \hline
Supersymmetry & $\frac{1}{2}$ BPS & None & None & $\frac{1}{2}$ BPS \\ \hline
Quench integrability & Yes & No & Yes & Yes \\ \hline
Determinant formula & Yes & No & Yes & Yes \\ \hline
Brane integrability & Yes & ? & ? & Yes \\ \hline
\end{tabular}\caption{Summary of holographic defects.\label{Table:HolographicDefects}}\end{center}\end{table}
\vspace{-.5cm}\indent A short summary of the probe-brane systems that we have examined can be found in table \ref{Table:HolographicDefects}. The integrable cases are characterized by classical reflection matrices $\U$ which satisfy the integrability condition \eqref{DbraneIntegrabilityCondition4} on the probe branes. It would be interesting to investigate the relation of these classical reflection matrices to the quantum reflection matrices $\textrm{K}$ which emerge on the dual gauge theory side and describe scattering in the open spin chain.\footnote{The author thanks C.\ Kristjansen for suggesting this problem.} For example, the classical reflection matrix of the integrable D3-D5 intersection is given by \eqref{ReflectionMatrices}, \eqref{ReflectionMatrixAdS5}, \eqref{ReflectionMatrixS5}. This matrix is dynamical and depends on the spacetime coordinates, the spectral parameter $\y$ and the flux number $\kappa$. However, the exact relation of this matrix to the quantum reflection matrix of the open $\N =4$ SYM spin chain (which has been obtained in \cite{DeWolfeMann04} for $\kk=0$) is not presently known. \\
\indent We have also examined two versions of the D3-D7 intersection. One has $SO(3)\times SO(3)$ global symmetry and the other $SO(5)$. Based on the study of the dual domain wall systems \cite{deLeeuwKristjansenLinardopoulos18a, deLeeuwGomborKristjansenLinardopoulosPozsgay19, deLeeuwKristjansenVardinghus19}, we expect that the former brane (wrapping AdS$_4\times\text{S}^2\times\text{S}^2$) will not be integrable, while the latter (wrapping AdS$_4\times\text{S}^4$) will be integrable. It would therefore be interesting if it could be shown that the set \eqref{D7braneStringBCsS1a}--\eqref{D7braneStringBCsS3} of boundary conditions (or equivalently \eqref{D7braneStringBCsS4a}--\eqref{D7braneStringBCsS5b}) on the probe $SO(3)\times SO(3)$ symmetric D7-brane cannot be integrable, or if a classical reflection matrix $\U$ (satisfying the integrability condition \eqref{DbraneIntegrabilityCondition4}) could be specified for the $SO(5)$ symmetric brane. \\
\indent Many more interesting holographic probe brane systems exist. The S-dual of the D3-D5 intersection is another half-BPS system, the D3-NS5 intersection \cite{GaiottoWitten08a}. This system is holographically dual to a defect CFT of the form \eqref{DefectAction}, i.e.\ $\N = 4$ SYM coupled to a $2+1$ dimensional SCFT on a codimension-1 defect \cite{Rapcak15}. Alternatively, we can $\beta$-deform the domain wall system that is dual to the D3-D5 intersection \cite{Widen18}. This should be similar to adding a probe D5-brane to the string theory dual of $\beta$-deformed $\N=4$ SYM \cite{LeighStrassler95a}, that is type IIB string theory on the Lunin-Maldacena (LM) background (AdS$_5\times\text{TsT transformed S}^5$) \cite{LuninMaldacena05}. For real values of the parameter $\beta$, the $\beta$-deformed AdS$_5$/CFT$_4$ correspondence is integrable \cite{Roiban03b, BerensteinCherkis04, BeisertRoiban05a, Frolov05}. On the other hand, the study of the two AdS/{\color{red}d}CFT setups (namely D3-NS5 and $\beta$-deformed D3-D5) at weak coupling suggests that they break integrability \cite{Rapcak15, Widen18}. However, we think that it would be interesting to support this conclusion with evidence from the string theory side by using the methods that we have described in the present section.\footnote{See the references \cite{PalRathiLalaRoychowdhury22, IshiiMurataYoshida23, GubarevMusaev24a} for some more methods which can be employed for the study of integrability of (brane) deformed systems. These methods could potentially complement the ones described here.} \\
\indent The branes that we have examined are mainly infinite-size flavor branes which have an AdS$_d$ geometry and a holographically dual dCFT$_d$. Even so, the methods that we have developed in this section may also apply to finite-size branes in AdS/CFT, such as giant gravitons (GGs) \cite{McGreevySusskindToumbas00}. Giant gravitons are supersymmetric branes which live in AdS space and are holographically dual to certain determinant and sub-determinant operators of the dual SCFT. The integrability of these configurations has been extensively studied on both sides of the AdS/CFT duality but many interesting questions remain unanswered.
\section[Holographic defect correlators]{Holographic defect correlators \label{Section:HolographicDefectCorrelators}}
\noindent The subject matter of the present section is the computation of correlation functions in strongly coupled CFTs and defect CFTs with semiclassical strings. As we have already mentioned in the introduction, the strong coupling regimes of quantum field theories (QFTs) are far beyond the reach of perturbation theory which is only applicable at weak gauge theory coupling. With holography \cite{Maldacena97, GubserKlebanovPolyakov98, Witten98a} we can probe the strongly coupled sides of QFTs, albeit at a relatively high price. Finiteness and increased (super) symmetry are present in a large number of holographic theories, making them highly unrealistic. \\
\indent What is more, the weak/strong coupling dilemma hinders our efforts to check the validity of holographic dualities. Whenever a holographic correspondence has an integrable structure however, its two disconnected ends (i.e.\ the weak and strong coupling regime) can be bridged. Apart from allowing high-precision tests of holography, integrability theoretically solves holographic dualities, that is it allows to compute all of their observables (spectrum, correlation functions, scattering amplitudes, form factors, Wilson loops, etc.) from weak to strong coupling. As of now, integrability has succeeded in solving the spectral problem in two main cases: the AdS$_5$/CFT$_4$ correspondence \eqref{AdS5CFT4duality} \cite{GromovKazakovLeurentVolin13} and the AdS$_4$/CFT$_3$ correspondence \eqref{AdS4CFT3duality} \cite{CavagliaFioravantiGromovTateo14}. The solution is generally not known in a closed form but the full set of equations which determine the spectrum has been specified. The computation of other observables of holographic theories is currently in progress. Correlation functions\footnote{That is three and higher-point functions. CFT two-point functions depend only on the scaling dimensions (or the spectrum), as we will see in subsection \ref{SubSection:ConformalCorrelationFunctions} below.} in particular can be computed within the so-called hexagon program which has been developed for AdS$_5$/CFT$_4$ in \cite{BassoKomatsuVieira15} and AdS$_3$/CFT$_2$ in \cite{EdenLePlatSfondrini21}. On the other hand, integrability and solvability usually depend on the presence of an infinite-dimensional symmetry (in accordance with the Liouville definition, see \S\ref{Section:StringIntegrabilityHolographicDefects}). Given that such a symmetry is quite rare, it is yet unknown whether integrability can be considered as a realistic property of quantum field theories. \\
\indent Another class of holographic theories, with reduced symmetry and supersymmetry, is significantly closer to realistic QFTs. One way to generate such holographies is by deforming the known holographic dualities. The deformation can take the form of an infinitely sized probe brane which is inserted on the gravitational (aka string theory) side \cite{KarchRandall01a, KarchRandall01b}. The probe brane induces a defect on the dual gauge theory side so that by starting with an AdS/CFT type of duality, we end up with an AdS/{\color{red}d}CFT duality of varying codimensionality. Because of the way these (AdS/{\color{red}d}CFT) theories are constructed, their spectrum usually coincides with that of the undeformed (AdS/CFT) duality. However, all the other observables are new, with correlation functions being especially interesting. \\
\indent At weak coupling, correlation functions can be computed perturbatively in the domain wall versions of all the codimension-1 defect CFTs we have described in section \ref{SubSection:StringIntegrabilitydCFT}. For the codimension-1 dCFT that is dual to the D3-D5 system (see \S\ref{SubSubSection:D3D5intersection}) tree-level one and two-point functions have been computed in \cite{NagasakiYamaguchi12, deLeeuwKristjansenZarembo15, Buhl-MortensenLeeuwKristjansenZarembo15, deLeeuwKristjansenMori16, deLeeuwKristjansenLinardopoulos18a, KristjansenMullerZarembo20a} and \cite{deLeeuwIpsenKristjansenVardinghusWilhelm17, Widen17, deLeeuwKristjansenLinardopoulosVolk23, BaermanChalabiKristjansen24}. At one-loop order, one-point functions have been computed in \cite{Buhl-MortensenLeeuwIpsenKristjansenWilhelm16a, Buhl-MortensenLeeuwIpsenKristjansenWilhelm16c}, while asymptotic all-loop proposals were made in \cite{Buhl-MortensenLeeuwIpsenKristjansenWilhelm17a, GomborBajnok20a, GomborBajnok20b, KristjansenMullerZarembo20b, KristjansenMullerZarembo21}. For the codimension-1 dCFTs that are dual to the D3-D7 system (see \S\ref{SubSubSection:D3D7intersection}), tree-level one-point functions have been computed in \cite{KristjansenSemenoffYoung12b, deLeeuwKristjansenLinardopoulos16, deLeeuwKristjansenVardinghus19, deLeeuwGomborKristjansenLinardopoulosPozsgay19, Linardopoulos25b}, while perturbative calculations were performed in \cite{GimenezGrauKristjansenVolkWilhelm18, GimenezGrauKristjansenVolkWilhelm19}. More recently, tree-level one-point functions have been computed for the codimension-1 dCFT that is dual to the D2-D4 intersection (described in \S\ref{SubSubSection:D2D4intersection}) in \cite{KristjansenVuZarembo21, Gombor21, GomborKristjansen22}, while one-point functions have been computed up to one-loop order in the codimension-3 dCFT that is dual to the D1-D3 intersection in \cite{KristjansenZarembo23, KristjansenZarembo24b}. As we have already mentioned in subsection \ref{SubSection:StringIntegrabilitydCFT}, closed-form determinant formulas for correlation functions is a smoking gun for the presence of integrability/solvability in a defect CFT. \\
\indent The computation of correlation functions in strongly coupled defect CFTs (i.e.\ in the regime where string theory is weakly coupled and therefore amenable to perturbative calculations) was recently put in a systematic framework \cite{GeorgiouLinardopoulosZoakos23}.\footnote{Conversely, there exist many computations of correlation functions which involve finite-sized branes such as giant gravitons. See e.g.\ \cite{BakChenWu11, BissiKristjansenYoungZoubos11, CaputaMelloKochZoubos12, HiranoKristjansenYoung12, Lin12, KristjansenMoriYoung15} for some interesting works.} One-point functions of protected chiral primary operators (CPOs) have been computed in \cite{NagasakiYamaguchi12} for the D3-D5 intersection (described in \S\ref{SubSubSection:D3D5intersection}), in \cite{KristjansenSemenoffYoung12b} for the D3-D7 intersection (described in \S\ref{SubSubSection:D3D7intersection}) and recently in \cite{KristjansenVuZarembo21} for the D2-D4 system (see \S\ref{SubSubSection:D2D4intersection}). See also the paper \cite{Buhl-MortensenLeeuwKristjansenZarembo15} for an interesting calculation of the one-point function of the vacuum state $\tr \left[Z^L\right]$ at strong coupling with strings. Correlation functions of strongly coupled dCFTs should also be computable with Witten diagrams in the supergravity approximation. \\
\indent Yet another powerful method which allows us to probe strongly coupled field theories is supersymmetric localization \cite{Pestun07} (more details can be found in the collection of articles \cite{Pestunetal16}). Even though the range of the method within AdS/CFT is somewhat restricted to free energies and expectation values of Wilson and 't Hooft loops \cite{Marino11, RussoZarembo13c, Zarembo16}, localization provides key insights about the validity of holography and the accuracy of string-theoretical predictions. Supersymmetric localization has also been introduced in AdS/{\color{red}d}CFT, mainly by Wang and Komatsu \cite{Wang20a, KomatsuWang20} (see also the paper \cite{BeccariaCaboBizet23} for a recent application), although the roots of the method should properly be traced back to the works of Robinson and Uhlemann \cite{RobinsonUhlemann17, Robinson17}. Once more, this method is a precious tool for the study of the AdS/{\color{red}d}CFT duality from weak to strong coupling, despite its crucial dependence on supersymmetry.
\subsection[Correlation functions in CFTs and dCFTs]{Correlation functions in CFTs and dCFTs \label{SubSection:ConformalCorrelationFunctions}}
\noindent Before continuing to the computation of correlation functions in strongly coupled CFTs and dCFTs, we will briefly revisit conformal and defect conformal field theories. We will only mention rudiments of these theories, namely correlation functions, operator expansions and bootstrapping. These will be necessary for the development of the holographic method for the computation of correlators in subsections \ref{SubSection:HolographicCorrelatorsCFTs} and \ref{SubSection:HolographicCorrelatorsdCFTs} below.
\subsubsection[Conformal field theories]{Conformal field theories \label{SubSubSection:ConformalFieldTheories}}
\noindent A well-known fact in conformal field theory (CFT) is that the form of CFT correlation functions is completely determined by conformal symmetry. One-point functions generally vanish, while scalar two and three-point functions are given by:\footnote{See any standard treatise of CFT, for instance \cite{DiFrancescoMathieuSenechal97}. Two-point functions will henceforth be normalized to unity. }
\begin{IEEEeqnarray}{c}
\left\langle\OO_1\left(\x_1\right)\right\rangle = 0, \quad (\text{except }\left\langle \mathbb{1}\right\rangle = 1) \label{OnePointFunctionsCFT} \\[6pt]
\left\langle\OO_1\left(\x_1\right)\OO_2\left(\x_2\right)\right\rangle = \frac{\delta_{12}}{\x_{12}^{\Delta_1 + \Delta_2}}, \quad \Delta \equiv \Delta_1 = \Delta_2 \label{TwoPointFunctionsCFT} \\
\left\langle\OO_1\left(\x_1\right)\OO_2\left(\x_2\right)\OO_3\left(\x_3\right)\right\rangle = \frac{\C_{123}}{\x_{12}^{\Delta_1 + \Delta_2 - \Delta_3} \x_{23}^{\Delta_2 + \Delta_3 - \Delta_1}\x_{31}^{\Delta_3 + \Delta_1 - \Delta_2}}, \label{ThreePointFunctionsCFT}
\end{IEEEeqnarray}
where $\Delta_{\mathfrak{i}}$ is the scaling dimension of the scalar operator $\OO_{\mathfrak{i}}$ ($\mathfrak{i} = 1,2,\ldots$), $\C_{123}$ is the three-point function structure constant and we have defined,
\begin{IEEEeqnarray}{l}
\x_{\mathfrak{ij}} \equiv \left|\x_{\mathfrak{i}} - \x_{\mathfrak{j}}\right|.
\end{IEEEeqnarray}
For fields with spin, such as conserved currents $V_{\mu}$ and the stress (aka energy-momentum) tensor $T_{\mu\nu}$,\footnote{In its improved version, that is (on-shell) traceless and symmetric.} similar results apply. Given that these fields generally obey,
\begin{IEEEeqnarray}{l}
\partial^{\mu}V_{\mu} = 0, \qquad \partial^{\mu}T_{\mu\nu} = 0, \qquad T_{\mu\nu} = T_{\nu\mu}, \qquad g^{\mu\nu} T_{\mu\nu} = 0, \label{SpinFieldsProperties}
\end{IEEEeqnarray}
the corresponding two-point functions take the following forms, in $d$-dimensional CFT \cite{OsbornPetkou93}:
\begin{IEEEeqnarray}{ll}
\left<V_{\mu}\left(\x_1\right) V_{\nu}\left(\x_2\right)\right> = \frac{C_{V}}{\x_{12}^{2(d-1)}} \cdot \I_{\mu\nu}\left(\x_1 - \x_2\right), \quad \left<T_{\mu\nu}\left(\x_1\right) T_{\rho\sigma}\left(\x_2\right)\right> = \frac{C_{T}}{\x_{12}^{2d}} \cdot \I_{\mu\nu\rho\sigma}\left(\x_1 - \x_2\right), \qquad
\end{IEEEeqnarray}
where the inversion tensors $\I_{\mu\nu}$, $\I_{\mu\nu\rho\sigma}$ are defined as
\begin{IEEEeqnarray}{l}
\I_{\mu\nu}\left(\x\right) \equiv g_{\mu\nu} - \frac{2\,\x_{\mu}\x_{\nu}}{\x^2}, \quad \I_{\mu\nu\rho\sigma}\left(\x\right) \equiv \frac{1}{2}\left(\I_{\mu\rho}\left(\x\right)\I_{\nu\sigma}\left(\x\right) + \I_{\mu\sigma}\left(\x\right)\I_{\nu\rho}\left(\x\right)\right) - \frac{1}{d}\,g_{\mu\nu}g_{\rho\sigma}. \qquad \label{InversionTensors}
\end{IEEEeqnarray}
If we have more than three points $n>3$, we may form $n(n-3)/2$ independent conformally invariant cross (aka anharmonic) ratios. The corresponding $n$-point correlation function ($n \geq 4$) will have a functional dependence on them, e.g.\ in the case of $n=4$ points there are 2 invariant cross ratios:
\begin{IEEEeqnarray}{c}
\frac{\x_{12}\x_{34}}{\x_{13}\x_{24}} \qquad \& \qquad \frac{\x_{12}\x_{34}}{\x_{14}\x_{23}}, \label{ConformalCrossRatios}
\end{IEEEeqnarray}
and the four-point scalar correlation function has an arbitrary dependence on them:
\begin{IEEEeqnarray}{c}
\left\langle\OO_1\left(\x_1\right)\OO_2\left(\x_2\right)\OO_3\left(\x_3\right)\OO_4\left(\x_4\right)\right\rangle = f\left(\frac{\x_{12}\x_{34}}{\x_{13}\x_{24}}, \frac{\x_{12}\x_{34}}{\x_{14}\x_{23}}\right) \cdot \prod_{\mathfrak{i}<\mathfrak{j}}^4 \x_{\mathfrak{ij}}^{\Delta/3 - \Delta_{\mathfrak{i}} - \Delta_{\mathfrak{j}}}, \quad \Delta \equiv \sum_{\mathfrak{i}=1}^4 \Delta_{\mathfrak{i}}. \qquad \label{FourPointFunctionsCFT}
\end{IEEEeqnarray}
\paragraph{Operator product expansion} One of the most important results of the Wightman formulation of quantum field theory is probably the reconstruction theorem \cite{Wightman56}. According to the reconstruction theorem, any QFT can be reformulated in terms of its local operators $\OO_{\mathfrak{i}}$ and their $n$-point correlators:
\begin{IEEEeqnarray}{ll}
\left\langle\OO_1\left(\x_1\right)\OO_2\left(\x_2\right)\ldots\OO_n\left(\x_n\right)\right\rangle. \label{CorrelationFunctionsQFT}
\end{IEEEeqnarray}
This way we can define a QFT even without a Lagrangian. In conformal field theories it is possible to reproduce the entire set of correlation functions \eqref{CorrelationFunctionsQFT} (and so the full theory \`{a} la Wightman) from the operator product expansion (OPE) and the conformal data. The latter is the set of all operator scaling dimensions $\Delta_{\mathfrak{i}}$, and two, three-point function structure constants $\left\{\C_{\mathfrak{ij}},\C_{\mathfrak{ijl}}\right\}$.\footnote{Scalar two-point functions have been normalized to unity, as we have just seen in \eqref{TwoPointFunctionsCFT}.} The conformal OPE reads, e.g.\ in the case of scalars:
\begin{IEEEeqnarray}{ll}
\OO_1\left(\x_1\right)\OO_2\left(\x_2\right) = \frac{\delta_{12}}{\x_{12}^{\Delta_1 + \Delta_2}} + \sum_{\mathfrak{j} > \mathbb{1}} \frac{\C_{12}^{\mathfrak{j}}}{\x_{12}^{\Delta_1 + \Delta_2 - \Delta_{\mathfrak{j}}}} \cdot \mathcal{P}_{\mathfrak{j}}\left(\x_{12}, \partial_2\right)\OO_{\mathfrak{j}}\left(\x_2\right), \qquad \label{OperatorProductExpansion}
\end{IEEEeqnarray}
where $\mathcal{P}_{\mathfrak{j}}$ are differential operators\footnote{Normalized as $\mathcal{P}_{\mathfrak{j}} = 1 + \OO\left(\x_{12}\right)$.} ($\mathfrak{j}$ is a group index) and the sum is over all the primary operators (with or without spin). In contrast to the usual OPE, which converges only in the short-distance limit $\x_{12} \rightarrow 0$, the conformal OPE \eqref{OperatorProductExpansion} converges even for finite values of the distance $\x_{12}$ \cite{PappadopuloRychkovEspinRattazzi12}. This implies that the correlators \eqref{CorrelationFunctionsQFT} can all be determined recursively from the conformal data $\left\{\Delta_{\mathfrak{i}}, \C_{\mathfrak{ij}}, \C_{\mathfrak{i}\mathfrak{j}\mathfrak{l}}\right\}$. For example, scalar $(n+2)$-point correlation functions can be expressed in terms of $(n+1)$-point correlation functions as:
\begin{IEEEeqnarray}{c}
\big\langle\OO_1\left(\x_1\right)\OO_2\left(\x_2\right)\prod_{\mathfrak{i}=3}^{n}\OO_{\mathfrak{i}}\left(\x_{\mathfrak{i}}\right)\big\rangle = \sum_{\mathfrak{j} \geq \mathbb{1}} \C_{12}^{\mathfrak{j}} \cdot \mathcal{P}_{\mathfrak{j}}\left(\x_{12},\partial_2\right)\big\langle\OO_{\mathfrak{j}}\left(\x_2\right)\prod_{\mathfrak{i}=3}^{n}
\OO_{\mathfrak{i}}\left(\x_{\mathfrak{i}}\right)\big\rangle. \label{ConformalCorrelationFunctions}
\end{IEEEeqnarray}
The other major implication of the convergent OPE \eqref{OperatorProductExpansion} is the conformal bootstrap program \cite{FerraraGrilloGatto73a, Polyakov74a}. By computing \eqref{ConformalCorrelationFunctions} in all possible channels of a CFT, the conformal bootstrap program aims in computing and constraining its conformal data.
\subsubsection[Defect conformal field theories]{Defect conformal field theories \label{SubSubSection:DefectConformalFieldTheories}}
\noindent Conformal field theory in the presence of a codimension-1 wall (boundary or defect) was studied long ago by Cardy \cite{Cardy84a}. Take a CFT in $d$ spacetime dimensions $x = \left(z,\textbf{\textit{x}}\right)$ and insert a $d-1$ dimensional planar boundary/defect at $z=0$. The subgroup of the $d$-dimensional (Euclidean) conformal group $SO\left(d+1,1\right)$ which leaves the $z=0$ plane invariant consists of:
\begin{itemize}
\item $(d-1)$-dimensional translations, $\textbf{\textit{x}}' = \textbf{\textit{x}} + \textbf{\textit{a}}$, where $\textbf{\textit{a}}$ is a vector in the $z = 0$ plane
\item $(d-1)$-dimensional rotations of $\textbf{\textit{x}}$ (i.e.\ invariance under the group $SO(d-1)$)
\item $d$-dimensional rescalings, $x_{\mu}' = c \cdot x_{\mu}$ (with $c$ a constant) and inversions, $x_{\mu}' = x_{\mu}/x^2$.
\end{itemize}
\noindent These transformations span the $d-1$ dimensional (Euclidean) conformal group $SO\left(d,1\right)$. The resulting theory is a boundary/defect CFT (BCFT/dCFT). Defects and boundaries of higher dimensionalities $p$ and codimensionalities $q$ (such that $p + q = d$) are also possible. In the present work we only examine BCFTs and dCFTs with codimension-1 ($q = 1$) boundaries and defects.
\begin{figure}[H]\begin{center}\includegraphics[scale=0.32]{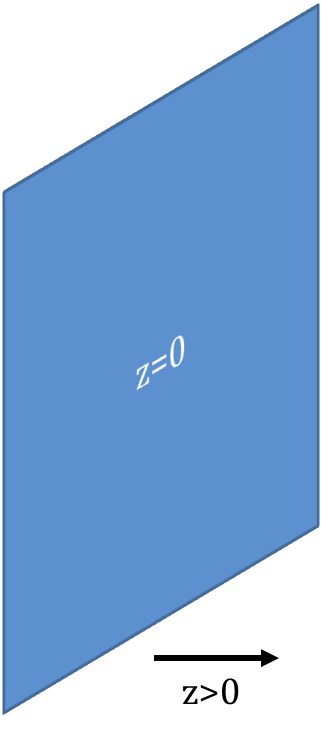}\caption{Boundary/defect CFT with a codimension-1 boundary/defect.}\label{Figure:Boundary}\end{center}\end{figure}
\vspace{-.4cm}\indent For simplicity, we only consider codimension-1 defects, i.e.\ $z$ can be both positive and negative. Extension of the results below to the case of boundaries ($z>0$) is quite straightforward. Because of the defect at $z=0$, we may form dCFT invariant ratios out of only 2 ambient points:\footnote{The term \quotes{ambient} denotes the space far away from the boundary/defect of BCFTs and dCFTs, while \quotes{bulk} will be used to describe the space where the gravity duals of holographic theories live.}
\begin{IEEEeqnarray}{c}
\xi = \frac{\x_{12}^2}{4\left|\z_1\right|\left|\z_2\right|}, \qquad v^2 = \frac{\xi}{\xi + 1} = \frac{\x_{12}^2}{\x_{12}^2 + 4\left|\z_1\right|\left|\z_2\right|}. \qquad \label{DefectInvariantRatios}
\end{IEEEeqnarray}
It follows that ambient one-point functions of scalar operators are generally nonzero and the only ones to be fully determined by defect symmetry:
\begin{IEEEeqnarray}{l}
\left\langle\OO_1\left(\z_1,\textbf{x}_1\right)\right\rangle = \frac{\C_1}{\left|\z_1\right|^{\Delta_1}}, \label{OnePointFunctionsDefectCFT}
\end{IEEEeqnarray}
where $\C_1$ is the one-point function structure constant. As expected, higher-point ambient correlation functions (involving $n$ points, with $n \geq 2$) contain an arbitrary dependence on the dCFT invariants \eqref{DefectInvariantRatios}. Two-point scalar correlation functions are in particular given by
\begin{IEEEeqnarray}{l}
\left\langle\OO_1\left(\z_1,\textbf{x}_1\right)\OO_2\left(\z_2,\textbf{x}_2\right)\right\rangle = \frac{f_{12}\left(\xi\right)}{\left|\z_1\right|^{\Delta_1}\left|\z_2\right|^{\Delta_2}}, \label{TwoPointFunctionsDefectCFT}
\end{IEEEeqnarray}
i.e.\ they no longer vanish when $\Delta_1 \neq \Delta_2$, cf.\ \eqref{TwoPointFunctionsCFT}. There is an obvious analogy between ambient $n$-point correlation functions of dCFTs and $(n+2)$-point correlation functions of CFTs ($n = 1,2,\ldots$), since they both depend on invariant ratios with the same number of points. This way, ambient one-point functions of dCFTs \eqref{OnePointFunctionsDefectCFT} are analogous to three-point functions of CFTs \eqref{ThreePointFunctionsCFT}, since they both contain the maximum set of points for which the correlator does not depend on any invariant ratio. On the other hand, the ambient two-point functions \eqref{TwoPointFunctionsDefectCFT} of dCFTs are very similar to the four-point functions \eqref{FourPointFunctionsCFT} of CFTs, because their forms depend on the simplest invariant ratios \eqref{DefectInvariantRatios} and \eqref{ConformalCrossRatios} respectively (i.e.\ having the minimal number of points). \\
\indent For fields with spin, such as conserved currents $V_{\mu}$ and the stress tensor $T_{\mu\nu}$ (obeying \eqref{SpinFieldsProperties}), one-point functions vanish \cite{McAvityOsborn93, McAvityOsborn95},
\begin{IEEEeqnarray}{ll}
\left<V_{\mu}\left(\x_1\right)\right> = \left<T_{\mu\nu}\left(\x_1\right)\right> = 0, \qquad \x_{\mathfrak{i}} = \left(\z_{\mathfrak{i}}, \textbf{x}_{\mathfrak{i}}\right),
\end{IEEEeqnarray}
whereas two-point functions are given by:
\begin{IEEEeqnarray}{c}
\left<V_{\mu}\left(\x_1\right)V_{\nu}\left(\x_2\right)\right> = \frac{1}{\x_{12}^{2(d-1)}} \Big[\I_{\mu\nu}C\left(v\right) - X_{\mu}X_{\nu}'D\left(v\right)\Big] \\[6pt]
\left<T_{\mu\nu}\left(\x_1\right)T_{\rho\sigma}\left(\x_2\right)\right> = \frac{1}{\x_{12}^{2d}} \cdot \Bigg\{\left(X_{\mu}X_{\nu} - \frac{g_{\mu\nu}}{d}\right)\left(X'_{\rho}X'_{\sigma} - \frac{g_{\rho\sigma}}{d}\right) A\left(v\right) + \Big(X_{\mu}X'_{\rho}I_{\nu\sigma}+ X_{\mu}X'_{\sigma}I_{\nu\rho} + \nonumber \\
+ X_{\nu}X'_{\sigma}I_{\mu\rho} + X_{\nu}X'_{\rho}I_{\mu\sigma} - \frac{4}{d} \, g_{\mu\nu}X'_{\rho}X'_{\sigma} - \frac{4}{d} \, g_{\rho\sigma}X_{\mu}X_{\nu} + \frac{4}{d^2}\,g_{\mu\nu}g_{\rho\sigma}\Big) B\left(v\right) + \I_{\mu\nu\rho\sigma} C\left(v\right)\Bigg\}, \qquad
\end{IEEEeqnarray}
where $A\left(v\right)$, $B\left(v\right)$, $C\left(v\right)$ are functions of the invariant $v$ (defined in \eqref{DefectInvariantRatios} above), while the inversion tensors $\I_{\mu\nu}$ and $\I_{\mu\nu\rho\sigma}$ are given by \eqref{InversionTensors}. We have also set,
\begin{IEEEeqnarray}{l}
X_{\mu} \equiv \z_1 \cdot \frac{v}{\xi} \frac{\partial\xi}{\partial \x_1^{\mu}} = v\left(\frac{2\z_1}{\x_{12}^2}\left(\x_{1\mu}-\x_{2\mu}\right) - n_{\mu}\right) \\
X'_{\rho} \equiv \z_2 \cdot \frac{v}{\xi} \frac{\partial\xi}{\partial \x_2^{\rho}} = -v\left(\frac{2\z_2}{x_{12}^2}\left(\x_{1\rho}-\x_{2\rho}\right) + n_{\rho}\right), \qquad
\end{IEEEeqnarray}
where $n \equiv \left(1,\textbf{0}\right)$ is the unit normal to the $z = 0$ defect. Notice that $X$, $X'$ obey
\begin{IEEEeqnarray}{l}
X_{\mu}X_{\mu} = X_{\rho}'X_{\rho}' = 1, \qquad X_{\rho}' = \I_{\rho\mu}X_{\mu}.
\end{IEEEeqnarray}
\paragraph{Boundary operator expansion} Conformal symmetry is intact on the $z = 0$ defect, so that the correlation functions of defect operators satisfy the usual relations of CFT$_{(d-1)}$. In the case of scalar operators,
\begin{IEEEeqnarray}{c}
\big\langle\widehat{\OO}_1\left(\x_1\right)\big\rangle = 0, \quad (\text{except }\big\langle \widehat{\mathbb{1}}\big\rangle = 1) \\[6pt]
\big\langle\widehat{\OO}_1\left(\textbf{x}_1\right)\widehat{\OO}_2\left(\textbf{x}_2\right)\big\rangle = \frac{\widehat{\B}_{12}}{\textbf{x}_{12}^{2\hat{\Delta}}}, \qquad \hat{\Delta} \equiv \hat{\Delta}_1 = \hat{\Delta}_2 \qquad \label{TwoPointFunctionsDefect} \\[6pt]
\big\langle\widehat{\OO}_1\left(\textbf{x}_1\right)\widehat{\OO}_2\left(\textbf{x}_2\right)\widehat{\OO}_3\left(\textbf{x}_3\right)\big\rangle = \frac{\widehat{\B}_{123}}{\textbf{x}_{12}^{\hat{\Delta}_1 + \hat{\Delta}_2 - \hat{\Delta}_3} \textbf{x}_{23}^{\hat{\Delta}_2 + \hat{\Delta}_3 - \hat{\Delta}_1} \textbf{x}_{31}^{\hat{\Delta}_3 + \hat{\Delta}_1 - \hat{\Delta}_2}},
\end{IEEEeqnarray}
where $\hat{\Delta}_{\mathfrak{i}}$ is the scaling dimension of the defect scalar operator $\widehat{\OO}_{\mathfrak{i}}$, and $\widehat{\B}_{12}$, $\widehat{\B}_{123}$ are the scalar structure constants of defect two and three-point functions. We have also defined
\begin{IEEEeqnarray}{l}
\textbf{x}_{12} \equiv \left|\textbf{x}_1 - \textbf{x}_2\right|.
\end{IEEEeqnarray}
Higher-point correlation functions of defect operators will again have an explicit dependence on the defect CFT$_{d-1}$ cross ratios. We may also consider mixed correlators of defect and ambient operators. For example, the ambient-defect two-point function of scalars has the following form \cite{McAvityOsborn95}:
\begin{IEEEeqnarray}{l}
\big\langle\OO_1\left(\z_1,\textbf{x}_1\right)\widehat{\OO}_2\left(\textbf{x}_2\right)\big\rangle = \frac{\B_{12}}{\left|\z_1\right|^{\Delta_1 - \hat{\Delta}_2}\x_{12}^{2\hat{\Delta}_2}}, \qquad \x_{12}^2 = \z_1^2 + \left(\textbf{x}_1 - \textbf{x}_2\right)^2,
\end{IEEEeqnarray}
where $\B_{12}$ is the scalar ambient-defect two point function structure constant. The fusion of ambient fields with the defect is described by the (convergent) boundary operator expansion (BOE) which reads:
\begin{IEEEeqnarray}{l}
\OO_1\left(\x_1\right) = \frac{\C_1}{\left|\z_1\right|^{\Delta_1}} + \sum_{\mathfrak{j}} \frac{\B_{1\mathfrak{j}}}{\left|\z_1\right|^{\Delta_1 - \hat{\Delta}_{\mathfrak{j}}}}\cdot\widehat{\mathcal{P}}_{\mathfrak{j}}\big(\z_1,\partial_{\textbf{x}_1}\big) \widehat{\OO}_{\mathfrak{j}}\left(\textbf{x}_1\right), \label{BoundaryOperatorExpansion}
\end{IEEEeqnarray}
where $\widehat{\mathcal{P}}_{\mathfrak{j}}$ are differential operators\footnote{Normalized once more as $\widehat{\mathcal{P}}_{\mathfrak{j}} = 1 + \OO\left(\z^2\right)$. Again, $\mathfrak{j}$ is a group index.} and the sum is over all the defect primary operators.
\paragraph{Boundary conformal bootstrap} Because of the way we constructed the defect CFT, that is by taking a pure CFT and adding a codimension-1 defect at $z = 0$, the conformal OPE (given by \eqref{OperatorProductExpansion} in the case of scalars) is obviously still valid for ambient dCFT operators independently of the presence of defects. It follows that the full set of defect correlation functions
\begin{IEEEeqnarray}{ll}
\big\langle\OO_1\left(\x_1\right)\ldots\widehat{\OO}_2\left(\x_2\right)\ldots\big\rangle, \label{CorrelationFunctionsDefectCFT}
\end{IEEEeqnarray}
can be determined recursively from the conformal data of the pure CFT$_d$ (i.e.\ scaling dimensions $\Delta_{\mathfrak{i}}$ and two, three-point function structure constants $\left\{\C_{\mathfrak{ij}}, \C_{\mathfrak{ijl}}\right\}$ of ambient operators $\OO_{\mathfrak{i}}$) and the defect CFT data, i.e.\ the one-point function structure constants $\C_{\mathfrak{i}}$ of ambient operators $\OO_{\mathfrak{i}}$, the conformal data of the defect CFT$_{d-1}$ (i.e.\ scaling dimensions $\hat{\Delta}_{\mathfrak{i}}$, two and three-point function structure constants $\{\widehat{\B}_{\mathfrak{ij}}, \widehat{\B}_{\mathfrak{ijk}}\}$ of defect operators $\widehat{\OO}_{\mathfrak{i}}$) and ambient-defect two-point function structure constants $\B_{\mathfrak{ij}}$. This time however, we have three operator expansions at our disposal, the ambient OPE (given by \eqref{OperatorProductExpansion} in the case of scalars), the defect OPE (for the defect CFT$_{(d-1)}$), and the BOE (given by \eqref{BoundaryOperatorExpansion} in the case of scalars). \\
\indent The bootstrap philosophy naturally carries over to defect CFTs \cite{LiendoRastellivanRees12}. The boundary conformal bootstrap program aims at solving (i.e.\ determining the minimal set of independent dCFT data) and constraining defect CFTs by applying the various product expansions in all possible channels. Take ambient two-point functions of scalar operators for example. In the ambient channel we may read off the ambient two-point function from the scalar conformal OPE \eqref{OperatorProductExpansion}:
\begin{IEEEeqnarray}{c}
\left\langle\OO_1\left(\x_1\right)\OO_2\left(\x_2\right)\right\rangle = \frac{\delta_{12}}{\x_{12}^{\Delta_1 + \Delta_2}} + \sum_{\mathfrak{j}} \frac{\C_{12}^{\mathfrak{j}}}{\x_{12}^{\Delta_1 + \Delta_2 - \Delta_{\mathfrak{j}}}} \cdot \mathcal{P}_{\mathfrak{j}}\left(\x_{12},\partial_2\right)\left\langle\OO_{\mathfrak{j}}\left(\x_2\right)\right\rangle,
\end{IEEEeqnarray}
so that by plugging the one and two-point functions \eqref{OnePointFunctionsDefectCFT}, \eqref{TwoPointFunctionsDefectCFT} we are led to\footnote{The extra factor $2^{\Delta_I}$ is necessary because our definition of one-point functions \eqref{OnePointFunctionsDefectCFT} does not include the standard factor of $2x_3$ in the denominator, cf.\ \cite{McAvityOsborn95, LiendoRastellivanRees12}.}
\begin{IEEEeqnarray}{ll}
f_{12}\left(\xi\right) = \left(4\xi\right)^{-\frac{\Delta_1 + \Delta_2}{2}} \left[\delta_{12} + \sum_{\mathfrak{j}} 2^{\Delta_{\mathfrak{j}}}\C_{12}^{\mathfrak{j}} \, \C_{\mathfrak{j}} \cdot F_{\text{ambient}}\left(\Delta_{\mathfrak{j}}, \delta\Delta,\xi\right)\right], \qquad \delta\Delta \equiv \Delta_1-\Delta_2, \label{TwoPointFunctionsAmbientChannel}
\end{IEEEeqnarray}
where the ambient conformal blocks $F_{\text{ambient}}$ can be determined from $\mathcal{P}_{\mathfrak{j}}\left(\x_{12},\partial_2\right) \left|\z_2\right|^{-\Delta_{\mathfrak{j}}}$ \cite{McAvityOsborn95, LiendoRastellivanRees12}:
\begin{IEEEeqnarray}{ll}
F_{\text{ambient}}\left(\Delta,\delta\Delta,\xi\right) = \xi^{\frac{\Delta}{2}} \, {_2}F_1\Big(\frac{\Delta + \delta\Delta}{2}, \frac{\Delta + \delta\Delta}{2},\Delta - 1;-\xi\Big). \label{AmbientConformalBlocks}
\end{IEEEeqnarray}
In the boundary/defect channel, we may compute the ambient two-point function of scalars from the scalar boundary operator expansion \eqref{BoundaryOperatorExpansion}:
\begin{IEEEeqnarray}{ll}
\left\langle\OO_1\left(\x_1\right)\OO_2\left(\x_2\right)\right\rangle = &\frac{\C_1\C_2}{\left|\z_1\right|^{\Delta_1} \left|\z_2\right|^{\Delta_2}} + \nonumber \\
&+ \sum_{\mathfrak{i},\mathfrak{j}} \frac{\B_{1\mathfrak{i}}\B_{2\mathfrak{j}}}{\left|\z_1\right|^{\Delta_1 - \hat{\Delta}_{\mathfrak{i}}} \left|\z_2\right|^{\Delta_2 - \hat{\Delta}_{\mathfrak{j}}}} \cdot \widehat{\mathcal{P}}_{\mathfrak{i}}\big(\z_1,\partial_{\textbf{x}_1}\big) \widehat{\mathcal{P}}_{\mathfrak{j}}\big(\z_2,\partial_{\textbf{x}_2}\big) \big\langle\widehat{\OO}_{\mathfrak{i}}\left(\textbf{x}_1\right) \widehat{\OO}_{\mathfrak{j}}\left(\textbf{x}_2\right)\big\rangle. \qquad
\end{IEEEeqnarray}
Plugging the two-point functions \eqref{TwoPointFunctionsDefectCFT}, \eqref{TwoPointFunctionsDefect} we find
\begin{IEEEeqnarray}{c}
f_{12}\left(\xi\right) = \C_1\C_2 + \sum_{\mathfrak{j}} \B_{1\mathfrak{j}}\B_{2}^{\mathfrak{j}} \cdot F_{\text{boundary}}\big(\hat{\Delta}_{\mathfrak{j}}, \xi\big), \label{TwoPointFunctionsBoundaryChannel}
\end{IEEEeqnarray}
where we have contracted the indices $\mathfrak{i},\mathfrak{j}$ inside the sum by $\widehat{\B}_{\mathfrak{i}\mathfrak{j}}$. The boundary/defect conformal blocks $F_{\text{boundary}}$ can be determined by computing $\widehat{\mathcal{P}}_{\mathfrak{i}}\big(\z_1,\partial_{\textbf{x}_1}\big) \widehat{\mathcal{P}}_{\mathfrak{j}}\big(\z_2,\partial_{\textbf{x}_2}\big) \textbf{x}_{12}^{-(\hat{\Delta}_{\mathfrak{i}} + \hat{\Delta}_{\mathfrak{j}})}$:
\begin{IEEEeqnarray}{c}
F_{\text{boundary}}\left(\Delta, \xi\right) = \xi^{-\Delta} \, {_2}F_1\left(\Delta, \Delta - 1, 2\Delta - 2;-\xi^{-1}\right). \label{BoundaryConformalBlocks}
\end{IEEEeqnarray}
Equating the two expressions for $f_{12}\left(\xi\right)$ that we have found in ambient space \eqref{TwoPointFunctionsAmbientChannel} and the boundary/defect channel \eqref{TwoPointFunctionsBoundaryChannel},
\begin{IEEEeqnarray}{ll}
f_{12}\left(\xi\right) &= \left(4\xi\right)^{-\frac{\Delta_1 + \Delta_2}{2}} \left[\delta_{12} + \sum_{\mathfrak{j}} 2^{\Delta_{\mathfrak{j}}}\C_{12}^{\mathfrak{j}} \, \C_{\mathfrak{j}} \cdot \xi^{\frac{\Delta_{\mathfrak{j}}}{2}} \, {_2}F_1\Big(\frac{\Delta_{\mathfrak{j}} + \delta\Delta}{2}, \frac{\Delta_{\mathfrak{j}} + \delta\Delta}{2},\Delta_{\mathfrak{j}} - 1;-\xi\Big)\right] = \nonumber \\[6pt]
& = \C_1\C_2 + \sum_{\mathfrak{j}} \B_{1\mathfrak{j}}\B_{2}^{\mathfrak{j}} \cdot \xi^{-\hat{\Delta}_{\mathfrak{j}}} \, {_2}F_1\Big(\hat{\Delta}_{\mathfrak{j}}, \hat{\Delta}_{\mathfrak{j}} - 1, 2\hat{\Delta}_{\mathfrak{j}} - 2;-\xi^{-1}\Big), \label{BootstrapEquations}
\end{IEEEeqnarray}
we may extract a set of bootstrap equations. These may be used to compute and/or constrain the defect as well as the pure CFT data $\{\Delta_{\mathfrak{i}}, \C_{\mathfrak{ijl}}, \C_{\mathfrak{i}}, \hat{\Delta}_{\mathfrak{i}}, \widehat{\B}_{\mathfrak{ij}}, \widehat{\B}_{\mathfrak{ijk}}, \B_{\mathfrak{ij}}\}$ \cite{LiendoRastellivanRees12, GliozziLiendoMeineriRago15, BilloGoncalvesLauriaMeineri16, LiendoMeneghelli16, Hogervorst17}. For the defect CFT that is dual to the D3-D5 intersection (see \S\ref{SubSubSection:D3D5intersection} above), we may employ its domain wall description \cite{NagasakiTanidaYamaguchi11, NagasakiYamaguchi12} in order to compute scalar two-point functions for various operators at weak 't Hooft coupling. Then, the bootstrap equations \eqref{BootstrapEquations} may be used in order to mine for (unknown) conformal data. This approach was put forward in \cite{deLeeuwIpsenKristjansenVardinghusWilhelm17}.

\begin{figure}[H]\begin{center}\begin{tikzpicture}\begin{feynman}
\fill [lightgray] (-.785,-1) rectangle (+0.785,-1.2);
\vertex (a) [dot] at (0,0) {};
\vertex (b) [empty dot] at (-.5,.7) {};
\vertex (c) [empty dot] at (+.5,.7) {};
\vertex (d) [dot] at (0,-.98) {};
\vertex (f1) at (-.9,-.98) {};
\vertex (f2) at (+.9,-.98) {};
\vertex (O1) at (-.7,1) {$\OO_1$};
\vertex (O2) at (+.7,1) {$\OO_2$};
\vertex (Oj) at (+.3,-.5) {\color{blue}$\OO_{\mathfrak{j}}$};
\vertex (sum) at (-1.4,-.5) {\huge$\sum\limits_{\scaleto{\mathfrak{j}\mathstrut}{10pt}}$};
\vertex (correlator) at (-2.7,-.25) {$\left\langle\OO_1\OO_2\right\rangle = $};
\vertex (equals) at (1.4,-.3) {$=$};
\diagram* [line width=0.3mm, black] {(c) -- (a) -- (b)};
\diagram* [line width=0.3mm, blue] {(a) -- (d)};
\diagram* [line width=0.3mm, black] {(f1) -- (f2)};
\fill [lightgray] (3.3-.785,-1) rectangle (3.3+0.785,-1.2);
\vertex (a) [dot] at (3.3+.5,-.98) {};
\vertex (b) [empty dot] at (3.3-.5,.7) {};
\vertex (c) [empty dot] at (3.3+.5,.7) {};
\vertex (d) [dot] at (3.3-.5,-.98) {};
\vertex (f1) at (3.3-.9,-.98) {};
\vertex (f2) at (3.3+.9,-.98) {};
\vertex (O1) at (3.3-.5,1) {$\OO_1$};
\vertex (O2) at (3.3+.5,1) {$\OO_2$};
\vertex (Oj) at (3.3,-.6) {\color{red}$\widehat{\OO}_{\mathfrak{j}}$};
\vertex (sum) at (3-1,-.5) {\huge$\sum\limits_{\scaleto{\mathfrak{j}\mathstrut}{10pt}}$};
\diagram* [line width=0.3mm, black] {(f1) -- (f2)};
\diagram* [line width=0.3mm, black] {(a) -- (c)};
\diagram* [line width=0.3mm, black] {(b) -- (d)};
\diagram* [line width=0.3mm, red] {(a) -- (d)};
\end{feynman}\end{tikzpicture}\caption{Bootstrapping ambient two-point functions of scalars in defect CFT.}\label{Figure:WittenDiagrams}\end{center}\end{figure}
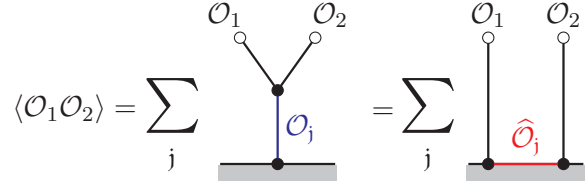
\vspace{-.4cm}\indent Let us now concentrate on the ambient channel and form the ratio of the ambient defect CFT two-point function \eqref{TwoPointFunctionsDefectCFT} of two scalar operators, over their pure CFT two-point function \eqref{TwoPointFunctionsCFT}:
\begin{IEEEeqnarray}{c}
\frac{\langle\OO_{1}\left(\x_1\right)\OO_{2}\left(\x_2\right)\rangle_{\text{{\color{red}d}CFT}}}{\langle\OO_{1}\left(\x_1\right)\OO_{2} \left(\x_2\right)\rangle_{\text{CFT}}} = \left(4\xi\right)^{\frac{\Delta_1 + \Delta_2}{2}} \frac{f_{12}\left(\xi\right)}{\delta_{12}} = 1 + \sum_{\mathfrak{j}} 2^{\Delta_{\mathfrak{j}}}\C_{12}^{\mathfrak{j}} \, \C_{\mathfrak{j}} \xi^{\frac{\Delta_{\mathfrak{j}}}{2}} \, {_2}F_1\Big(\frac{\Delta_{\mathfrak{j}}}{2}, \frac{\Delta_{\mathfrak{j}}}{2},\Delta_{\mathfrak{j}} - 1;-\xi\Big), \qquad
\end{IEEEeqnarray}
where we have replaced $f_{12}$ with its ambient channel value \eqref{TwoPointFunctionsAmbientChannel}--\eqref{AmbientConformalBlocks}. In the case of a single exchanged scalar primary operator $\OO_I$, of dimension $\Delta_I = L = 2j$, we get
\begin{IEEEeqnarray}{c}
\frac{\langle\OO_{1}\left(\x_1\right)\OO_{2}\left(\x_2\right)\rangle_{\text{{\color{red}d}CFT}}}{\langle\OO_{1}\left(\x_1\right)\OO_{2} \left(\x_2\right)\rangle_{\text{CFT}}} = 1 + 2^L \C_{12}^{I} \, \C_I \, \xi^{\frac{L}{2}} \cdot {_2}F_1\Big(\frac{L}{2}, \frac{L}{2},L - 1;-\xi\Big),
\end{IEEEeqnarray}
so that by expanding the hypergeometric function around $\xi = 0$, we are led to
\begin{IEEEeqnarray}{c}
\frac{\langle\OO_{1}\left(\x_1\right)\OO_{2}\left(\x_2\right)\rangle_{\text{{\color{red}d}CFT}}}{\langle\OO_{1}\left(\x_1\right)\OO_{2} \left(\x_2\right)\rangle_{\text{CFT}}} = 1 + 2^{L} \C_{12}^{I} \, \C_I \, \xi^j \cdot \Bigg\{1 - \frac{j^2}{2j-1} \cdot\xi + \frac{j(j+1)^2}{4(2j-1)} \cdot\xi^2 + \ldots\Bigg\}. \label{TwoPointFunctionRatio1}
\end{IEEEeqnarray}
In the following we will verify that \eqref{TwoPointFunctionRatio1} holds at strong 't Hooft coupling in the dCFT that is dual to the D3-D5 intersection, to leading order in $\xi$ for any two heavy operators and to all orders in $\xi$ for two heavy BMN operators \cite{GeorgiouLinardopoulosZoakos23}. In order to be able to proceed, we must first set up the computation of generic defect CFT correlation functions at strong 't Hooft coupling with semiclassical strings.
\subsection[Holographic correlators in CFTs]{Holographic correlators in CFTs \label{SubSection:HolographicCorrelatorsCFTs}}
\noindent Holography is an invaluable resource for the study QFTs at strong coupling, despite the obvious drawback that it is usually only possible when the theory has enhanced symmetry such as (super) conformal or defect (super) conformal symmetry. On the other hand, the form of correlation functions in CFTs and dCFTs is very concrete, while these theories can also be bootstrapped by means of converging OPEs (see \S\ref{SubSubSection:ConformalFieldTheories}--\S\ref{SubSubSection:DefectConformalFieldTheories} above). The presence of integrability greatly facilitates and sometimes even solves the problem of structure constants in the planar limit. Before turning to the computation of correlation functions in strongly coupled (holographic) defect CFTs, let us briefly go through the computation of correlation functions in AdS/CFT.
\subsubsection[Weak coupling]{Weak coupling \label{SubSection:HolographicCorrelatorsCFTsWeak}}
\noindent Correlation functions can be computed at any given perturbative order in quantum field theory by Wick contracting fields and applying Feynman rules. Typically this is a combinatorial problem of increasing complexity, depending on the number of performed contractions. In integrable (2 dimensional) IQFTs, correlation functions can be computed with the form factor/TBA approach \cite{Zamolodchikov91a, Smirnov92e}. \\
\indent The computation of correlation functions in holographic theories (such as the gauge theories which appear on the lhs of the dualities \eqref{AdS5CFT4duality}, \eqref{AdS4CFT3duality}) is also amenable to huge simplifications. This is due to the fact that the dilatation operators of these theories (which control operator scaling dimensions as we have seen) are given by the Hamiltonians of integrable spin chains. The spin chains can be diagonalized by the Bethe ansatz so that the objects which partake in correlation functions are gauge invariant operators of definite scaling dimensions $\Delta$. The form of these operators is fully specified by the Bethe ansatz. This allows to set up a very systematic framework for the computation of correlation functions.
\paragraph{Two-point functions} Scalar two-point functions have been normalized to unity while they are only nonzero when the operator scaling dimensions are identical (cf.\ \eqref{TwoPointFunctionsCFT}). As a consequence, their computation is essentially equivalent to that of computing operator scaling dimensions, i.e.\ the theory's spectrum. To compute the spectrum in theories such as $\N = 4$ SYM in \eqref{LagrangianSYM} and ABJM in \eqref{LagrangianABJM}, the corresponding dilatation operator (which is given by the Hamiltonian of an integrable spin chain as we have explained) must be diagonalized. This is accomplished by the Bethe ansatz. \\
\indent The generic form of the perturbative expansion for the scaling dimensions of scalar gauge invariant single-trace operators in planar weakly-coupled $\N = 4$ SYM theory is:
\begin{IEEEeqnarray}{l}
\Delta = \Delta^{(0)} + \gamma^{(1)}\lambda + \gamma^{(2)}\lambda^2 + \ldots, \qquad \lambda \rightarrow 0. \label{ScalingWeakCouplingExpansionSYM}
\end{IEEEeqnarray}
The coupling constant coefficients in the above expansion generally depend on the geometric characteristics of the operator (spin chain), such as its (finite) size and momenta of the various excitations (Bethe roots). For operators of finite size, spin chain interactions start to wrap around the operator when the loop order becomes equal to the operator length. Wrapping corrections settle in the spectrum at this critical loop order (and above it). They cannot be accounted for by the Bethe ansatz, which enters its so-called asymptotic regime (see \cite{Gombor22a} for a recent study of wrapping effects in long-range spin chains). Wrapping corrections are encoded in the TBA and the QSC however, both of which include Lüscher corrections, i.e.\ the effect of virtual particles circulating around the spin chain/worldsheet cylinder (more can be found in the reviews \cite{Sieg10b, Janik10, Bajnok10}). \\
\indent As we have already mentioned, the spectral problems of both $\N = 4$ SYM \eqref{LagrangianSYM} and ABJM theory \eqref{LagrangianABJM} have been solved by the quantum spectral curve method (QSC) in the planar limit \cite{GromovKazakovLeurentVolin13, CavagliaFioravantiGromovTateo14}. More recently a QSC was proposed for the AdS$_3\times\text{S}^3\times\text{T}^4$ formulation of AdS$_3$/CFT$_2$ duality \cite{CavagliaGromovStefanskiTorrielli21, EkhammarVolin21b}. The QSC \cite{GromovKazakovLeurentVolin11, GromovKazakovLeurentVolin15} is an integrability-based non-perturbative approach which evolved from the TBA/Y-system and describes the spectra of quantum integrable systems.\footnote{See also the review \cite{LevkovichMaslyuk19}.} The application of QSC to the spectral problem of $\N = 4$ SYM has led to a number of spectacular results, such as the calculation of anomalous dimensions of twist-2 operators up to 7 loops \cite{MarboeVelizhaninVolin14, MarboeVelizhanin16} and the Konishi multiplet up to 11 loops \cite{MarboeVolin14, MarboeVolin17, MarboeVolin18}. Early reviews on the application of integrability to the spectral problems of $\N = 4$ SYM and ABJM theory can be found in the collections of articles \cite{Beisertetal10, KristjansenStaudacherTseytlin09, DoreyMinahanTseytlin11}. \\
\indent The generic form of the perturbative expansion of operator scaling dimensions in ABJM theory is quite different from that of $\N = 4$ SYM, \eqref{ScalingWeakCouplingExpansionSYM}. In fact, it has been observed that many quantities in AdS$_4$/CFT$_3$ can be related to the ones in AdS$_5$/CFT$_4$ by simply replacing the 't Hooft coupling constant $\lambda$ by an interpolating function $h(\lambda)$. The asymptotics of the ABJM interpolating function $h(\lambda)$ are different at weak and strong coupling. At weak coupling, the generic form of $h(\lambda)$ reads:
\begin{IEEEeqnarray}{l}
h(\lambda) = \lambda + c_1\lambda^3 + c_2\lambda^5 + \cdots, \qquad \lambda \rightarrow 0,
\end{IEEEeqnarray}
see \cite{GromovSizov14, CavagliaGromovLevkovichMaslyuk16} for the precise expression. As a consequence, odd loop orders vanish and the perturbative expansion \eqref{ScalingWeakCouplingExpansionSYM} only contains even powers in the case of ABJM. State of the art calculations of scaling dimensions at weak coupling with the ABJM quantum spectral curve can be found in \cite{AnselmettiBombardelliCavagliaTateo15, LeeOnishchenko17, BombardelliCavagliaContiTateo18, LeeOnishchenko18, LeeOnishchenko19a}.
\paragraph{Three and higher-point functions} The general form of scalar three-point functions in CFTs is \eqref{ThreePointFunctionsCFT}. In planar weakly coupled $\N = 4$ SYM, the structure constants of single-trace operators have the following perturbative expansion:
\begin{IEEEeqnarray}{l}
\C_{123} = \C_{123}^{(0)} + \C_{123}^{(1)}\lambda + \C_{123}^{(2)}\lambda^2 + \ldots, \qquad \lambda \rightarrow 0. \label{CorrelatorWeakCouplingExpansionSYM}
\end{IEEEeqnarray}
The computation of the first (tree level) term in this series can be greatly facilitated by an integrability-based technique called tailoring.\footnote{Details can be found in the thesis \cite{Escobedo12}.} Tailoring essentially carries out all the Wick contractions by cutting, flipping and sewing each of the three closed spin chains that make up the correlator. Apart from the $SU(2)$ sector \cite{EscobedoGromovSeverVieira10}, more sectors have been explored in \cite{EscobedoGromovSeverVieira11, Georgiou11, VieiraWang13, CaetanoFleury14, KazamaKomatsuNishimura14}. One-loop computations (i.e.\ the second term in \eqref{CorrelatorWeakCouplingExpansionSYM}) have been considered in \cite{GromovVieira12a, GromovVieira12b, VieiraWang13, CaetanoFleury14}. Four and higher-point correlation functions have been studied in \cite{CaetanoEscobedo11}. Tailoring was directly succeeded by another integrability-based non-perturbative framework called hexagonalization \cite{BassoKomatsuVieira15}. Hexagonalization is thought to be so powerful as to solve the structure constants problem of planar $\N = 4$ SYM. More recently, it was also introduced for the case of the AdS$_3$/CFT$_2$ correspondence \cite{EdenLePlatSfondrini21}.
\subsubsection[Strong coupling]{Strong coupling \label{SubSection:HolographicCorrelatorsCFTsStrong}}
\noindent We have time and again emphasized that holography provides access to the nonperturbative regimes of gauge theories. These are generally inaccessible by other methods,\footnote{Supersymmetric localization \cite{Pestun07} is another nonperturbative method which can be used for the study of strongly coupled gauge theories, but is obviously dependent on the presence of supersymmetry.} and certainly far outside the realm of perturbation theory. The price to pay however is the high degree of symmetry (e.g.\ conformal symmetry and supersymmetry) that is present in holographic theories. Some holographic theories may even have an underlying integrability structure (i.e.\ infinitely many conserved quantities) which is manifested by the description of their spectra by integrable spin chains. \\
\indent Introducing defects takes us away from idealized holography by breaking as many symmetries as possible in a controllable way. The presence of integrability is generally a pro, despite the limited availability of integrability-based non-perturbative techniques for theories with defects. Actually, we would rather keep integrability just in case methods similar to the TBA/Y-system/QSC/hexagon become available for AdS/{\color{red}d}CFT. In the next subsection we will review the computation of correlation functions in strongly coupled AdS/{\color{red}d}CFT duality with semiclassical strings. \\
\indent String theory computations are generally possible in holographic dualities (with or without boundaries/defects) thanks to their weak/strong coupling property. The weak/strong coupling property implies that the gauge and string theory coupling constants are inversely proportional to each other, so that when the former side of the duality is strongly coupled, the latter is weakly coupled (and vice versa). Although the weak/strong coupling property allows us to probe the strongly coupled regimes of gauge theories with perturbative string theory computations, in practice it cuts off the two perturbative regimes from one another. This dilemma of computing observables at weak and strong coupling with two different theories, but without knowing whether the two theories are actually equivalent (insofar holography is still an unproven conjecture) can be overcome by integrability. \\
\indent As we have already explained, integrability-based spectral methods such as the TBA/Y-system and the QSC are genuinely non-perturbative and have been developed for planar $\N = 4$ SYM \eqref{LagrangianSYM} and ABJM theory \eqref{LagrangianABJM}. Another non-perturbative integrability-based framework for the computation of correlation function structure constants is hexagonalization which has been developed for planar $\N = 4$ SYM. These methods are designed to dial the spectrum and structure constants for arbitrary values of the coupling. Their validity is usually supported by a large number of nontrivial tests, both at weak and strong coupling. At strong coupling they reproduce the results which are obtained by supergravity and string theory calculations. In what follows, we briefly discuss string theory computations in planar $\N = 4$ SYM theory. \\
\indent While at weak coupling the perturbative expansions of scaling dimensions and structure constants have similar forms (cf.\ \eqref{ScalingWeakCouplingExpansionSYM}--\eqref{CorrelatorWeakCouplingExpansionSYM}), at strong coupling they can be very different for different operators. In fact, gauge invariant single-trace operators in planar $\N = 4$ SYM can be classified based on their lengths at weak coupling and the leading behavior of their scaling dimensions at strong coupling. The classification of single-trace operators in planar $\N = 4$ SYM is summarized in table \ref{Table:OperatorTypes} below. \\
\renewcommand{\arraystretch}{1.5}\setlength{\tabcolsep}{5pt}
\begin{table}[H]\begin{center}\begin{tabular}{|l||c|c|c|}
\hline
\multirow{2}{*} & Weak coupling & Strong coupling & \multirow{2}{*}{Example} \\[-6pt]
& ($\lambda \rightarrow 0$) & ($\lambda \rightarrow \infty$) & \\ \hline\hline
Light operators & $\Delta \sim 1$ & $\Delta \sim 1 $ & CPOs \\ \hline
Medium operators & $L \sim 1$ & $\Delta \sim \lambda^{1/4}$ & Konishi operator \\ \hline
Heavy operators & $L \gg 1 $ & $\Delta \sim \lambda^{1/2}$ & long GKP strings \\ \hline
\end{tabular}\caption{Types of single-trace operators in planar $\N = 4$ SYM. \label{Table:OperatorTypes}}\end{center}\end{table}
\vspace{-.4cm}\indent Light operators of $\N = 4$ SYM are BPS operators (i.e.\ they have zero anomalous dimensions to all orders of perturbation theory), so that their scaling dimensions are equal to their classical dimensions for all values of the coupling constant. The only single-trace BPS operators of $\N = 4$ SYM are chiral primary operators (CPOs). Their definition and basic properties can be found in appendix \ref{Appendix:ChiralPrimaryOperators}. CPOs are holographically dual to supergravity modes (aka massless string modes); for example the vacuum state of the $\N = 4$ SYM spin chain, $\tr \left[Z^L\right]$ is a CPO that is holographically dual to a pointlike string rotating on the equator of an S$^2 \subset \text{AdS}_5\times\text{S}^5$. Obviously two, but also three-point correlation functions of CPOs are not renormalized, i.e.\ they are the same for all values of the coupling constant (just like scaling dimensions) \cite{LeeMinwallaRangamaniSeiberg98}. \\
\indent Medium operators of $\N = 4$ SYM (e.g.\ the Konishi operator, $\tr\left[\varphi_i\varphi_i\right]$) have short lengths ($L \sim 1$) and nonzero anomalous dimensions. It is quite challenging to study their spectra at high loop orders with the Bethe ansatz due to wrapping which commences at a very low (critical) loop order (equal to their short lengths). Medium operators are dual to the lightest massive string states. To leading order, their scaling dimensions behave as $\lambda^{1/4}$ at strong coupling. On the other hand, heavy operators also have nonzero anomalous dimensions and are dual to (arbitrarily) massive string states, but have very long lengths ($L \gg 1$). Their leading-order scaling dimensions behave as $\lambda^{1/2}$ at strong coupling. \\
\indent Take for example GKP strings in AdS$_5$ and S$^5$ \cite{GubserKlebanovPolyakov02}. These strings are respectively dual to the single-trace operators of $\N = 4$ SYM, $\tr\left[D_+^{S}Z^2\right]$ (\quotes{twist-2}) and $\tr\left[Z^{J}X^2\right]$ (2 magnon).\footnote{Plus all possible permutations of the fields/impurities. $Z \equiv \varphi_1 + i\varphi_2$, $X \equiv \varphi_3 + i\varphi_4$, $Y \equiv \varphi_5 + i\varphi_6$ are the three complex scalar fields of $\N =4$ SYM, made up from the six real scalars $\varphi_i$, ($i=1,\ldots,6$). $\mathcal{D}_+ = \mathcal{D}_0 + \mathcal{D}_3$, $\mathcal{D}_- = \mathcal{D}_1 + \mathcal{D}_2$ are the light-cone derivatives. Typically, $Z$ corresponds to the vacuum state of the $\N = 4$ SYM spin chain, $\tr \left[Z^L\right]$.} GKP operators can be either medium or heavy, depending on whether their charges are respectively small ($S,J \rightarrow 0$) or large ($S,J \rightarrow \infty$). CPOs with large charges can also be made to behave as heavy, e.g.\ by taking their charges to scale as $S,J \sim \lambda^{1/2}\rightarrow\infty$ at strong coupling. In fact, this particular type of scaling is known as BMN scaling and the corresponding heavy operators are known as BMN operators. Below, we will specifically compute correlation functions involving BMN operators in both the AdS/CFT and the AdS/{\color{red}d}CFT correspondence. \\
\indent Scaling/anomalous dimensions can be computed at strong coupling by using semiclassical string dynamics. According to the dictionary of AdS/CFT, operator scaling dimensions are equal to the energies of their dual string states. This means that by computing the energies of classical string configurations, we obtain the scaling dimensions of their dual gauge theory operators at strong coupling. Quantum corrections can be obtained by string perturbation theory (aka semiclassical analysis). This way, string energies in AdS$_5$/CFT$_4$ have been computed up to two loops for very short or infinitely long twist-2 operators, and up to one loop for long twist-2 operators of finite size.\footnote{Short operators include members of the Konishi multiplet. Long string configurations include the giant magnon, the single spike and the pulsating string. More can be found in the references \cite{Tseytlin10, McLoughlin10} which review semiclassical analysis in AdS$_5\times\text{S}^5$. See also \cite{Linardopoulos15b}.} When the dust settles, the perturbative expansion of heavy operators in strongly-coupled $\N = 4$ SYM has the following form (see e.g.\ \cite{FloratosGeorgiouLinardopoulos13, DimovMladenovRashkov14} for short/long GKP/folded strings in AdS$_5\times\text{S}^5$ and AdS$_4\times\CP^3$):
\begin{IEEEeqnarray}{l}
\Delta_{\text{H}} = \Delta^{(0)} + \gamma^{(1)} \, \lambda^{\frac{1}{2}} + \gamma^{(2)} \, \lambda^{-\frac{1}{2}} + \ldots, \qquad \lambda \rightarrow \infty, \label{ScalingStrongCouplingExpansionH}
\end{IEEEeqnarray}
where again the expansion coefficients depend on the geometric characteristics of the string, such as its (finite) size and its angular extent (momenta). The expansion of medium sized operators is similar, albeit with a different overall functional dependence on the coupling constant:
\begin{IEEEeqnarray}{l}
\Delta_{\text{M}} = \Delta^{(0)} + \gamma^{(1)} \, \lambda^{\frac{1}{4}} + \gamma^{(2)} \, \lambda^{-\frac{1}{4}} + \gamma^{(3)} \, \lambda^{-\frac{3}{4}} + \ldots, \qquad \lambda \rightarrow \infty. \label{ScalingStrongCouplingExpansionM}
\end{IEEEeqnarray}
\indent The perturbative expansions of operator scaling dimensions in strongly coupled ABJM theory are very similar to the corresponding expansions \eqref{ScalingStrongCouplingExpansionH}--\eqref{ScalingStrongCouplingExpansionM} of $\N = 4$ SYM. All the same, the asymptotic form of the ABJM interpolating function $h(\lambda)$ at strong coupling reads:
\begin{IEEEeqnarray}{l}
h(\lambda) = \sqrt{\frac{1}{2}\left(\lambda - \frac{1}{24}\right)} - \frac{\log2}{2\pi} + \cdots, \qquad \lambda \rightarrow \infty,
\end{IEEEeqnarray}
see \cite{GromovSizov14, CavagliaGromovLevkovichMaslyuk16} for the subleading terms. Because the equivalence of the string sigma model \eqref{MetsaevTseytlinAction} on the supercoset space \eqref{AdS4CP3supercoset} to the full type IIA GS superstring action on AdS$_4\times\CP^3$ is partial, only one-loop energies can, and have been computed by semiclassical analysis in ABJM theory. \\
\indent A very elegant geometric framework which describes the spectrum of string energies in AdS/CFT is the spectral curve \cite{KazakovMarshakovMinahanZarembo04, GromovVieira08b}.\footnote{More details can be found in the reviews \cite{Zarembo10b, SchaferNameki10}.} For any classically integrable system with monodromy matrix $\M(\y)$ and Lax connection $L(\y)$, the spectral curve is defined in terms of the eigenvalue equation:
\begin{IEEEeqnarray}{l}
\text{sdet}\left(\mathfrak{m}\, I - \M(\y)\right) = 0,
\end{IEEEeqnarray}
where $\mathfrak{m}$ are the eigenvalues of the monodromy matrix $\M$ and $\y$ is the spectral parameter. Finite-genus (aka finite-gap) solutions of the spectral curve are algebraic curves. Algebraic curves are defined in terms of the eigenvalues $\mathfrak{m}_i(\y) \equiv e^{i \, \mathfrak{p}_i(\y)}$ of the monodromy matrix, or equivalently by the quasi-momenta $\mathfrak{p}_i(\y)$. Each classical string solution in AdS$_5\times\text{S}^5$ is characterized by a set of quasi-momenta. String energies, along with their (1-loop) quantum corrections, can then be extracted from the classical algebraic curve and its semiclassical quantization. This way, one-loop corrections to the energies of a wide range of string configurations (including finite-size corrections) have been computed. \\
\indent Nonperturbative integrability methods (TBA/Y-system/QSC) are also capable of providing operator scaling dimensions at strong coupling. For example the scaling dimensions of the (medium/short) Konishi operator have been computed up to 3 loops by the QSC method \cite{GromovLevkovichMaslyukSizovValatka14, GromovHegedusJuliusSokolova24}. On the other hand (to the best of our knowledge), no computations of the spectrum for heavy/long operators at strong coupling have been carried out by using these methods (let alone finite-size corrections, see e.g.\ \cite{AxenidesFloratosLinardopoulos15a}). \\
\indent Like scaling dimensions, the strong coupling expansion of correlation functions differs significantly from the corresponding expansion at weak coupling (which is given by e.g.\ \eqref{CorrelatorWeakCouplingExpansionSYM} for $\N = 4$ SYM three-point functions). Moreover, the generic form of the expansion depends on the contracted operators. For example, the correlation function of three heavy operators (HHH) in $\N = 4$ SYM (similar expansions can be written for higher-point correlation functions, as well as ABJM theory) scales as
\begin{IEEEeqnarray}{l}
\C_{123}^{\bullet\bullet\bullet} = \C_{123}^{(0)} + \sqrt{\lambda} \, \C_{123}^{(1)} + \C_{123}^{(2)} + \frac{1}{\sqrt{\lambda}} \, \C_{123}^{(3)} + \ldots, \qquad \lambda \rightarrow \infty, \label{CorrelatorStrongCouplingExpansion}
\end{IEEEeqnarray}
where it is customary to denote BPS operators with $\circ$ and non-BPS operators with $\bullet$. One can imagine all possible three-point combinations of light (L), medium (M) and heavy (H) operators, with generically different perturbative expansions for their three-point functions at strong coupling. Once more, the expansion coefficients generally depend on the geometric characteristics of the involved operators/string states, such as their (finite) sizes and angular extents. We will now go through the computation of correlation functions in strongly coupled holographic CFTs. We start with the supergravity approximation and the computation of correlation functions with Witten diagrams. \\
\paragraph{Supergravity approximation} According to the state/operator correspondence of AdS/CFT \cite{GubserKlebanovPolyakov98, Witten98a},
\begin{IEEEeqnarray}{c}
\big\langle e^{\int d^4 x \phi_0\left(x\right)\OO\left(x\right)}\big\rangle_{\text{CFT}} = \mathcal{Z}_{\text{string}}\left[\big.\phi\left(x,z\right)\big|_{z = 0} \sim \phi_0\left(x\right)\right], \label{StateOperatorCorrespondence}
\end{IEEEeqnarray}
any local gauge invariant CFT operator $\OO\left(x\right)$ has a holographically dual bulk field $\phi\left(x,z\right)$, the boundary value $\phi_0\left(x\right)$ of which sources the CFT operator $\OO\left(x\right)$. In other words, every closed superstring state in the bulk of AdS/CFT has a dual gauge theory operator (on the AdS boundary, $z = 0$). It follows that the two partition functions of AdS/CFT should be equal:
\begin{IEEEeqnarray}{c}
\mathcal{Z}_{\text{CFT}} = \mathcal{Z}_{\text{string}},
\end{IEEEeqnarray}
which implies that every boundary observable (e.g.\ scaling dimensions, correlation functions, etc.) has a dual observable in the bulk too. \\
\indent The Gubser-Klebanov-Polyakov-Witten (GKPW) prescription for computing connected correlation functions in the AdS/CFT correspondence reads:
\begin{IEEEeqnarray}{c}
\left\langle\OO_1\left(x_1\right)\ldots\OO_n\left(x_n\right)\right\rangle = (-1)^{n+1}\frac{\delta}{\delta \phi_0^1}\ldots\frac{\delta}{\delta \phi_0^n} \log\mathcal{Z}_{\text{string}}\bigg|_{\phi_0^1,\ldots, \phi_0^n = 0}, \label{GKPWprescription}
\end{IEEEeqnarray}
which amounts to differentiating the logarithm of both sides of the state/operator correspondence \eqref{StateOperatorCorrespondence}. In the supergravity approximation ($\lambda, N_c = \infty$), the on-shell supergravity action becomes the generating functional of connected CFT correlation functions since,
\begin{IEEEeqnarray}{c}
e^{-W_{\text{CFT}}} = \mathcal{Z}_{\text{CFT}} = \mathcal{Z}_{\text{string}} \approx e^{-S_{\text{sugra}}}, \qquad N_c,\lambda = \infty. \label{GeneratingFunctional}
\end{IEEEeqnarray}
By virtue of the state/operator correspondence \eqref{StateOperatorCorrespondence}, the identity \eqref{GeneratingFunctional} leads to
\begin{IEEEeqnarray}{c}
W_{\text{gauge}}\left[\phi_0\left(x\right)\right] = -\log\left\langle e^{\int d^4x \phi_0\left(x\right)\OO\left(x\right)}\right\rangle_{\text{CFT}} = \text{extremum}\Big\{S_{\text{sugra}}\left[\phi\left(x,z\right)\right]\Big|_{\phi = \phi_0}\Big.\Big\}. \label{WittenDiagrams}
\end{IEEEeqnarray}
The calculation of correlation functions to leading order in strongly coupled CFTs (i.e.\ in the supergravity approximation, $\lambda, N_c = \infty$), boils down to the computation of Witten diagrams \cite{FreedmanMathurMatusisRastelli98a, ChalmersNastaseSchalmSiebelink98, LeeMinwallaRangamaniSeiberg98, ArutyunovFrolov99b, Lee99}.
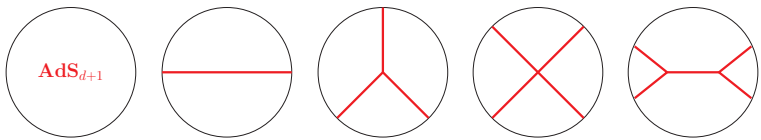
\begin{figure}[H]\begin{center}\resizebox{10cm}{!}{\begin{tikzpicture}
\draw (0,0) circle(2);
\node (0,0) {\LARGE\color{red}\textbf{AdS$_{d+1}$}};
\end{tikzpicture}\qquad
\begin{tikzpicture}
\draw (0,0) circle(2);\draw[line width=0.7mm, red] (-2,0) -- (2,0);
\end{tikzpicture}\qquad
\begin{tikzpicture}
\draw (0,0) circle(2);
\draw[line width=0.7mm, red] (-1.41,-1.41) -- (0,0) -- (1.41,-1.41);
\draw[line width=0.7mm, red] (0,0) -- (0,2);
\end{tikzpicture}\qquad
\begin{tikzpicture}
\draw (0,0) circle(2);
\draw[line width=0.7mm, red] (-1.41,1.41) -- (1.41,-1.41);
\draw[line width=0.7mm, red] (1.41,1.41) -- (-1.41,-1.41);
\end{tikzpicture}\qquad
\begin{tikzpicture}
\draw (0,0) circle(2);
\draw[line width=0.7mm, red] (-1.8,0.8) -- (-.8,0) -- (-1.8,-0.8);
\draw[line width=0.7mm, red] (-.8,0) -- (.8,0);
\draw[line width=0.7mm, red] (1.8,0.8) -- (.8,0) -- (1.8,-0.8);
\end{tikzpicture}}\caption{Witten diagrams.}\label{Figure:WittenDiagrams}\end{center}\end{figure}
\vspace{-.4cm}\noindent Without going into too many details or specific examples of computations, let us provide the rules for computing correlation functions with Witten diagrams in the supergravity approximation:\footnote{For a pedagogical introduction, we refer the interested reader to the references \cite{DHokerFreedman02, FreedmanVanProeyen12}.}
\begin{itemize}[leftmargin=*]
\item For each bulk point $\left(x,z\right)$ in AdS$_{d+1}$ include a (Euclidean AdS) integration $\int d^d x \, dz \sqrt{g\left(x,z\right)}$.
\item For each bulk-to-bulk line between $(x,z)$ and $(y,w)$ insert a bulk-to-bulk propagator $G_{\Delta}$, and for each bulk-to-boundary line between $(x,z)$ and $(y,0)$ insert a bulk-to-boundary propagator $\G_{\Delta}$.
\item Vertex factors follow from the interaction terms of the bulk Lagrangian $\LL = \frac{1}{3}b\phi^3 + \frac{1}{4}c\phi^4 + \ldots$
\end{itemize}
For the classical on-shell action only tree-level diagrams should can be computed (i.e.\ no loops). Two-point functions are special because they do not contain a bulk integration and require renormalization (typically a cutoff). As a result, two-point functions cannot be computed with the above procedure, but need to be treated separately. Refer to the works \cite{DHokerFreedman02, FreedmanVanProeyen12} for more information and further references.
\paragraph{Beyond supergravity} As we have explained, the problem of computing CFT two-point functions is essentially equivalent to computing scaling dimensions. The theory's dilatation operator must be diagonalized, and for theories whose dilatation operators are given by the Hamiltonians of integrable spin chains, the spectral problem can be fully solved from weak to strong coupling. Therefore two-point functions are in principle known exactly for all values of the coupling.\\
\indent Integrability also facilitates perturbative (i.e.\ Feynman diagram) calculations of the spectrum/two-point functions at weak coupling. In strongly coupled holographic theories, one equivalently evaluates semiclassical string energies. These have by and large been found to agree with the QSC predictions in both $\N = 4$ SYM and ABJM theory. Take for example a heavy $\N = 4$ SYM operator $\OO$. The perturbative expansion of its scaling dimensions at strong coupling is given by \eqref{ScalingStrongCouplingExpansionH}. Plugging \eqref{ScalingStrongCouplingExpansionH} into the generic form \eqref{TwoPointFunctionsCFT} of CFT two-point functions we get,
\begin{IEEEeqnarray}{l}
\left\langle\OO\left(\x_1\right)\OO\left(\x_2\right)\right\rangle = \frac{1}{\x_{12}^{2\Delta_{0}}}\left[\sqrt{\lambda} \, \C_{12}^{(1)} + \frac{1}{\sqrt{\lambda}} \, \C_{12}^{(2)} + \frac{1}{\sqrt{\lambda^3}} \, \C_{12}^{(3)} + \ldots\right], \qquad \lambda \rightarrow \infty, \label{TwoPointFunctionsAdSCFT}
\end{IEEEeqnarray}
where the expansion coefficients $\C_{12}^{(i)}$ are known functions of $\log\x_{12}$ and the $\gamma^{(i)}$ coefficients of scaling dimensions in \eqref{ScalingStrongCouplingExpansionH}. Similarly for medium operators, their two-point functions at weak and strong coupling are fully determined by scaling dimensions. \\
\indent Even so, it should in principle be possible to reproduce the full form \eqref{TwoPointFunctionsCFT} of two-point functions in strongly coupled AdS/CFT by using string theory. String computations are in fact perturbative in $\sqrt{\lambda}$, so that the result (e.g.\ in the case of heavy operators in AdS$_5$/CFT$_4$) should take the form \eqref{TwoPointFunctionsAdSCFT}. Indeed, by considering a class of $\N = 4$ SYM operators with large charges which are dual to classical string solutions spinning inside AdS$_5\times\text{S}^5$ (\quotes{fattened} Witten diagrams), the authors of \cite{JanikSurowkaWereszczynski10} managed to reproduce the full functional form of first term in \eqref{TwoPointFunctionsAdSCFT}. The same was accomplished in \cite{Buchbinder10, BuchbinderTseytlin10a, Tseytlin03a} by using vertex operators.
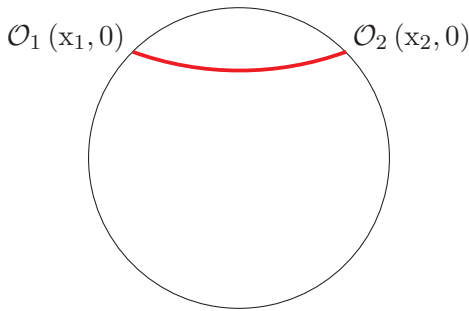
\begin{figure}[H]\begin{center}\begin{tikzpicture}
\draw (0,0) circle(2); \draw[line width=0.5mm, red] (-1.414,1.414) arc (250:290:4.15);
\node (O1) at (-2.3,1.6) {$\OO_1\left(\x_1,0\right)$};
\node (O2) at (2.3,1.6) {$\OO_2\left(\x_2,0\right)$};
\end{tikzpicture}\caption{Holographic two-point function in AdS/CFT.}\label{Figure:TwoPointFunction}\end{center}\end{figure}
\vspace{-.4cm}\indent The computation of three and higher-point correlation functions in strongly coupled AdS/CFT, when one of the operators is dual to a supergravity mode, was set up in \cite{Zarembo10c, CostaMonteiroSantosZoakos10} following the earlier work \cite{BerensteinCorradoFischlerMaldacena99}. Take a nonlocal gauge invariant CFT operator $\W$ (e.g.\ a Wilson loop or a product of local operators) which is dual to a classical string worldsheet, and a local gauge invariant CFT operator $\OO_I\left(y\right)$ which is dual to the scalar supergravity field $\phi_I\left(y,w\right)$. Defining the ratio of expectation values,
\begin{IEEEeqnarray}{l}
\left\langle\OO_I\left(y\right)\right\rangle_{\W} \equiv \frac{\left\langle\W \OO_I\left(y\right)\right\rangle}{\left\langle\W\right\rangle}, \label{CorrelationFunction0}
\end{IEEEeqnarray}
the recipe for computing it at strong coupling takes the following form:
\begin{IEEEeqnarray}{l}
\left\langle\OO_I\left(y\right)\right\rangle_{\W} = \lim_{w\rightarrow 0}\left[\frac{\pi}{w^{\Delta_I}}\sqrt{\frac{2}{\Delta_I - 1}} \cdot \big\langle\phi_I\left(y,w\right)\cdot \frac{1}{Z_{\text{str}}}\int D\X\,e^{-S_{\text{str}}\left[\X\right]}\big\rangle_{\text{bulk}}\right], \qquad \label{CorrelationFunction1}
\end{IEEEeqnarray}
where the prefactor multiplying the expectation value is related to the proper normalization of the scalar field $\phi_I$ (scaling dimension $\Delta_I$). Moreover, $S_{\text{str}}\left[\X\right]$ stands for the classical string action,
\begin{IEEEeqnarray}{l}
S_{\text{str}} = \frac{\sqrt{\lambda}}{4\pi}\int d^2\sigma\sqrt{-\gamma}\gamma^{ab}\partial_a\X^M\partial_b\X^N g_{MN} + \ldots, \label{PolyakovAction2}
\end{IEEEeqnarray}
where we have omitted the fermions, the dilaton and the Kalb-Ramond field. $\X$ are the target-space coordinates, aka the embedding coordinates of the string worldsheet. For example, when $\W$ is a product of $n$ local operators, the recipe \eqref{CorrelationFunction0}--\eqref{CorrelationFunction1} allows to compute the $n+1$ point correlation function of the $n$ local operators (which make up the nonlocal operator $\W$) and the local operator $\OO_I$ (which is dual to the scalar supergravity field $\phi_I$). \\
\indent To compute the correlation function \eqref{CorrelationFunction1}, we note that the string action $S_{\text{str}}$ depends indirectly on the bulk supergravity modes $\phi_I$ via a disturbance which is induced on the supergravity fields by a local operator insertion. In the case of AdS$_5$/CFT$_4$, the relevant type IIB supergravity fields are the graviton $g_{MN}$ and the 4-form Ramond-Ramond (RR) potential $C_{MNPQ}$ which we expand as follows:
\begin{IEEEeqnarray}{l}
g_{MN} = \hat{g}_{MN} + \delta g_{MN}, \qquad C_{MNPQ} = \hat{C}_{MNPQ} + \delta C_{MNPQ}. \label{FieldPerturbationIIB}
\end{IEEEeqnarray}
The corresponding background solution consists of the AdS$_5\times\text{S}^5$ metric $\hat{g}_{MN}$ (whose precise expression can be found in \eqref{MetricAdS}--\eqref{MetricAdS5xS5}) and the 4-form potential $\hat{C}_{MNPQ}$ (written out in \eqref{RamondRamondPotentialAdS5xS5}).\footnote{Throughout the present section, we will be using a Euclidean signature for the AdS$_{d+1}$ metric; cf.\ \eqref{MetricAdS}.} To see how the string action $S_{\text{str}}$ can indirectly depend on the bulk supergravity modes $\phi_I$, as we briefly alluded above, we express the leading-order perturbations in \eqref{FieldPerturbationIIB} as linear combinations of the bulk supergravity modes $\phi_I$ and their derivatives:
\begin{IEEEeqnarray}{l}
\delta g_{MN} = V^I_{MN} \cdot \phi_I, \qquad \delta C_{MNPQ} = v^I_{MNPQ} \cdot \phi_I, \label{VertexOperatorsIIB}
\end{IEEEeqnarray}
where $V^I_{MN}$ and $v^I_{MNPQ}$ are differential (vertex) operators whose coefficients depend on the target-space/embedding coordinates $\X$. By expanding the string action around $\phi_I = 0$ we obtain
\begin{IEEEeqnarray}{l}
S_{\text{str}}\left[\X\right] = S_{\text{str}}\left[\X\right]\Big|_{\phi_I = 0} + \left.\frac{\partial S_{\text{str}}\left[\X\right]}{\partial\phi_I}\right|_{\phi_I = 0}\cdot\phi_I\left(x,z\right) + \ldots \label{PolyakovAction3}
\end{IEEEeqnarray}
\indent In the strong coupling regime ($\lambda\to\infty$), the path integral in \eqref{CorrelationFunction1} will be dominated by a saddle point that corresponds to classical solutions $\X_{\text{cl}}$.\footnote{Actually, the proper way to carry out the saddle point approximation goes by the name orbit averaging. More information can be found in the original papers \cite{BajnokJanikWereszczynski14, YangJiangKomatsuWu21b, HolguinWeng22}.} The first term in the expansion \eqref{PolyakovAction3} cancels the partition function which shows up in the denominator of \eqref{CorrelationFunction1} and the correlation function becomes:
\begin{IEEEeqnarray}{ll}
\left\langle\OO_I\left(y\right)\right\rangle_{\W} &= \lim_{w \rightarrow 0}\frac{\pi}{w^{\Delta_I}}\sqrt{\frac{2}{\Delta_I - 1}} \cdot \langle\phi_I\left(y,w\right) \cdot \exp\left\{\left.\frac{\partial S_{\text{str}}\left[\X_{\text{cl}}\right]}{\partial\phi_I}\right|_{\phi_I = 0}\phi_I\left(x,z\right) + \ldots\right\}\rangle_{\text{bulk}} = \nonumber \\[6pt]
&= \lim_{w \rightarrow 0}\frac{\pi}{w^{\Delta_I}}\sqrt{\frac{2}{\Delta_I - 1}} \cdot\langle\phi_I\left(y,w\right)\cdot\left[1 + \left.\frac{\partial S_{\text{str}}\left[\X_{\text{cl}}\right]}{\partial\phi_I}\right|_{\phi_I = 0}\phi_I\left(x,z\right) + \ldots \right]\rangle_{\text{bulk}}. \qquad
\end{IEEEeqnarray}
In conformal field theories, scalar fields have vanishing expectation values (a result which can also be confirmed by a holographic computation), so that $\langle\phi_I\left(y,w\right)\rangle = 0$. Plugging the vanishing expectation value of the scalar mode $\phi_I$ into the above expansion we obtain, to leading order:
\begin{IEEEeqnarray}{c}
\left\langle\OO_I\left(y\right)\right\rangle_{\W} = \lim_{w \rightarrow 0} \frac{\pi}{w^{\Delta_I}}\sqrt{\frac{2}{\Delta_I - 1}} \cdot \left.\frac{\partial S_{\text{str}}\left[\X_{\text{cl}}\right]}{\partial\phi_I}\right|_{\phi_I = 0} \left\langle\phi_I\left(y,w\right)\cdot\phi_I\left(x,z\right)\right\rangle_{\text{bulk}} + \ldots, \qquad \ \label{CorrelationFunction2}
\end{IEEEeqnarray}
where the two-point function $\left\langle\phi_I\left(y,w\right)\cdot\phi_I\left(x,z\right)\right\rangle_{\text{bulk}}$ is just the bulk-to-bulk propagator \eqref{PropagatorBulkToBulk1}--\eqref{PropagatorBulkToBulk2} of the massive scalar field $\phi_I$ in Euclidean AdS$_{d+1}$. \\
\indent In the case of the AdS$_5$/CFT$_4$ duality and AdS$_5\times\text{S}^5$ target space, we plug the expression for the classical string action \eqref{PolyakovAction2} and the perturbations \eqref{FieldPerturbationIIB}--\eqref{VertexOperatorsIIB} of type IIB supergravity fields in \eqref{CorrelationFunction2} which leads to
\begin{IEEEeqnarray}{ll}
\left\langle\OO_I\left(y\right)\right\rangle_{\W} = \frac{1}{4}\sqrt{\frac{2\lambda}{\Delta_I - 1}} \int& d^2\sigma\,\partial_a\X^M\partial^a\X^N\,V^I_{MN}\left(\X,\partial_x,\partial_z\right) \G_{\Delta_I}\left(x,z;y\right) + \ldots, \qquad \label{CorrelationFunction3}
\end{IEEEeqnarray}
in the conformal gauge, $\gamma_{ab} = \text{diag}\left(-,+\right)$. The boundary limit $\G_{\Delta_I}\left(x,z;y\right)$ of the bulk-to-bulk propagator \eqref{PropagatorBulkToBulk1}--\eqref{PropagatorBulkToBulk2} is given by \eqref{PropagatorBulkToBoundary1}.
\begin{figure}[H]\begin{center}\begin{tikzpicture}
\draw (0,0) circle(2); \draw[line width=0.5mm, red] (-1.414,1.414) arc (250:290:4.15);
\node (O1) at (-2.3,1.6) {$\OO_1\left(\x_1,0\right)$};
\node (O2) at (2.3,1.6) {$\OO_2\left(\x_2,0\right)$};
\node (O3) at (-0.1,1.5) {$\left(x,z\right)$};
\node (O4) at (-0.1,-2.3) {$\OO_I\left(y,0\right)$};
\begin{feynman}
\vertex (a) at (-.1,1.15);
\vertex (b) at (0.1,-2);
\diagram*{(a) -- [photon, edge label=$\phi_{I} (y{,}w)$] (b)};
\end{feynman}
\end{tikzpicture}\caption{Holographic heavy-heavy-light (HHL) correlation function in AdS/CFT.}\label{Figure:HeavyHeavyLight}\end{center}\end{figure}
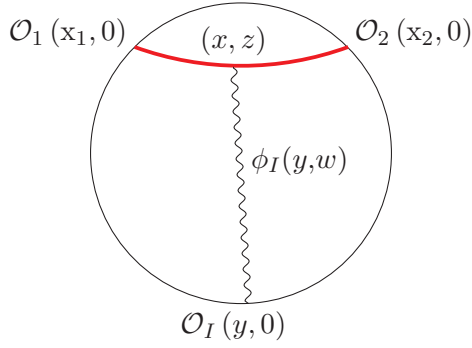

\vspace{-.4cm}\indent Now take $\OO_I$ to be a (light) chiral primary operator (CPO) of $\N = 4$ SYM with length $L$.\footnote{See \eqref{ChiralPrimaryOperators} of appendix \ref{Appendix:ChiralPrimaryOperators} for the definition of CPO's in $\N = 4$ SYM.} The dual supergravity modes of CPO's have been identified and the fluctuations of the background type IIB supergravity fields \eqref{VertexOperatorsIIB} are given by \eqref{FluctuationsMetric1}--\eqref{FluctuationsPotential1}. Expanding the terms in \eqref{CorrelationFunction3} we obtain
\begin{IEEEeqnarray}{ll}
\left\langle\OO_I^{\text{CPO}}\left(y\right)\right\rangle_{\W} = \frac{1}{4}\sqrt{\frac{2\lambda}{L - 1}} \cdot\int d^2\sigma \cdot \Big\{&\partial_a\X^i\partial^a\X^j V^I_{ij} + 2\partial_a\X^i\partial^a\X^z V^I_{iz} + \partial_a\X^z\partial^a\X^z V^I_{zz} + \nonumber \\
& + \partial_a\X^{\mu}\partial^a\X^{\nu} V^I_{\mu\nu}\Big\}\cdot \G_L + \ldots, \label{CorrelationFunction4}
\end{IEEEeqnarray}
so that by using the fluctuations \eqref{FluctuationsMetric2a}--\eqref{FluctuationsMetric2b} we find
\begin{IEEEeqnarray}{l}
\left\langle\OO_I^{\text{CPO}}\left(y\right)\right\rangle_{\W} = \frac{\ell^2 L\sqrt{2(L-1)\lambda}}{4\pi^2 \N_L} \int d^2\sigma \, Y_I\left(x_{\mu}\right) \cdot \Bigg\{\partial_a\X^i\partial^a\X^j \left[-\frac{\delta_{ij}}{z^2} + \frac{8r_i r_j}{K_z^2}\right] + \nonumber \\
+ \partial_a\X^i\partial^a\X^z\left[\frac{16r_i z}{K_z^2} - \frac{8r_i}{zK_z}\right] + \partial_a\X^z\partial^a\X^z \left[\frac{1}{z^2} - \frac{8}{K_z} + \frac{8z^2}{K_z^2}\right] + \partial_a\X^{\mu}\partial^a\X^{\nu} \hat{g}_{\mu\nu}\ell^{-2}\Bigg\} \cdot \frac{z^L}{K_z^L}, \qquad \label{CorrelationFunction5}
\end{IEEEeqnarray}
where $r_i = x_i - y_i$, for $i,j = 0,\ldots,3$ and $K_z = r^2 + z^2$. The definition of $\N_L$ can be found in \eqref{NormalizationFactor}. Notice that the correlation function \eqref{CorrelationFunction5} simplifies significantly in the $y_i\rightarrow\infty$ limit:
\begin{IEEEeqnarray}{ll}
\left\langle\OO_I^{\text{CPO}}\left(y\right)\right\rangle_{\W} = \frac{\ell^2 L\sqrt{2(L-1)\lambda}}{4\pi^2 \N_L y^{2L}} \int d^2\sigma \, Y_I\left(x_{\mu}\right) \cdot \bigg\{&-z^{L-2}\partial_a\X^i\partial^a\X^i + z^{L-2}\partial_a\X^z\partial^a\X^z + \nonumber \\
& + z^L\ell^{-2}\partial_a\X^{\mu}\partial^a\X^{\nu} \hat{g}_{\mu\nu}\bigg\}. \qquad \label{CorrelationFunction6}
\end{IEEEeqnarray}
\paragraph{Two BMN operators} At this point we will take the composite nonlocal gauge invariant operator $\W(\x_1,\x_2)$ to be given by $\W(\x_1,\x_2) \equiv \OO^{\dag}_{1}(\x_1)\OO_{2}(\x_2)$, where $\OO_{i}(\x_i) \ (i=1,2)$ are heavy local gauge invariant single-trace operators with length $L_i$,
\begin{IEEEeqnarray}{c}
\OO_{i} = \frac{1}{\sqrt{L_i}} \left(\frac{4\pi^2}{\lambda}\right)^{\frac{L_i}{2}} \text{tr}\left[Z^{L_i}\right], \qquad L_i \sim \sqrt{\lambda} \rightarrow \infty, \qquad i = 1,2, \label{BMNoperators}
\end{IEEEeqnarray}
but also BMN chiral primaries, such that $L = L_1 - L_2$ is small. By dubbing the operators BMN, we mean that the operators $\OO_{i}$ in \eqref{BMNoperators} are CPOs/single-trace BPS operators of $\N = 4$ SYM (defined by \eqref{ChiralPrimaryOperators} in appendix \ref{Appendix:ChiralPrimaryOperators}) which scale like heavy BMN operators, i.e.\ their scaling dimensions behave as $L_i \sim \sqrt{\lambda}$. There is a classical (pointlike) string solution which is holographically dual to the heavy nonlocal operator $\W(\x_1,x_2)$. The string solution is given by
\begin{IEEEeqnarray}{l}
x_3 = \bar\x + R \tanh\omega\tau, \quad z = \frac{R}{\cosh\omega\tau}, \quad \psi = 0, \quad \chi = i\omega\tau, \quad \theta = \frac{\pi}{2}, \qquad \label{BMNstring1}
\end{IEEEeqnarray}
with $\omega = L_2/\sqrt{\lambda}$. The two heavy BMN operators $\OO_{1,2}$ are located at the points $\x_{1,2}$ on the $x_3$ axis and a small distance from each other. In other words, $R = \x_{12}/2$ is small:
\begin{IEEEeqnarray}{l}
R = \frac{|\x_1 - \x_2|}{2} = \frac{\x_{12}}{2}, \qquad \bar \x = \frac{\x_1 + \x_2}{2}. \qquad \label{BMNstring2}
\end{IEEEeqnarray}
Plugging the AdS$_5\times\text{S}^5$ solution \eqref{BMNstring1}--\eqref{BMNstring2} into the expression for the correlation function \eqref{CorrelationFunction6}, we are led to the heavy-heavy-light (HHL) three-point function $\left\langle\OO_I^{\text{CPO}}\left(y\right)\right\rangle_{\W}$ in the large-distance limit $y_i\rightarrow\infty$:
\begin{IEEEeqnarray}{ll}
\left\langle\OO_I^{\text{CPO}}\left(y\right)\right\rangle_{\W} &= -\frac{\ell^2 L\sqrt{2(L-1)\lambda} \, \omega^2 \CC_{L/2}}{2\pi^2 \N_L}\cdot\frac{R^L}{y^{2L}}\int_{0}^{2\pi}\int_{-\infty}^{+\infty} d\sigma \, d\tau \, \text{sech}^{L+2}\omega\tau = \nonumber \\
& = -\frac{(-1)^{L/2} \ell^2}{2^{L+\frac{3}{2}} N_c} \cdot L_2\sqrt{L(L+1)(L+2)} \cdot B\Big(\frac{L}{2}+1,\frac{1}{2}\Big) \cdot \frac{\x_{12}^L}{y^{2L}}, \qquad \label{ThreePointFunctionBMN}
\end{IEEEeqnarray}
where the normalization constant $\CC_{L/2}$ has been defined in \eqref{SphericalHarmonicsSO3normalization2}. For reasons that will become apparent below, when evaluate the OPE in defect CFTs with $SO(3)\times SO(3)$ global symmetry, we have used the $SO(3)\times SO(3) \subset SO(5)$ invariant spherical harmonics \eqref{SphericalHarmonicsSO3}, evaluated at the point $\psi = 0$ of the five-sphere parametrization \eqref{MetricS5so3so3}. Put differently, we have imposed an $SO(3)\times SO(3)$ global symmetry for the light bulk mode $\phi_I$. \\
\indent Comparing the final result \eqref{ThreePointFunctionBMN} for the HHL three-point function we have just computed with the generic form \eqref{ThreePointFunctionsCFT} of CFT three-point functions (for $\Delta_1 + \Delta_2 - \Delta_3 = L_1 + L_2 - L$, $\Delta_2 + \Delta_3 - \Delta_1 = 0$, $\Delta_3 + \Delta_1 - \Delta_2 = 2L$), and using the definition \eqref{CorrelationFunction0} for the ratio of correlation functions $\left\langle\OO_I^{\text{CPO}}\left(y\right)\right\rangle_{\W}$, along with the CFT two-point function \eqref{TwoPointFunctionsCFT},
\begin{IEEEeqnarray}{ll}
\left\langle\W\left(\x_1,\x_2\right)\right\rangle = \langle\OO^{\dag}_{1}\left(\x_1\right)\OO_{2}\left(\x_2\right)\rangle = \frac{\delta_{12}}{\x_{12}^{L_1+L_2}}, \quad L = L_1 - L_2 \rightarrow 0, \qquad
\end{IEEEeqnarray}
we extract the following structure constant
\begin{IEEEeqnarray}{ll}
\C^{\bullet\bullet\circ}_{12I} = -\frac{(-1)^{L/2} \ell^2}{2^{L+\frac{3}{2}}N_c} \cdot L_2\sqrt{L(L+1)(L+2)} \cdot B\Big(\frac{L}{2}+1,\frac{1}{2}\Big). \qquad \label{ThreePointFunctionStructureConstant}
\end{IEEEeqnarray}
for the HHL three-point function (see figure \ref{Figure:HeavyHeavyLight}) of two heavy BMN chiral primaries \eqref{BMNoperators} (with lengths $L_{1,2} \sim \sqrt{\lambda}$) and a light $SO(3)\times SO(3)$ symmetric CPO \eqref{ChiralPrimaryOperators} (with length $L = L_1-L_2 \rightarrow 0$). Because two and three-point functions of CPOs are protected from receiving quantum corrections, the HHL structure constant $\C^{\bullet\bullet\circ}$ which we computed at strong coupling is expected to be same at weak coupling. It would be interesting to reproduce its strong coupling value \eqref{ThreePointFunctionStructureConstant} from a field-theoretic computation, i.e.\ by performing all the Wick contractions between the involved operators.
\subsection[Holographic correlators in dCFTs]{Holographic correlators in dCFTs \label{SubSection:HolographicCorrelatorsdCFTs}}
\noindent The stage is now set for the computation of correlation functions in strongly coupled holographic dCFTs. The dCFTs emerge on the gauge theory side of holographic dualities when a probe brane is inserted on their string theory side. To compute dCFT correlation functions at strong coupling we can use semiclassical strings and Witten diagrams, just like in the undeformed case (cf.\ \S\ref{SubSection:HolographicCorrelatorsCFTs}). This time however there is a probe brane in the bulk which interacts with the strings and their light modes. Nonperturbative integrability methods (such as the TBA, QSC, hexagons etc.) have not been developed for dCFTs, so semiclassical analysis is still the most widely available tool for the computation of correlation functions at strong coupling. Besides, we have seen in \S\ref{Section:StringIntegrabilityHolographicDefects} that, even when the AdS/{\color{red}d}CFT duality descends from \quotes{solvable} holographic dualities like \eqref{AdS5CFT4duality}, \eqref{AdS4CFT3duality}, it might still break integrability. Supersymmetric localization \cite{RobinsonUhlemann17, Robinson17, Wang20a, KomatsuWang20, BeccariaCaboBizet23} is another powerful method for the computation of correlation functions in strongly coupled dCFTs, yet it is only available for supersymmetric defects. \\
\indent The methods for computing correlation functions that we will describe below are generally independent from the presence of supersymmetry and integrability. They are also generally independent from codimensionality, although the present work mainly discusses codimension-1 dCFTs. Scalar one-point functions are very important observables in defect CFTs (as we have seen for codimension-1 defects in \S\ref{SubSubSection:DefectConformalFieldTheories} above), since their two and higher point correlation functions can generally be obtained from one-point function structure constants, the CFT data and the OPE. On the other hand, one-point functions of operators with spin vanish in codimension-1 dCFTs, while their two and higher point correlation functions are generally nonzero. As a consequence, their structure constants show up as independent dCFT data in the OPE. In what follows however we will only compute scalar two-point functions for codimension-1 defects and verify their agreement with the OPE.
\subsubsection[Weak coupling]{Weak coupling \label{SubSection:HolographicCorrelatorsdCFTsWeak}}
The study of correlation functions in holographic defect CFTs by a combination of integrability and perturbative methods has grown significantly as a field over the past few years. The present review focuses only on the strongly coupled sides of holographic dCFTs (and string theory methods which have been developed to study them), so that we will only attempt to summarize the results which have been obtained at weak coupling. The interested reader may find more details about weak coupling computations in the reports \cite{deLeeuwIpsenKristjansenWilhelm17, deLeeuw19, Linardopoulos20, KristjansenZarembo24a}, as well as in the PhD theses \cite{BuhlMortensen17, Vardinghus19, Widen19, Gombor20, Volk21}.\footnote{See also the master theses \cite{Guo17, Li24}.} \\
\indent Correlation functions can be computed in weakly coupled dCFTs by means of \quotes{interfaces} or domain walls which separate two different, or even two copies of the same QFT. Interfaces are described by means of classical solutions of the ambient equations of motion, aka \quotes{fuzzy funnel} solutions \cite{ConstableMyersTafjord99, ConstableMyersTafjord01a}. To be valid descriptions of a probe-brane system, the classical solutions must share its global symmetry structure. For supersymmetric configurations they must also satisfy the Nahm equations \cite{Nahm79c}. Classical solutions for the scalar fields of $\N = 4$ SYM and ABJM theory (\eqref{LagrangianSYM}, \eqref{LagrangianABJM}) corresponding to the D3-D5, D3-D7, and D2-D4 defects have been obtained in \cite{Diaconescu96, GiveonKutasov98, KristjansenSemenoffYoung12b, Terashima08b}. \\
\indent Fuzzy funnel solutions provide the vevs which are acquired by the ambient fields in the presence of the boundary/defect. In fact we only consider boundaries (not defects) at weak coupling, although we usually call them defects (and dCFTs). The vevs vanish right on the boundary but are generally nonzero as we move away from it. Conformal symmetry is restored very far from the defect where all the field vevs vanish (they typically scale as $1/z$, where $z$ is the distance from the boundary at $z=0$). To compute correlation functions of local gauge invariant operators in the defect CFT, we must replace each field entry in the operator by the corresponding vev (fuzzy funnel solution) \cite{NagasakiTanidaYamaguchi11, NagasakiYamaguchi12}. \\
\indent Take for example scalar single-trace operators in a codimension-1 domain wall version of $\N = 4$ SYM. The generic form of (scalar) one-point functions in codimension-1 dCFTs is given by \eqref{OnePointFunctionsDefectCFT}. To compute them, we substitute the corresponding vevs of the ambient fields,
\begin{IEEEeqnarray}{c}
\OO\left(\z,\textbf{x}\right) \equiv \Psi^{\mu_1 \ldots \mu_L}\tr\left[\varphi_{\mu_1}\ldots\varphi_{\mu_L}\right] \xrightarrow[\text{ interface }]{\varphi_i = t_{i}/{\z}} \frac{1}{{\z}^L} \cdot \Psi^{\mu_1 \ldots \mu_L}\tr\left[t_{\mu_1}\ldots t_{\mu_L}\right], \label{HiggsMechanism}
\end{IEEEeqnarray}
where $\Psi^{\mu_1 \ldots \mu_L}$ is an $SO\left(6\right)$ symmetric tensor. One-point function structure constants $\C$ are given by
\begin{IEEEeqnarray}{c}
\C = \frac{1}{\sqrt{L}}\left(\frac{8\pi^2}{\lambda}\right)^{L/2} \cdot \frac{\left\langle\text{MPS}|\Psi\right\rangle}{\left\langle\Psi|\Psi\right\rangle^{\frac{1}{2}}}, \qquad \left\langle\Psi|\Psi\right\rangle \equiv \Psi^{\mu_1 \ldots \mu_L} \Psi_{\mu_1 \ldots \mu_L},
\end{IEEEeqnarray}
where $\left\langle\text{MPS}|\Psi\right\rangle \equiv \Psi^{\mu_1 \ldots \mu_L}\cdot\tr\left[t_{\mu_1}\ldots t_{\mu_L}\right]$ is the overlap of the ambient operator $\OO$ (in the form of a quantum state $|\Psi\rangle$) with the matrix product state (MPS). The latter encodes the fuzzy funnel solution, i.e.\ the Higgs vevs of the ambient fields. The normalization of the operator $\OO$ as $\OO \rightarrow (2\pi)^L \left(L\lambda^L\right)^{-1/2}\OO$ ensures that its ambient 2-point function will be normalized to unity, i.e.\ it will have the form \eqref{TwoPointFunctionsCFT}. \\
\indent Obviously, the ambient operators $\OO$ have to have definite scaling dimensions, that is they should be eigenstates of the dilatation operator in the ambient theory. For $\N = 4$ SYM (and ABJM theory), the dilatation operator is an integrable spin chain (periodic because of the trace), so that the operators $\OO$ (and the corresponding states $|\Psi\rangle$) can be Bethe (highest-weight) states or their descendants. On the other hand, matrix product states $|\text{MPS}\rangle$ are common forms of spin chain boundary states $|\text{B}\rangle$ (another type is valence bond states). In turn, boundary states can be associated with quantum reflection matrices $\textrm{K}$ which describe scattering off the ends of open spin chains (i.e.\ operators which contain defect fields or host appropriate BCs for the ambient fields). Thus, another characterization of integrable boundary states and matrix product states (besides quench criteria such as \eqref{QuenchIntegrabilityCriterion}) requires that the $\textrm{K}$-matrix satisfies the boundary Yang-Baxter equation. \\
\indent By associating holographic defect CFTs to interfaces and boundary states in the way we just described, we can compute tree-level correlation functions at weak coupling by means of the recipe \eqref{HiggsMechanism}. To go beyond tree-level, we generally have to take into account the fluctuations of the ambient Lagrangian, higher-loop corrections to the dilatation operator, as well as the renormalization of the operators under study. Tree-level one and two-point functions of scalar and tensor single-trace local operators have been computed for the codimension-1 dCFT in four dimensions (based on $\N = 4$ SYM) that is dual to the D3-D5 system in \cite{NagasakiYamaguchi12, deLeeuwKristjansenZarembo15, Buhl-MortensenLeeuwKristjansenZarembo15, deLeeuwKristjansenMori16, deLeeuwKristjansenLinardopoulos18a, KristjansenMullerZarembo20a} and \cite{deLeeuwIpsenKristjansenVardinghusWilhelm17, Widen17, deLeeuwKristjansenLinardopoulosVolk23, BaermanChalabiKristjansen24}. The Coulomb branch of this theory was recently explored in \cite{IvanovskiyKomatsuMishnyakovTerzievZaigraevZarembo24}. At one-loop order, one-point functions of scalar single-trace operators were computed in \cite{Buhl-MortensenLeeuwIpsenKristjansenWilhelm16a, Buhl-MortensenLeeuwIpsenKristjansenWilhelm16c}. Asymptotic all-loop proposals for the one-point functions of scalar operators can be found in \cite{Buhl-MortensenLeeuwIpsenKristjansenWilhelm17a, GomborBajnok20a, GomborBajnok20b, KristjansenMullerZarembo20b, KristjansenMullerZarembo21}. \\
\indent For the two (codimension-1) dCFTs in four dimensions (based on $\N = 4$ SYM) which are dual to the D3-D7 system, tree-level one and two-point functions of scalar and tensor single-trace operators have been computed in \cite{KristjansenSemenoffYoung12b, deLeeuwKristjansenLinardopoulos16, deLeeuwKristjansenVardinghus19, deLeeuwGomborKristjansenLinardopoulosPozsgay19, Linardopoulos25b}. One-loop corrections to scalar correlation functions appeared in \cite{GimenezGrauKristjansenVolkWilhelm18, GimenezGrauKristjansenVolkWilhelm19}. For the codimension-1 dCFT in three dimensions (based on ABJM theory) that is dual to the D2-D4 intersection, tree-level one-point functions of scalar operators have been studied in \cite{KristjansenVuZarembo21, Gombor21, GomborKristjansen22}. For the codimension-3 dCFT in one dimension (based on $\N = 4$ SYM) that is dual to the D1-D3 intersection, scalar one-point functions have been computed up to one-loop order in \cite{KristjansenZarembo23, KristjansenZarembo24b}. One-point functions in the 4-dimensional, codimension-1 dCFTs (based on $\N = 4$ SYM) that are dual to the D3-NS5 system and the $\beta$-deformed D3-D5 system have been studied in \cite{Rapcak15, Widen18}.
\subsubsection[Strong coupling]{Strong coupling \label{SubSection:HolographicCorrelatorsdCFTsStrong}}
\noindent At strong coupling, the computation of correlation functions in holographic defect CFTs with probe branes is mainly carried out with semiclassical analysis. One-point functions of (protected) chiral primary operators (CPOs) \eqref{ChiralPrimaryOperators} have been worked out in the supergravity approximation with the GKPW prescription \eqref{GKPWprescription}--\eqref{WittenDiagrams}. The results agree with the corresponding (matrix product state) calculations at weak coupling as we will see below. For the D3-D5 intersection, one-point functions of CPOs have been computed in \cite{NagasakiYamaguchi12} (see also the paper \cite{Buhl-MortensenLeeuwKristjansenZarembo15}), for the D3-D7 intersection in \cite{KristjansenSemenoffYoung12b} and more recently for the D2-D4 system in \cite{KristjansenVuZarembo21} and for the D1-D3 system in \cite{KristjansenZarembo23}. The computation of two and higher-point correlation functions in probe-brane systems was recently set up in \cite{GeorgiouLinardopoulosZoakos23}. The above calculations involve infinite-sized (probe) branes in AdS which stretch from one boundary to the other. They were preceded by a wide set of computations with finite-sized (probe) branes, such as giant gravitons. See e.g.\ \cite{BakChenWu11, BissiKristjansenYoungZoubos11, CaputaMelloKochZoubos12, HiranoKristjansenYoung12, Lin12, KristjansenMoriYoung15} for some interesting works in AdS$_5\times\text{S}^5$ and AdS$_4\times\CP^3$. Supersymmetric localization \cite{RobinsonUhlemann17, Robinson17, Wang20a, KomatsuWang20, BeccariaCaboBizet23} is a powerful nonperturbative method for the computation of correlation functions from weak to strong coupling. The results obtained by localization are in full agreement with the corresponding supergravity calculations, however we will not discuss them here. Let us now describe semiclassical methods in more detail. 
\paragraph{One-point functions} One-point functions of defect CFT operators that are dual to light (supergravity) modes $\phi_I$ can be computed in strongly coupled AdS/{\color{red}d}CFT by using the recipe \eqref{CorrelationFunction1}. This time, instead of the nonlocal gauge invariant CFT operator $\W$ (which was dual to a classical string worldsheet in \eqref{CorrelationFunction1}), there is a probe Dp-brane in the bulk of AdS, so that: 
\begin{IEEEeqnarray}{l}
\left\langle\OO_I\left(x\right)\right\rangle_{\text{Dp}} = \lim_{z\rightarrow 0}\left[\frac{\pi}{z^{\Delta_I}}\sqrt{\frac{2}{\Delta_I - 1}} \cdot\big\langle\phi_I\left(x,z\right)\cdot \frac{1}{Z_{\text{Dp}}}\int D\Y\,e^{-S_{\text{Dp}}\left[\Y\right]}\big\rangle_{\text{bulk}}\right], \qquad \label{OnePointFunctionsDpbrane}
\end{IEEEeqnarray}
where $\Delta_I$ are the scaling dimensions of the light scalar mode $\phi_I$. The Dp-brane action $S_{\text{Dp}}$ is given by \eqref{D5braneAction} for the probe D5-brane, \eqref{D7braneAction} for the probe D7-brane and \eqref{D4braneAction} for the probe D4-brane. As before, it depends indirectly on the bulk supergravity modes $\phi_I$ via a disturbance which is induced on the supergravity fields by a local operator insertion. By expanding the Dp-brane action around $\phi_I = 0$ we get,
\begin{IEEEeqnarray}{l}
S_{\text{Dp}}\left[\Y\right] = S_{\text{Dp}}\left[\Y\right]\Big|_{\phi_I = 0} + \left.\frac{\partial S_{\text{Dp}}\left[\Y\right]}{\partial\phi_I}\right|_{\phi_I = 0}\cdot\phi_I\left(y,w\right) + \ldots \label{DpbraneAction}
\end{IEEEeqnarray}
In the strong coupling regime ($\lambda\to\infty$), the Dp-brane path integral is dominated by a saddle point which corresponds to classical Dp-brane solutions $\Y_{\text{cl}}$. The one-point function becomes:
\begin{IEEEeqnarray}{c}
\left\langle\OO_I\left(x\right)\right\rangle_{\text{Dp}} = \lim_{z\rightarrow 0}\frac{\pi}{z^{\Delta_I}}\sqrt{\frac{2}{\Delta_I - 1}}\cdot \left.\frac{\partial S_{\text{Dp}}\left[\Y_{\text{cl}}\right]}{\partial\phi_I}\right|_{\phi_I = 0} \cdot \left\langle\phi_I\left(x,z\right)\cdot\phi_I\left(y,w\right)\right\rangle_{\text{bulk}} + \ldots, \qquad
\end{IEEEeqnarray}
where $\left\langle\phi_I\left(x,z\right)\phi_I\left(y,w\right)\right\rangle$ is the bulk-to-bulk propagator in AdS$_{d+1}$ which can be found in \eqref{PropagatorBulkToBulk1}--\eqref{PropagatorBulkToBulk2}.
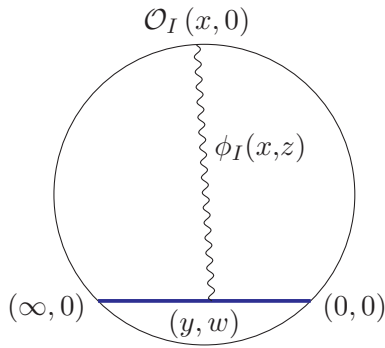
\begin{figure}[H]\begin{center}\begin{tikzpicture}
\draw (0,0) circle(2);\draw[line width=0.5mm, blue] (-1.414,-1.414) -- (1.414,-1.414);
\node (O3) at (-0.1,2.3) {$\OO_I\left(x,0\right)$};
\node (O4) at (2,-1.5) {$\left(0,0\right)$};
\node (O5) at (-2.1,-1.5) {$\left(\infty,0\right)$};
\node (O6) at (0,-1.72) {$\left(y,w\right)$};
\begin{feynman}
\vertex (a) at (-.1,2);
\vertex (b) at (0.1,-1.414);
\diagram*{(a) -- [photon, edge label=\(\phi_{I} (x{,}z)\)] (b)};
\end{feynman}
\end{tikzpicture}\caption{Holographic one-point function in AdS/{\color{red}d}CFT.}\label{Figure:OnePointFunction-dCFT}\end{center}\end{figure}
\vspace{-.4cm} Let us consider the case of the probe D5-brane. By varying the probe D5-brane action $S_{\text{D5}}$ in \eqref{D5braneAction} and applying the variations \eqref{FieldPerturbationIIB}--\eqref{VertexOperatorsIIB} of type IIB supergravity fields, we arrive at \cite{NagasakiYamaguchi12, KristjansenSemenoffYoung12b}:
\begin{IEEEeqnarray}{ll}
\left\langle\OO_I\left(x\right)\right\rangle_{\text{D5}} = \frac{\pi}{z^{\Delta_I}}\sqrt{\frac{2}{\Delta_I - 1}} \cdot \frac{T_5}{g_s}\int d^6\zeta \left(\delta \LL_{\text{DBI}} + \delta \LL_{\text{WZ}}\right) \cdot \G_{\Delta_I}\left(y,w;x\right), \qquad \label{OnePointFunctionsD5brane}
\end{IEEEeqnarray}
where $\G_{\Delta_I}\left(y,w;x\right)$ is the boundary limit of the bulk-to-bulk propagator in \eqref{PropagatorBulkToBulk1}--\eqref{PropagatorBulkToBulk2}. The variations of the DBI and WZ terms take the following forms:
\begin{IEEEeqnarray}{c}
\delta \LL_{\text{DBI}} \equiv \sqrt{h} h^{ab} \partial_a \Y^M \partial_b \Y^N V^I_{MN}(\Y,\partial_y,\partial_w), \qquad \delta \LL_{\text{WZ}} \equiv 2\pi\alpha' \left(F\wedge v^I(\Y,\partial_y,\partial_w)\right), \qquad \label{FluctuationsDBI+WZ}
\end{IEEEeqnarray}
where $V^I$ and $v^I$ are the vertex operators which describe the interaction of the D5-brane with the light modes $\phi_I$. \\
\indent Now take $\OO_I$ to be a chiral primary operator (CPO) of $\N = 4$ SYM with length $L$. The definition of $\N = 4$ SYM CPOs was given in \eqref{ChiralPrimaryOperators}. The CPO is placed at a distance $x_3 >0$ from the planar boundary of $\N = 4$ SYM at $x_3 = 0$.\footnote{Setting $z = 0$ in \eqref{D5braneEmbedding1}, it is easily seen that the boundary is just the plane $x_3 = 0$ in 4d flat space $x = (x_0,\ldots,x_3)$.} More precisely, the CPO is located at
\begin{IEEEeqnarray}{c}
x_0 = x_1 = x_2 = x_4 = 0, \quad x_3 > 0.
\end{IEEEeqnarray}
Not all CPOs of $\N = 4$ SYM are expected to have non-vanishing one-point functions. The only CPOs with nonzero one-point functions are those which share the $SO(3)\times SO(3)$ global symmetry of the defect CFT. Likewise, the dual supergravity modes $\phi = \phi_I Y_I$ should also share the $SO(3)\times SO(3)$ global symmetry of the probe D5-brane. As a consequence, the spherical harmonics on S$^5$ must be $SO(3)\times SO(3)$ invariant (see appendix \ref{Appendix:SphericalHarmonicsSO3}). As it turns out, $SO(3)\times SO(3)$ symmetric spherical harmonics depend on a single quantum number $j$ which is related to the length of the associated CPO as $L = 2j$. \\
\indent Plugging the induced metric \eqref{InducedMetricD5-2} and the vertex operators \eqref{FluctuationsMetric2a}--\eqref{FluctuationsMetric2b} into the expression \eqref{OnePointFunctionsD5brane}--\eqref{FluctuationsDBI+WZ} for the one-point function we find, for the DBI part
\begin{IEEEeqnarray}{ll}
\delta \LL_{\text{DBI}}\cdot\G_L\left(y,w;x\right) = -\frac{2 \ell^6 L (L-1) \sin\tilde{\theta}}{\pi^2 \N_L w^4 K_w^2} \cdot Y_{L/2}(\tilde{\psi}) \cdot \bigg\{&2 r_3^2 w^2 - 2 \kappa r_3 w \left(K_w - 2w^2\right) + \nonumber \\
& + \kappa^2\left(K_w^2 - 2 r^2 w^2\right)\bigg\}\cdot\frac{w^L}{K_w^L}, \qquad \label{FluctuationsDBI}
\end{IEEEeqnarray}
where $r_i \equiv x_i - y_i$ and $K_w \equiv r^2 + w^2$. The WZ part of the one-point function \eqref{OnePointFunctionsD5brane} becomes:
\begin{IEEEeqnarray}{l}
\delta \LL_{\text{WZ}}\cdot\G_L\left(y,w;x\right) = -\frac{2 \ell^6 L(L-1) \sin\tilde{\theta}}{\pi^2 \N_L w^4 K_w^2} \cdot Y_{L/2}(\tilde{\psi}) \cdot \Big\{2\kappa r_3 w K_w - \kappa^2\left(K_w^2 - 2w^2 K_w\right)\Big\} \cdot \frac{w^L}{K_w^L}. \qquad \ \label{FluctuationsWZ}
\end{IEEEeqnarray}
Adding up the two contributions \eqref{FluctuationsDBI}--\eqref{FluctuationsWZ}, we arrive at:
\begin{IEEEeqnarray}{l}
\left\langle\OO_I^{\text{CPO}}\left(x_3\right)\right\rangle_{\text{D5}} = -\frac{4 \ell^6 T_5}{\pi \N_L g_s} \cdot L\sqrt{2(L-1)} \cdot \int d^6\zeta \, Y_{L/2}(\tilde{\psi}) \sin\tilde{\theta} \left(y_3 - x_3 - \kappa w\right)^2 \cdot \frac{w^{L-2}}{K_w^{L+2}}, \qquad \
\end{IEEEeqnarray}
so that the one-point function becomes, for the D5-brane parametrization \eqref{D5braneEmbedding1}:
\begin{IEEEeqnarray}{ll}
\left\langle\OO_I^{\text{CPO}}\left(x_3\right)\right\rangle_{\text{D5}} = -\frac{4 \ell^6 T_5}{\pi \N_L g_s} &\cdot L\sqrt{2(L-1)} \, \CC_{L/2} \, x_3^2 \cdot \int d^6\zeta \, \frac{\sin\tilde{\theta} \, w^{L-2}}{\left[w^2 + (x_3 - \kappa w)^2 + y_py_p\right]^{L+2}}, \qquad \
\end{IEEEeqnarray}
for $p = 0,1,2$. The worldvolume parametrization of probe D5-brane reads:
\begin{IEEEeqnarray}{l}
\zeta_0 = y_0, \quad \zeta_1 = y_1, \quad \zeta_2 = y_2, \quad \zeta_3 = w, \quad \zeta_4 = \tilde{\theta}, \quad \zeta_5 = \tilde{\chi}, \label{D5braneParametrization}
\end{IEEEeqnarray}
so that by performing the angular integrations we obtain
\begin{IEEEeqnarray}{l}
\left\langle\OO_I^{\text{CPO}}\left(x_3\right)\right\rangle_{\text{D5}} = -\frac{16 \ell^6 T_5}{\N_L g_s} L\sqrt{2(L-1)} \CC_{L/2} x_3^2 \cdot \int\displaylimits_{0}^{\infty} dw \int\displaylimits_{-\infty}^{+\infty} d^3y \frac{w^{L-2}}{\left[w^2 + (x_3 - \kappa w)^2 + y_py_p\right]^{L+2}}. \qquad \
\end{IEEEeqnarray}
Performing the $y$ integrations and changing variables $w = x_3 u$, we obtain the one-point function,
\begin{IEEEeqnarray}{c}
\left\langle\OO_I^{\text{CPO}}\left(x_3\right)\right\rangle_{\text{D5}} = \frac{\C_{I}^{\circ}}{x_3^L},
\end{IEEEeqnarray}
which has the generic form \eqref{OnePointFunctionsDefectCFT} of one-point functions. The structure constant $\C_{I}^{\circ}$ is given by:
\begin{IEEEeqnarray}{c}
\C_{I}^{\circ} = -\frac{(-1)^{L/2}\sqrt{\lambda}}{\pi^{3/2}}\sqrt{\frac{L+2}{2L(L+1)}}\cdot \frac{\Gamma\left(L + \frac{1}{2}\right)}{\Gamma\left(L\right)} \cdot \I_{L-2,L+\frac{1}{2}}, \qquad \label{OnePointFunctionStructureConstantD5}
\end{IEEEeqnarray}
where $L = 2j$ should be even and $j = 0,1,\ldots$ is a nonnegative integer. The structure constant \eqref{OnePointFunctionStructureConstantD5} includes the integral
\begin{IEEEeqnarray}{c}
\I_{a,b}\left(\kappa\right) \equiv \int\displaylimits_{0}^{\infty} du \, \frac{u^a}{\left[u^2 + (1 - \kappa u)^2\right]^b}, \label{IntegralsI}
\end{IEEEeqnarray}
which has been computed in appendix \ref{Appendix:Integrals}. For $a = L-2$, $b = L + 1/2$ ($L=2j$) the integral reads:
\begin{IEEEeqnarray}{c}
\I_{2j-2,2j+\frac{1}{2}} = \kappa^{2j+1}B\big(2j,\frac{1}{2}\big) \cdot \left\{1 + \left[\frac{3}{2} + \frac{\left(2j-3\right)\left(j-1\right)}{2\left(2j-1\right)}\right] \frac{1}{\kappa^2} + \ldots\right\}. \label{IntegralD3D5}
\end{IEEEeqnarray}
The strong coupling value ($\lambda \rightarrow \infty$) of the one-point function structure constant \eqref{OnePointFunctionStructureConstantD5}--\eqref{IntegralD3D5} can be compared with the result of the corresponding calculation at weak coupling ($\lambda \rightarrow 0$), order by order in the perturbative expansion of $1/\kappa^2 = \lambda/(\pi^2 k^2)\rightarrow 0$. Complete agreement has been reported at tree level \cite{NagasakiYamaguchi12} and one-loop \cite{Buhl-MortensenLeeuwIpsenKristjansenWilhelm16a, Buhl-MortensenLeeuwIpsenKristjansenWilhelm16c}.
\paragraph{Two and higher-point functions} To compute leading-order two and higher-point correlation functions in strongly coupled AdS/{\color{red}d}CFT, we can combine the previous two approaches, namely the recipe \eqref{CorrelationFunction1} for computing holographic three and higher-point correlation functions in AdS/CFT, and the recipe \eqref{OnePointFunctionsDpbrane} for computing holographic one-point functions in AdS/{\color{red}d}CFT. Let $\W$ be a nonlocal gauge-invariant CFT operator (e.g.\ a Wilson loop or a product of local operators) that is dual to a classical string worldsheet. Suppose also that there is a probe Dp-brane in the bulk of AdS/CFT which interacts with the semiclassical string state via a light scalar (supergravity) mode $\phi_I$ whose scaling dimension is $\Delta_I$ and its mass is $m$. The light mode $\phi_I$ is essentially emitted from the probe Dp-brane and absorbed by the semiclassical string.\footnote{Two (and higher) point functions in the presence of codimension-1 defects only receive contributions from scalar mediator fields. These are the only ones which can contribute to the OPE \eqref{TwoPointFunctionRatio1}, via their nonzero defect one-point functions $\C_I$. Fields with spin have zero one-point functions and cannot contribute to the OPE \eqref{TwoPointFunctionRatio1}. The author would like to thank G.\ Georgiou for pointing this out.} \\
\indent The ratio of the expectation value $\left\langle\W\right\rangle_{\text{Dp}}$ of the operator $\W$ in the dCFT which is holographically dual to a probe Dp-brane system over its expectation value $\langle\W\rangle$ in the CFT which is holographically dual to the same bulk system in the absence of the probe brane is given at strong coupling by the following expression:
\begin{IEEEeqnarray}{l}
\langle\widetilde{\W}\rangle_{\text{Dp}} \equiv \frac{\langle\W\rangle_{\text{Dp}}}{\langle\W\rangle} = \big\langle\frac{1}{Z_{\text{str}}}\int D\X\,e^{-S_{\text{str}}\left[\X\right]} \cdot \frac{1}{Z_{\text{Dp}}}\int D\Y\,e^{-S_{\text{Dp}}\left[\Y\right]}\big\rangle_{\text{bulk}}. \qquad \label{DefectCorrelator1}
\end{IEEEeqnarray}
By expanding the string and Dp-brane actions around $\phi_I = 0$ (as it was done in \eqref{PolyakovAction3}, \eqref{DpbraneAction} above) and using the fact that when the coupling is strong ($\lambda\to\infty$), the path integrals which show up in \eqref{DefectCorrelator1} are dominated by their saddle points (corresponding to classical solutions $\X_{\text{cl}}$, $\Y_{\text{cl}}$ of the string and Dp-brane equations of motion), we arrive at
\begin{IEEEeqnarray}{ll}
\langle\widetilde{\W}\rangle_{\text{Dp}} &= \langle\exp\left\{\left.\frac{\partial S_{\text{str}}\left[\X_{\text{cl}}\right]}{\partial\phi_I}\right|_{\phi_I = 0}\phi_I + \ldots\right\}\cdot \exp\left\{\left.\frac{\partial S_{\text{Dp}}\left[\Y_{\text{cl}}\right]}{\partial\phi_I}\right|_{\phi_I = 0}\phi_I + \ldots\right\}\rangle_{\text{bulk}} = \nonumber \\[6pt]
& = \langle\left[1 + \left.\frac{\partial S_{\text{str}}\left[\X_{\text{cl}}\right]}{\partial\phi_I}\right|_{\phi_I = 0}\phi_I + \ldots \right]\cdot \left[1 + \left.\frac{\partial S_{\text{Dp}}\left[\Y_{\text{cl}}\right]}{\partial\phi_I}\right|_{\phi_I = 0}\phi_I + \ldots \right]\rangle_{\text{bulk}}. \qquad
\end{IEEEeqnarray}
Because the bulk expectation values of the light modes vanish (i.e.\ $\langle\phi_I\rangle = 0$ in the bulk dual of the CFT without a probe brane) we obtain, to leading order:
\begin{IEEEeqnarray}{c}
\langle\widetilde{\W}\rangle_{\text{Dp}} = 1 + \left.\frac{\partial S_{\text{str}}\left[\X_{\text{cl}}\right]}{\partial\phi_I}\cdot\frac{\partial S_{\text{Dp}}\left[\Y_{\text{cl}}\right]}{\partial\phi_I}\right|_{\phi_I = 0} \cdot \left\langle\phi_I\left(x,z\right)\phi_I\left(y,w\right)\right\rangle_{\text{bulk}} + \ldots, \qquad\label{DefectCorrelator2}
\end{IEEEeqnarray}
where the two-point function $\left\langle\langle\phi_I\left(x,z\right)\cdot\phi_I\left(y,w\right)\right\rangle$ is just the bulk-to-bulk propagator for massive scalar fields $\phi_I$ in Euclidean AdS$_{d+1}$, which has been spelled out in \eqref{PropagatorBulkToBulk1}--\eqref{PropagatorBulkToBulk2}.
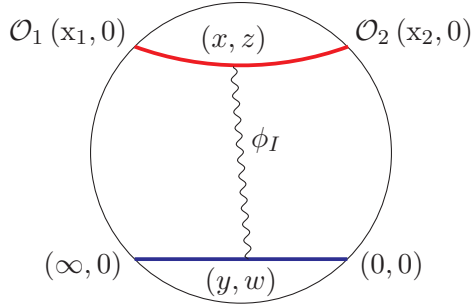
\begin{figure}[H]\begin{center}\begin{tikzpicture}
\draw (0,0) circle(2); \draw[line width=0.5mm, red] (-1.414,1.414) arc (250:290:4.15);\draw[line width=0.5mm, blue] (-1.414,-1.414) -- (1.414,-1.414);
\node (O1) at (-2.3,1.6) {$\OO_1\left(\x_1,0\right)$};
\node (O2) at (2.3,1.6) {$\OO_2\left(\x_2,0\right)$};
\node (O3) at (-0.1,1.5) {$\left(x,z\right)$};
\node (O4) at (2,-1.5) {$\left(0,0\right)$};
\node (O5) at (-2.1,-1.5) {$\left(\infty,0\right)$};
\node (O6) at (0,-1.72) {$\left(y,w\right)$};
\begin{feynman}
\vertex (a) at (-.1,1.15);
\vertex (b) at (0.1,-1.414);
\diagram*{(a) -- [photon, edge label=\(\phi_I\)] (b)};
\end{feynman}
\end{tikzpicture}\caption{Holographic two-point function in a AdS/{\color{red}d}CFT.}\label{Figure:TwoPointFunction}\end{center}\end{figure}
\vspace{-.4cm}\indent Now take the probe brane to be a single D5. The defect correlation function of a nonlocal gauge-invariant operator $\W$ that is dual to a semiclassical string state in AdS$_5\times\text{S}^5$ becomes:
\begin{IEEEeqnarray}{c}
\langle\widetilde{\W}\rangle_{\text{D5}} = 1 + \frac{T_2T_5}{2g_s} \int d^2\sigma \, d^6\zeta \cdot \bigg\{\delta\LL_{\text{str}}\left(\sigma,x,z\right) \delta\LL_{\text{D5}}\left(\zeta,y,w,x\right) G_{\Delta_I}\left(x,z;y,w\right)\bigg\}, \qquad \label{DefectCorrelator3}
\end{IEEEeqnarray}
where $G_{\Delta_I}$ is the bulk-to-bulk propagator \eqref{PropagatorBulkToBulk1}--\eqref{PropagatorBulkToBulk2} of a massive scalar field (scaling dimension $\Delta_I$, mass $m$) in AdS$_5$. The variations of the string and D5-brane Lagrangians read
\begin{IEEEeqnarray}{l}
\delta\LL_{\text{str}} = \partial_a\X^M\partial^a\X^N\,V^I_{MN}\left(\X,\partial_x,\partial_z\right) + \ldots, \qquad \label{FluctuationsString} \\[6pt]
\delta\LL_{\text{D5}} = \sqrt{h} h^{ab} \partial_a \Y^M \partial_b \Y^N V^I_{MN}(\Y,\partial_y,\partial_w) + 2\pi\alpha' \left(F\wedge v^I(\Y,\partial_y,\partial_w)\right), \qquad \label{FluctuationsD5brane}
\end{IEEEeqnarray}
where again, $V^I$ and $v^I$ are the vertex operators which describe the interaction of the light mode $\phi_I$ with the string worldsheet and the D5-brane worldvolume. \\
\indent As before, we take the light mode $\phi_I$ to be dual to a CPO, $\OO^{\text{CPO}}_I$ of $\N = 4$ SYM. Let the length of the CPO be equal to $L$. In order for the correlation function of the operator $\W$ to be nonzero, the CPO and its dual supergravity field $\phi = \phi_I Y_I$ must share the $SO(3)\times SO(3)$ global symmetry of the defect. This means that the S$^5$ spherical harmonics $Y_I$ should be $SO(3)\times SO(3)$ invariant. As such, they depend on a single quantum number $j$ which again fixes the length of the CPO to be equal to $L = 2j$. To simplify the calculations which follow, we assume that the string worldsheet lies very close to the AdS boundary, $z\rightarrow 0$. The near-boundary expansion of the bulk-to-bulk propagator of the massive scalar field $\phi_I$ in AdS$_5$ (mass $m^2\ell^2 = L(L-4)$) is given by \eqref{PropagatorBulkToBulk2}, for $d = 4$:
\begin{IEEEeqnarray}{c}
G_L\left(x,z;y,w\right) = \frac{L - 1}{2\pi^2}\cdot\left\{1 + \frac{L\Lambda_w z^2}{\left(L - 1\right) K_w^2} + \OO\left(z^4\right)\right\}\cdot \left(\frac{z w}{K_w}\right)^L, \qquad \label{PropagatorBulkToBulk3}
\end{IEEEeqnarray}
where $K_w \equiv w^2 + \left(x - y\right)^2$ and $\Lambda_w \equiv 2 w^2 - \left(L - 1\right)\left(x - y\right)^2$. The integrations over the worldvolume coordinates $\zeta$ of the D5-brane in \eqref{DefectCorrelator3} are very similar to those in the defect one-point function \eqref{OnePointFunctionsD5brane}--\eqref{OnePointFunctionStructureConstantD5}. Acting with the vertex operators \eqref{FluctuationsMetric1}--\eqref{FluctuationsPotential1} on the propagator \eqref{PropagatorBulkToBulk3} and using the D5-brane parametrization \eqref{D5braneParametrization},
\begin{IEEEeqnarray}{ll}
\delta\LL_{\text{D5}}\left(\zeta,y,w,x\right) G_L\left(x,z;y,w\right) = &-\frac{4\CC_{L/2}\ell^6 L\left(L-1\right)}{\pi^2\N_L}\,\sin\tilde\theta \cdot \Bigg\{x_3^2 w^{L-2}K_w^{-L-2}z^L - \nonumber \\[6pt]
& - \frac{z^{L+2}}{2(L-1)}\bigg[w^{L-2} K_w^{-L-2} + 2(L+2)\Big[(L-1) x_3^2 w^{L-2} + \nonumber \\[6pt]
& + 2\kappa x_3 w^{L-1}\Big] K_w^{-L-3} - 2(L+2)(L+3) x_3^2 w^L K_w^{-L-4}\bigg]\Bigg\}, \qquad
\end{IEEEeqnarray}
where $K_w = w^2 + (x_3 - \kappa w)^2 + r_p r_p$ and $r_p = x_p - y_p$, for $p = 0,1,2$. For the worldvolume integrals,
\begin{IEEEeqnarray}{l}
\int d^6\zeta \, \delta\LL_{\text{D5}} \, G_L = -\frac{16\CC_{L/2}\ell^6 L\left(L-1\right)}{\pi\N_L}\cdot \int\displaylimits_{0}^{\infty} dw \int\displaylimits_{-\infty}^{+\infty} d^3r \cdot \Bigg\{\ldots\Bigg\} = \nonumber \\[6pt]
= -\frac{16\pi^{1/2}\CC_{L/2}\ell^6 L\left(L-1\right)}{\N_L} \int\displaylimits_{0}^{\infty} dw \, \Bigg\{\Bigg[\frac{\Gamma\left(L + \frac{1}{2}\right)}{\Gamma\left(L + 2\right)} \frac{x_3^2 w^{L-2}}{K_w^{L+\frac{1}{2}}}\Bigg] z^L - \Bigg[\frac{\Gamma\left(L+\frac{1}{2}\right)}{\Gamma(L+2)}\frac{w^{L-2}}{K_w^{L+\frac{1}{2}}} + \nonumber \\[6pt]
+ 2(L+2) \frac{\Gamma\left(L+\frac{3}{2}\right)}{\Gamma(L+3)} \left[(L-1)x_3^2 w^{L-2} + 2\kappa x_3 w^{L-1}\right] \frac{1}{K_w^{L+\frac{3}{2}}} - \nonumber \\[6pt]
- 2(L+2)(L+3) \frac{\Gamma\left(L+\frac{5}{2}\right)}{\Gamma(L+4)} \frac{x_3^2 w^L}{K_w^{L+\frac{5}{2}}}\Bigg] \frac{z^{L+2}}{2(L-1)} + \ldots\Bigg\}, \qquad
\end{IEEEeqnarray}
where $K_w = w^2 + (x_3 - \kappa w)^2$ and we also assumed that $x_{0,1,2} \ll \infty$. Changing variables $w = x_3 u$,
\begin{IEEEeqnarray}{l}
\int d^6\zeta \, \delta\LL_{\text{D5}} \, G_L = -\frac{16\pi^{1/2}\CC_{L/2}\ell^6 L\left(L-1\right)}{\N_L}\cdot \sum_{n=0}^{\infty} \F_n \cdot \frac{z^{L+2n}}{x_3^{L+2n}}, \qquad \label{D5braneIntegral}
\end{IEEEeqnarray}
where the first two coefficients $\F_n$ have the following forms, for $L = 2j$:
\begin{IEEEeqnarray}{ll}
\F_0 &= \frac{\Gamma\left(2j + \frac{1}{2}\right)}{\Gamma\left(2j + 2\right)} \cdot \I_{2j-2,2j+\frac{1}{2}} = \frac{\sqrt{\pi}\,\kappa^{2j+1}}{2j(2j+1)}\cdot\bigg\{1 + \frac{j(2j+1)}{2(2j-1)}\cdot\frac{1}{\kappa^2} + \ldots\bigg\} \label{CoefficientF0} \\[6pt]
\F_1 &= \frac{-1}{2(2j-1)}\Bigg[\frac{\Gamma\left(2j+\frac{1}{2}\right)}{\Gamma(2j+2)} \cdot \I_{2j-2,2j+\frac{1}{2}} + 2(2j+2)\cdot \frac{\Gamma\left(2j+\frac{3}{2}\right)}{\Gamma(2j+3)}\Big[2\kappa \cdot \I_{2j-1,2j+\frac{3}{2}} + \nonumber \\[6pt]
& + (2j-1)\cdot \I_{2j-2,2j+\frac{3}{2}}\Big] - 2(2j+2)(2j+3) \cdot \frac{\Gamma\left(2j+\frac{5}{2}\right)}{\Gamma(2j+4)} \cdot \I_{2j,2j+\frac{5}{2}}\Bigg] = \nonumber \\[6pt]
& = -\frac{\sqrt{\pi}\,\kappa^{2j+1}}{4(2j-1)} \cdot \bigg\{1 + \frac{j(2j+1)}{2(2j-1)}\cdot\frac{1}{\kappa^2} + \ldots \bigg\}. \qquad \label{CoefficientF1}
\end{IEEEeqnarray}
Analytic expressions for the integrals $\I_{a,b}$, for various values of $a$ and $b$, can be found in appendix \ref{Appendix:Integrals}. The integrals are given as perturbative expansions in $1/\kappa^2 \rightarrow 0$. See equations \eqref{IntegralI1}--\eqref{IntegralI7} for the first few terms of the integrals which appear in the expressions of $\F_0$ and $\F_1$ above. \\
\indent Once the $\zeta$ integrations have been carried out, and in order to find the value of the correlation function \eqref{DefectCorrelator3}, we integrate over the string worldsheet coordinates $\sigma$. We again apply the vertex operators \eqref{FluctuationsMetric1} on the D5-brane integral \eqref{D5braneIntegral},
\begin{IEEEeqnarray}{l}
\delta\LL_{\text{str}}\left(\sigma,x,z\right) \int d^6\zeta \, \delta\LL_{\text{D5}}\left(\zeta,y,w,x\right) G_L\left(x,z;y,w\right).
\end{IEEEeqnarray}
For the leading term of \eqref{D5braneIntegral} we find
\begin{IEEEeqnarray}{ll}
\delta\LL_{\text{str}}\left(\sigma,x,z\right) \cdot \frac{z^L}{x_3^L} &= \frac{2\ell^2}{\N_L} \cdot \CC_{L/2} \cdot \bigg\{\left[-\frac{L\delta_{i,j}}{z^2} + \frac{2L\delta_{i,3}\delta_{j,3}}{x_3^2}\right] \partial_a\X^i\partial^a\X^j + \frac{L}{z^2} \, \partial_a\X^z\partial^a\X^z - \nonumber \\
& \quad - \frac{4L\delta_{i,3}}{z x_3} \, \partial_a\X^i\partial^a\X^z + L\ell^{-2}\partial_a\X^{\mu}\partial^a\X^{\nu} \hat{g}_{\mu\nu}\bigg\} \cdot \frac{z^L}{x_3^L} = \nonumber \\
& = \frac{2L\ell^2\CC_{L/2}}{\N_L}\,\frac{1}{x_3^L} \cdot \bigg\{(-\partial_a\X^i\partial^a\X^i + \partial_a\X^z\partial^a\X^z) z^{L-2} - 4x_3^{-1}\partial_a\X^3\partial^a\X^z z^{L-1} + \nonumber \\
& \quad + (2x_3^{-2} \partial_a\X^3\partial^a\X^3 + \ell^{-2}\partial_a\X^{\mu}\partial^a\X^{\nu} \hat{g}_{\mu\nu}) z^L\bigg\},
\end{IEEEeqnarray}
and so on for all the next-to-leading terms, $n = 0,1,2,\ldots$,
\begin{IEEEeqnarray}{l}
\delta\LL_{\text{str}}\left(\sigma,x,z\right) \cdot \frac{z^{L+2n}}{x_3^{L+2n}} = \frac{2\ell^2\CC_{L/2}}{\left(L+1\right)\N_L}\,\frac{1}{x_3^{L+2n}} \cdot \bigg\{\Big[-\left(L^2 + L + 4n\right)\,\partial_a\X^i\partial^a\X^i + \nonumber \\[6pt]
+ \left(L^2 + (8n+1)L + 8n^2\right)\partial_a\X^z\partial^a\X^z\Big]z^{L+2n-2} - 4\left(L + 2n\right)\left(L + 2n +1\right) x_3^{-1}\partial_a\X^3\partial^a\X^z z^{L+2n-1} + \nonumber \\[6pt]
+ \Big[2 \left(L + 2n\right)\left(L + 2n +1\right) \,x_3^{-2} \partial_a\X^3\partial^a\X^3 + L\left(L+1\right)\ell^{-2}\partial_a\X^{\mu}\partial^a\X^{\nu} \hat{g}_{\mu\nu}\Big] z^{L+2n}\bigg\}.
\end{IEEEeqnarray}
Putting together the contributions from $\delta\LL_{\text{D5}}$ and $\delta\LL_{\text{str}}$, we obtain the correlation function \eqref{DefectCorrelator1}:
\begin{IEEEeqnarray}{ll}
\langle\widetilde{\W}\rangle_{\text{D5}} = & 1 + \frac{(-1)^L (L+2) \lambda}{16N_c\pi^{5/2}} \int_{0}^{2\pi}\int_{-\infty}^{+\infty} d\sigma \, d\tau \cdot \sum_{n=0}^{\infty} \F_n \cdot \frac{z^{L+2n}}{x_3^{L+2n}}\cdot \nonumber \\[6pt]
&\cdot\bigg\{\Big[\left(L^2 + L + 4n\right)\left(\partial_a\X^i\partial^a\X^i\right) - \left(L^2 + (8n+1)L + 8n^2\right)\left(\partial_a\X^z\partial^a\X^z\right)\Big] z^{-2} \nonumber \\[6pt]
& - L\left(L+1\right)\left(\ell^{-2}\partial_a\X^{\mu}\partial^a\X^{\nu} \hat{g}_{\mu\nu}\right) + 4\left(L + 2n\right)\left(L + 2n +1\right)\left(\partial_a\X^3\partial^a\X^z\right)x_3^{-1}z^{-1} \nonumber \\[6pt]
&- 2 \left(L + 2n\right)\left(L + 2n +1\right)\left(\partial_a\X^3\partial^a\X^3\right)x_3^{-2}\bigg\}. \qquad \label{DefectTwoPointFunction}
\end{IEEEeqnarray}
\paragraph{Two BMN operators} We can take again the nonlocal operator $\W$ to be given by $\W \equiv \OO^{\dag}_{1}\OO_{2}$, where $\OO_{i}$ ($i = 1,2$) are two BMN chiral primaries, defined in \eqref{BMNoperators}. The primaries are located at the points $\x_{1,2}$ on the $x_3$ axis, at a small distance $x_{12}$ from each other. As before (see \S\ref{SubSection:HolographicCorrelatorsCFTs}), when the length difference of the two operators $L = L_1 - L_2$ is small, the operator $\W$ is holographically dual to a classical (pointlike) string solution which is given in \eqref{BMNstring1}--\eqref{BMNstring2}, and $R = \x_{12}/2 \rightarrow 0$. By using the definition of $R$ in \eqref{BMNstring2}, we can write the defect invariant ratio $\xi$ as follows:
\begin{IEEEeqnarray}{ll}
\xi \equiv \frac{\x_{12}^2}{4\x_1\x_2} = \frac{R^2}{\bar\x^2 - R^2} \Rightarrow \frac{1}{\uu^2} \equiv \frac{R^2}{\bar\x^2} = \frac{\xi}{1 + \xi}.
\end{IEEEeqnarray}
By plugging the ansatz \eqref{BMNstring1}--\eqref{BMNstring2} into the formula \eqref{DefectTwoPointFunction} which gives the defect correlation function $\langle\widetilde{\W}\rangle_{\text{D5}}$ (now a two-point function), we are led to the following value for the ratio of the defect two-point function over the CFT two-point function (i.e.\ in the absence of the defect):
\begin{IEEEeqnarray}{ll}
\frac{\langle\OO^{\dag}_{1}\left(\x_1\right)\OO_{2}\left(\x_2\right)\rangle_{\text{D5}}}{\langle\OO^{\dag}_{1}\left(\x_1\right)\OO_{2} \left(\x_2\right)\rangle} =& 1 - \frac{(-1)^L(L+2)\omega\lambda}{8N_c\pi^{3/2}}\sum_{n=0}^{\infty} \F_n \cdot \bigg\{2\Big[L^2 + 2n + L - \nonumber \\[6pt]
&- y^2(L + 2n)(L + 2n + 1)\Big]\J_{\frac{L}{2} + n, L + 2n + 2} + 8n(L+n)\Big[2y\J_{\frac{L}{2} + n - 1, L + 2n + 1} - \nonumber \\[6pt]
&- \left(y^2 - 1\right)\J_{\frac{L}{2} + n - 1, L + 2n + 2}\Big]\bigg\}. \qquad \label{TwoPointFunctionBMN1}
\end{IEEEeqnarray}
The definitions of the integrals $\J_{a,b}$ can be found in appendix \ref{Appendix:Integrals}, see equation \eqref{IntegralsJ}. For small values of the defect ratios $\xi \sim y^{-2} \rightarrow 0$, the integrals $\J_{a,b}$ which appear in \eqref{TwoPointFunctionBMN1} can be expanded in power series. The analytic expressions of these power series can be found in \eqref{IntegralJ1}--\eqref{IntegralJ3}. Plugging these expressions into the formula \eqref{TwoPointFunctionBMN1} for the two-point function we find, for $L=2j$:
\begin{IEEEeqnarray}{ll}
\frac{\langle\OO^{\dag}_{1}\left(\x_1\right)\OO_{2}\left(\x_2\right) \cdot \rangle_{\text{D5}}}{\langle\OO^{\dag}_{1}\left(\x_1\right)\OO_{2} \left(\x_2\right)\rangle} = 1 &+ \frac{2j^2(j+1)L_2\sqrt{\lambda}}{N_c\pi^{3/2}} \cdot B\left(j,1/2\right)\,\F_0 \, \xi^{j} \Bigg\{1 + \frac{2j}{(2j+1)}\cdot\frac{\F_1}{\F_0} \cdot \xi + \nonumber \\[6pt]
&+ \frac{4j(j+1)}{(2j+1)(2j+3)}\cdot\frac{\F_2}{\F_0} \cdot \xi^2 + \ldots \Bigg\}, \qquad \quad \label{TwoPointFunctionBMN2}
\end{IEEEeqnarray}
so that by plugging the coefficients $\F_{0,1}$ that we found above in \eqref{CoefficientF0}--\eqref{CoefficientF1}, we reproduce the expected result from the OPE \eqref{TwoPointFunctionRatio1}:
\begin{IEEEeqnarray}{c}
\frac{\langle\OO_{1}\left(\x_1\right)\OO_{2}\left(\x_2\right)\rangle_{\text{{\color{red}d}CFT}}}{\langle\OO_{1}\left(\x_1\right)\OO_{2} \left(\x_2\right)\rangle_{\text{CFT}}} = 1 + 2^{L} \C_{12}^{I} \, \C_I \, \xi^j \cdot \Bigg\{1 - \frac{j^2}{2j-1} \cdot\xi + \frac{j(j+1)^2}{4(2j-1)} \cdot\xi^2 + \ldots\Bigg\}. \label{TwoPointFunctionRatio2}
\end{IEEEeqnarray}
We have checked agreement of the two-point function to the OPE prediction up to next-to-next-to-leading order ($\xi^2$), but it should be quite straightforward to obtain higher-order terms in $\xi$ and show that they agree too. Therefore the full conformal block in the ambient channel \eqref{AmbientConformalBlocks} has been reproduced for two BMN operators given by \eqref{BMNoperators}. Of course, in this case and all cases of codimension-1 defects, the correlation function \eqref{TwoPointFunctionBMN2} of any two CFT operators (and in fact any higher-order correlation function) which interact with a probe-brane via a light mode $\phi_I$ could have been predicted from the sole knowledge of the defect one-point function structure constant $\C^{\circ}$ (given in \eqref{OnePointFunctionStructureConstantD5} for the D3-D5 brane), the HHL three-point function structure constant $\C_{123}^{\bullet\bullet\circ}$ \eqref{ThreePointFunctionStructureConstant} (i.e.\ in the absence of a defect) and the ambient OPE \eqref{OperatorProductExpansion}.
\subsection[Discussion and outlook]{Discussion and outlook \label{SubSection:CorrelatorOutlook}}
\noindent Based on the recent paper \cite{GeorgiouLinardopoulosZoakos23}, we have reviewed the computation of correlation functions with semiclassical analysis in strongly coupled holographic defect CFTs which are dual to probe-brane systems. Our method is based on the works \cite{BerensteinCorradoFischlerMaldacena99} and \cite{Zarembo10c, CostaMonteiroSantosZoakos10} which set up semiclassical analysis for strongly coupled CFTs. A series of calculations with finite-sized branes in CFTs, such as giant gravitons \cite{BakChenWu11, BissiKristjansenYoungZoubos11, CaputaMelloKochZoubos12, HiranoKristjansenYoung12, Lin12, KristjansenMoriYoung15}, appeared shortly after these works. Computations of one-point functions with supergravity in strongly coupled dCFTs (still dual to probe D-brane systems) were carried out in \cite{NagasakiYamaguchi12, KristjansenSemenoffYoung12b, KristjansenVuZarembo21, KristjansenZarembo23}. Here we have shown that the computation of one-point functions in supergravity can be part of a broader and more systematic framework. \\
\indent Although we have focused on the case of infinitely sized codimension-1 defects, where scalar two and higher-point functions are uniquely determined by dCFT one-point functions, CFT data and the OPE, our methods apply to dCFTs of higher codimensionalities, as well as non-scalar correlation functions. Operators with spin have vanishing one-point functions in codimension-1 dCFTs, so that their two-point function structure constants are independent dCFT data. As such, they cannot be determined in terms of the OPE and the other CFT and dCFT data; semiclassical analysis is the only way to proceed. Semiclassical methods are also generally independent from the presence of integrability and supersymmetry. \\
\indent Localization \cite{Pestun07} is a powerful nonperturbative method which was recently applied to the computation of (D3-D5) defect correlation functions of scalar operators in \cite{RobinsonUhlemann17, Robinson17, Wang20a, KomatsuWang20, BeccariaCaboBizet23}. Supersymmetric localization can in principle be used for the computation of scalar one-point functions at strong coupling in all supersymmetric dCFTs which are dual to probe-brane systems (such as e.g\ D3-D5, D2-D4), but not in non-supersymmetric setups (e.g.\ D3-D7). Yet another method for the computation of correlation functions in strongly coupled AdS/CFT is the geodesic approximation method \cite{BalasubramanianRoss99, BalasubramanianBernamontiCrapsKeranenKeskiVakkuriMullerThorlaciusVanhoof12} which was recently applied to probe-brane systems in \cite{KastikainenShashi21, Park24}. In the supergravity limit, boundary and defect CFT correlation functions can also be computed with the so-called Mellin transform formalism \cite{RastelliZhou17a, GoncalvesItsios18, KavirajPaulos18, MazacRastelliZhou18, ChenZhou23, GimenezGrau23}. However, the precise relation of these methods to the semiclassical analysis we have been discussing in the present section remains unclear. It would be of course very interesting to explore it further. \\
\indent In \eqref{TwoPointFunctionBMN2}--\eqref{TwoPointFunctionRatio2}, we essentially reproduced the ambient conformal block \eqref{AmbientConformalBlocks} in the case of two BMN operators \eqref{BMNoperators}. Our treatment was perturbative in the dCFT invariant ratio $\xi$ so that it would be interesting to work out a nonperturbative proof which is valid to all orders in $\xi$, possibly by using the Mellin transform approach \cite{RastelliZhou17a}. It would also be interesting to consider other heavy and possibly more complicated operators than the simple CPOs we have examined here. In principle though, the full conformal block \eqref{AmbientConformalBlocks} shows up in the correlation functions of any two scalar operators, so that it should in principle be possible (if only relatively hard) to demonstrate this fact explicitly. Indeed, it was shown in \cite{GeorgiouLinardopoulosZoakos23} that the leading term in the expansion \eqref{TwoPointFunctionBMN2}--\eqref{TwoPointFunctionRatio2} is independent from the involved operators. It would be interesting if an all-loop argument could be worked out. \\
\indent In the present work, we have only considered defect correlation functions for relatively simple string states with no conserved charges. However, we could just as well consider setups which involve more complicated string states which do carry conserved charges. In this context, the inclusion of finite-size effects in defect correlation functions would be particularly interesting and it would have to follow closely the corresponding calculations in strongly coupled AdS/CFT (see e.g.\ \cite{AhnBozhilov11a, LeePark11}). Also, for a wide set of holographic dCFTs of various codimensions which were analyzed in \cite{deLeeuwHolguin24}, it would be interesting to study their correlation functions at strong coupling with the semiclassical methods which were described in the present work, and examine whether the results match with the corresponding calculations at weak coupling. For this, one would have to properly identify the dual probe-brane configurations.
\section[Acknowledgements]{Acknowledgements}
\noindent The author is thankful to C.\ Ahn, M.\ Axenides, C.\ Bachas, Z.\ Bajnok, G.\ Georgiou, C.\ Kristjansen, H.-B.\ Nielsen, C.\ Park, M.\ Roberts, K.\ Zarembo, and D.\ Zoakos for discussions. This work was supported by the National Development Research and Innovation Office (NKFIH) research grant K134946. The work of G.L.\ was supported in part by the National Research Foundation of Korea (NRF) grant funded by the Korea government (MSIT) (No.\ 2023R1A2C1006975), as well as by an appointment to the JRG Program at the APCTP through the Science and Technology Promotion Fund and Lottery Fund of the Korean Government.
\newpage\appendix
\section[Gamma, T \& R matrices]{Gamma, T \& R matrices}
\subsection{Gamma matrices in $3+1$ dimensions \label{Appendix:4dGammaMatrices}}
\paragraph{Weyl (chiral) basis} In $d = 3+1$ dimensions, gamma matrices are 4-dimensional. In the Weyl or chiral basis, the gamma matrices are defined as follows:
\begin{IEEEeqnarray}{l}
\gamma^0 = \left(\begin{array}{cc}
0 & \sigma_0 \\
\sigma_0 & 0 \\\end{array}\right), \quad
\gamma^i = i\sigma_2 \otimes \sigma_i = \left(\begin{array}{cc}
0 & \sigma_i \\
-\sigma_i & 0 \\\end{array}\right), \quad
\gamma_5 = \left(\begin{array}{cc}
-\sigma_0 & 0 \\
0 & \sigma_0 \\\end{array}\right) = i \gamma^0\gamma^1\gamma^2\gamma^3, \qquad \label{GammaMatrices1}
\end{IEEEeqnarray}
where $i = 1, 2, 3$ and the $\gamma^0$ matrix is antidiagonal. The Pauli matrices $\sigma_{\mu}$ are as usual defined as:
\begin{IEEEeqnarray}{llll}
\sigma_0 = \left(\begin{array}{cc}
1 & 0 \\
0 & 1 \\\end{array}\right) = I_2, \qquad &
\sigma_1 = \left(\begin{array}{cc}
0 & 1 \\
1 & 0 \\\end{array}\right), \qquad &
\sigma_2 = \left(\begin{array}{cc}
0 & -i \\
i & 0 \\\end{array}\right), \qquad &
\sigma_3 = \left(\begin{array}{cc}
1 & 0 \\
0 & -1 \\\end{array}\right). \qquad \label{PauliMatrices}
\end{IEEEeqnarray}
The gamma matrices \eqref{GammaMatrices1} obey the Minkowski Clifford algebra with a mostly minus signature:
\begin{IEEEeqnarray}{l}
\gamma^{\mu}\gamma^{\nu} + \gamma^{\nu}\gamma^{\mu} = -2 \eta^{\mu\nu} = 2 \times \text{diag}\left(+,-,-,-\right), \qquad
\gamma^{\mu}\gamma_{5} + \gamma_{5}\gamma^{\mu} = 2\, \delta^{\mu}_5, \qquad
\end{IEEEeqnarray}
where $\mu, \nu = 0,\ldots,3$. The matrices $G^i$ that show up in the Lagrangian density \eqref{LagrangianSYM} of $\N = 4$ SYM are given by (see e.g.\ \cite{BuhlMortensen17, Buhl-MortensenLeeuwIpsenKristjansenWilhelm16c} for more):
\begin{IEEEeqnarray}{lll}
G^1 = \left(\begin{array}{cc}
0 & -i\sigma_3 \\
i\sigma_3 & 0 \\\end{array}\right), \qquad &
G^2 = \left(\begin{array}{cc}
0 & i\sigma_1 \\
-i\sigma_1 & 0 \\\end{array}\right), \qquad &
G^3 = \left(\begin{array}{cc}
\sigma_2 & 0 \\
0 & \sigma_2 \\\end{array}\right) \label{G-Matrices1} \\[6pt]
G^4 = \left(\begin{array}{cc}
0 & -i\sigma_2 \\
-i\sigma_2 & 0 \\\end{array}\right), \qquad &
G^5 = \left(\begin{array}{cc}
0 & -\sigma_0 \\
\sigma_0 & 0 \\\end{array}\right), \qquad &
G^6 =\left(\begin{array}{cc}
i\sigma_2 & 0 \\
0 & -i\sigma_2 \\\end{array}\right). \label{G-Matrices2}
\end{IEEEeqnarray}
\paragraph{Dirac basis} Alternatively, 4-dimensional gamma matrices (in $d = 3+1$ dimensions) can be defined in the Dirac basis, in which the $\gamma_0$ matrix is diagonal:
\begin{IEEEeqnarray}{ll}
\gamma_0 = i \, \sigma_3 \otimes I_2 = \left(\begin{array}{cccc}
 i & 0 & 0 & 0 \\
 0 & i & 0 & 0 \\
 0 & 0 & -i & 0 \\
 0 & 0 & 0 & -i \\\end{array}\right), \qquad &
\gamma_1 = \sigma_2 \otimes \sigma_3 = \left(\begin{array}{cccc}
 0 & 0 & -i & 0 \\
 0 & 0 & 0 & i \\
 i & 0 & 0 & 0 \\
 0 & -i & 0 & 0 \\\end{array}\right) \qquad \label{GammaMatrices2} \\[6pt]
\gamma_2 = -\sigma_2 \otimes \sigma_1 = \left(\begin{array}{cccc}
 0 & 0 & 0 & i \\
 0 & 0 & i & 0 \\
 0 & -i & 0 & 0 \\
 -i & 0 & 0 & 0 \\\end{array}\right), \qquad &
\gamma_3 = \sigma_2 \otimes \sigma_2 = \left(\begin{array}{cccc}
 0 & 0 & 0 & -1 \\
 0 & 0 & 1 & 0 \\
 0 & 1 & 0 & 0 \\
 -1 & 0 & 0 & 0 \\\end{array}\right). \qquad \label{GammaMatrices3}
\end{IEEEeqnarray}
The Dirac matrices \eqref{GammaMatrices2}--\eqref{GammaMatrices3} satisfy the Minkowski Clifford algebra $\left\{\gamma_{\mu},\gamma_{\nu}\right\} = 2\eta_{\mu\nu}$, where $\eta_{\mu\nu} = (-+++)$. We may also define the matrix $K_4 \equiv \gamma_{12}$ and the charge conjugation matrix $C_4 \equiv i\gamma_{03}$ which satisfy the following relations (see \eqref{GeneratorsAdS4} below for the definition of $\gamma_{\mu\nu}$):
\begin{IEEEeqnarray}{c}
\gamma_{\mu}^t = K_4^{-1}\gamma_{\mu} K_4 = -C_4\gamma_{\mu} C_4^{-1}, \qquad \gamma_{\mu\nu}^t = -K_4^{-1}\gamma_{\mu\nu} K_4, \qquad
\end{IEEEeqnarray}
for $\mu,\nu = 0,\ldots,3$, as well as
\begin{IEEEeqnarray}{c}
K_4\gamma_2 = \gamma_1, \qquad K_4\gamma_1 = -\gamma_2, \qquad \gamma_2 K_4 = -\gamma_1, \qquad \gamma_1 K_4 = \gamma_2 \qquad \\[6pt]
K_4^2 = C_4^2 = -1, \qquad K_4^t = -K_4, \qquad \left[K_4,C_4\right] = 0. \qquad
\end{IEEEeqnarray}
\paragraph{Bosonic generators} The bosonic subalgebra of $\mathfrak{osp}\left(2,2|6\right)$ is $\mathfrak{usp}\left(2,2\right) \oplus \mathfrak{so}\left(6\right)$. The 10 generators of the conformal algebra $\mathfrak{so}\left(3,2\right) \sim \mathfrak{usp}\left(2,2\right)$ are:
\begin{IEEEeqnarray}{c}
D \equiv \frac{\gamma_3}{2}, \quad P_{\mu} \equiv \Pi_{+}\gamma_{\mu}, \quad K_{\mu} \equiv \Pi_{-}\gamma_{\mu}, \quad L_{\mu\nu} \equiv \gamma_{\mu\nu} \equiv \frac{1}{2}\left[\gamma_{\mu},\gamma_{\nu}\right], \quad \Pi_{\pm} \equiv \frac{1}{2}\left(1 \pm \gamma_3\right), \qquad \label{GeneratorsAdS4}
\end{IEEEeqnarray}
for $\mu,\nu = 0,1,2$. The expressions for the 15 generators of $\mathfrak{so}\left(6\right)$ that will be used for the parametrization of $\CP^3$ are given in \S\ref{Appendix:TR-Matrices} below. The conformal generators \eqref{GeneratorsAdS4} satisfy the following set of commutation relations:
\begin{IEEEeqnarray}{l}
\left[P_{\mu},P_{\nu}\right] = \left[K_{\mu},K_{\nu}\right] = 0, \ \left[P_{\mu},K_{\nu}\right] = 2 \eta_{\mu\nu}D + L_{\mu\nu}, \quad \left[D,P_{\mu}\right] = P_{\mu}, \ \left[D,K_{\mu}\right] = -K_{\mu} \qquad \quad \label{CommutationRelations1} \\[6pt]
\left[D,L_{\mu\nu}\right] = 0, \quad \left[P_{\mu}, L_{\rho\sigma}\right] = 2\left(\eta_{\mu\rho}P_{\sigma} - \eta_{\mu\sigma}P_{\rho}\right), \quad \left[K_{\mu},L_{\rho\sigma }\right] = 2\left(\eta_{\mu\rho} K_{\sigma} - \eta_{\mu\sigma} K_{\rho}\right) \qquad \quad \label{CommutationRelations2} \\[6pt]
\left[L_{\mu\nu},L_{\rho\sigma}\right] = 2\left(\eta_{\mu\sigma} L_{\nu\rho} + \eta_{\nu\rho} L_{\mu\sigma} - \eta_{\mu\rho}L_{\nu\sigma} - \eta_{\nu\sigma} L_{\mu\rho}\right). \label{CommutationRelations3}
\end{IEEEeqnarray}
\subsection[T \& R matrices]{T \& R matrices \label{Appendix:TR-Matrices}}
\noindent The Lie algebra of $\mathfrak{so}\left(6\right)$ is spanned by 15 antisymmetric $6 \times 6$ matrices $M_{ij}$ which are defined in terms of the standard unity ($6 \times 6$) matrices $E_{ij}$ as follows
\begin{IEEEeqnarray}{c}
M_{ij} \equiv E_{ij} - E_{ji}, \qquad E_{ij} = \left\{\begin{array}{ll} 1, \ &\text{at position} \ (i,j) \\[6pt] 0, \ &\text{elsewhere} \end{array}\right., \qquad i,j = 1, \ldots 6.
\end{IEEEeqnarray}
The generators $M_{ij}$ obey the $\mathfrak{so}\left(6\right)$ commutation relations:
\begin{IEEEeqnarray}{c}
\left[M_{ij},M_{kl}\right] = \delta_{il} M_{jk} + \delta_{jk} M_{il} - \delta_{ik} M_{jl} - \delta_{jl} M_{ik}. \label{CommutationRelations4}
\end{IEEEeqnarray}
A $\mathfrak{u}\left(3\right)$ subalgebra within $\mathfrak{so}\left(6\right)$ is generated by the 9 antisymmetric R-matrices,
\begin{IEEEeqnarray}{lll}
R_1 = \frac{1}{2}\left(M_{13} + M_{24}\right), \quad & R_2 = \frac{1}{2}\left(M_{23} - M_{14}\right), \quad & R_3 = \frac{1}{2}\left(M_{15} + M_{26}\right) \qquad \label{Rmatrices1} \\[6pt]
R_4 = \frac{1}{2}\left(M_{25} - M_{16}\right), \quad & R_5 = \frac{1}{2}\left(M_{35} + M_{46}\right), \quad & R_6 = \frac{1}{2}\left(M_{45} - M_{36}\right) \qquad \label{Rmatrices2} \\[12pt]
R_7 = M_{12}, \quad & R_8 = M_{34}, \quad & R_9 = M_{56}, \qquad \label{Rmatrices3}
\end{IEEEeqnarray}
which define the graded-0 generators of $\mathfrak{so}\left(6\right)$ with respect to $\mathfrak{u}\left(3\right)$. The 6 antisymmetric T-matrices
\begin{IEEEeqnarray}{lll}
T_1 = \frac{1}{2}\left(M_{13} - M_{24}\right), \quad & T_2 = \frac{1}{2}\left(M_{14} + M_{23}\right), \quad & T_3 = \frac{1}{2}\left(M_{15} - M_{26}\right) \qquad \label{Tmatrices1} \\[6pt]
T_4 = \frac{1}{2}\left(M_{16} + M_{25}\right), \quad & T_5 = \frac{1}{2}\left(M_{35} - M_{46}\right), \quad & T_6 = \frac{1}{2}\left(M_{36} + M_{45}\right), \qquad \label{Tmatrices2}
\end{IEEEeqnarray}
are the graded-2 generators belonging to the orthogonal space of $\mathfrak{u}\left(3\right)$ within $\mathfrak{so}\left(6\right)$. The R-matrices commute, while the T-matrices anticommute with the matrix $K_6 \equiv \textrm{I}_3 \otimes \left(i\sigma_{2}\right)$:
\begin{IEEEeqnarray}{c}
\left[R_a, K_6\right] = \left\{T_b, K_6\right\} = 0 \quad (a = 1,\ldots,9, \ b = 1,\ldots,6), \qquad K_6^2 = -1.
\end{IEEEeqnarray}
\subsection{Gamma matrices in $4+1$ dimensions \label{Appendix:5dGammaMatrices}}
\noindent In $d = 5$ dimensions, gamma matrices are again 4-dimensional. They can be defined as follows:
\begin{IEEEeqnarray}{ll}
\gamma_0 = i \, \sigma_3 \otimes \sigma_0 = \left(\begin{array}{cccc}
 i & 0 & 0 & 0 \\
 0 & i & 0 & 0 \\
 0 & 0 & -i & 0 \\
 0 & 0 & 0 & -i \\\end{array}\right), \qquad &
\gamma_1 = \sigma_2 \otimes \sigma_3 = \left(\begin{array}{cccc}
 0 & 0 & -i & 0 \\
 0 & 0 & 0 & i \\
 i & 0 & 0 & 0 \\
 0 & -i & 0 & 0 \\\end{array}\right) \label{GammaMatrices4} \\[6pt]
\gamma_2 = \sigma_1 \otimes \sigma_0 = \left(\begin{array}{cccc}
 0 & 0 & 1 & 0 \\
 0 & 0 & 0 & 1 \\
 1 & 0 & 0 & 0 \\
 0 & 1 & 0 & 0 \\\end{array}\right), \qquad &
\gamma_3 = -\sigma_2 \otimes \sigma_1 = \left(\begin{array}{cccc}
 0 & 0 & 0 & i \\
 0 & 0 & i & 0 \\
 0 & -i & 0 & 0 \\
 -i & 0 & 0 & 0 \\\end{array}\right) \label{GammaMatrices5} \\[6pt]
\gamma_4 = \sigma_2 \otimes \sigma_2 = \left(\begin{array}{cccc}
 0 & 0 & 0 & -1 \\
 0 & 0 & 1 & 0 \\
 0 & 1 & 0 & 0 \\
 -1 & 0 & 0 & 0 \\\end{array}\right), \qquad &
\gamma_5 = -\gamma_{1234} = \left(
\begin{array}{cccc}
 1 & 0 & 0 & 0 \\
 0 & 1 & 0 & 0 \\
 0 & 0 & -1 & 0 \\
 0 & 0 & 0 & -1 \\\end{array} \right) = -i\gamma_0. \qquad \label{GammaMatrices6}
\end{IEEEeqnarray}
The gamma matrices \eqref{GammaMatrices4}--\eqref{GammaMatrices6} can be packed into two sets, namely $\gamma_a$, for $a = 0,\ldots,4$ and $\gamma_{\grave{a}}$, for $\grave{a} = 1,\ldots,5$. The two sets satisfy the $(-++++)$ (Minkowski) and $(+++++)$ (Euclidean) Clifford algebras, providing representations of the algebras $\mathfrak{so}\left(4,2\right) \sim \mathfrak{su}\left(2,2\right)$ and $\mathfrak{so}\left(6\right) \sim \mathfrak{su}\left(4\right)$. We also define the matrix $K_4 \equiv \gamma_{13}$ and the charge conjugation matrix $C_4 \equiv -\gamma_{24}$ which satisfy the following identities:
\begin{IEEEeqnarray}{c}
\gamma_{a}^t = K_4^{-1}\gamma_{a} K_4, \qquad \gamma_{ab}^t = -K_4^{-1}\gamma_{ab} K_4,
\end{IEEEeqnarray}
for $a,b = 0,\ldots,5$. In addition,
\begin{IEEEeqnarray}{c}
K_4\gamma_3 = \gamma_1, \qquad K_4\gamma_1 = -\gamma_3, \qquad \gamma_3 K_4 = -\gamma_1, \qquad \gamma_1 K_4 = \gamma_3 \\[6pt]
K_4^2 = C_4^2 = -1, \qquad K_4^t = -K_4.
\end{IEEEeqnarray}
\paragraph{Bosonic generators} The bosonic subalgebra of $\mathfrak{psu}\left(2,2|4\right)$ is $\mathfrak{su}\left(2,2\right) \oplus \mathfrak{su}\left(4\right) \oplus \mathfrak{u}\left(1\right)$. The 15 generators of the conformal algebra $\mathfrak{su}\left(2,2\right) \sim \mathfrak{so}\left(4,2\right)$ are:
\begin{IEEEeqnarray}{c}
D \equiv \frac{\gamma_4}{2}, \quad P_{\mu} \equiv \Pi_{+}\gamma_{\mu}, \quad K_{\mu} \equiv \Pi_{-}\gamma_{\mu}, \quad L_{\mu\nu} \equiv \gamma_{\mu\nu} \equiv \frac{1}{2}\left[\gamma_{\mu},\gamma_{\nu}\right], \quad \Pi_{\pm} \equiv \frac{1}{2}\left(1 \pm \gamma_4\right), \qquad \label{GeneratorsAdS5}
\end{IEEEeqnarray}
for $\mu,\nu = 0,\ldots,3$. These generators satisfy the commutation relations \eqref{CommutationRelations1}--\eqref{CommutationRelations3}. The other 15 generators of $\mathfrak{su}\left(4\right) \sim \mathfrak{so}\left(6\right)$ are:
\begin{IEEEeqnarray}{c}
M_{\grave{a}\grave{b}} \equiv \frac{1}{2}\,\gamma_{\grave{a}\grave{b}} = \frac{1}{4}\left[\gamma_{\grave{a}},\gamma_{\grave{b}}\right], \qquad M_{\grave{a}6} \equiv -M_{6\grave{a}} = \frac{i}{2}\,\gamma_{\grave{a}}, \label{GeneratorsS}
\end{IEEEeqnarray}
for $\grave{a}, \grave{b} = 1,\ldots,5$. The generators \eqref{GeneratorsS} satisfy the commutation relations \eqref{CommutationRelations4}.
\section[Conventions]{Conventions \label{Appendix:Conventions}}
The metric of $d+1$ dimensional anti-de Sitter space AdS$_{d+1}$, in the Poincar\'{e} coordinate system reads:
\begin{IEEEeqnarray}{c}
ds^2_{\text{AdS}} = \frac{\ell^2}{z^2} \left(\pm dx_0^2 + dx_1^2 + \ldots + dx_{d-1}^2 + dz^2\right), \label{MetricAdS}
\end{IEEEeqnarray}
where $\ell$ is the AdS radius. The sign in front of the time differential is generally negative unless we are working in Euclidean AdS in which case it must be positive. Apart from the coordinate set $(x,z)$ showing up in \eqref{MetricAdS}, we will occasionally also employ the set $(y,w)$ to denote a second point inside AdS. See e.g.\ \eqref{PropagatorBulkToBulk1}--\eqref{PropagatorBulkToBoundary2} below. The former set $(x,z)$ will usually be combined with untilded coordinates in the compact space, while the latter set $(y,w)$ will usually be combined with tilded coordinates. For example, $(x,z)$ combines with $(\psi, \theta, \chi, \vartheta, \varrho)$, and $(y,w)$ combines with $(\tilde{\psi}, \tilde{\theta}, \tilde{\chi}, \tilde{\vartheta}, \tilde{\varrho})$ on the 5-sphere \eqref{MetricS5so3so3}. Similarly for S$^4$ and $\CP^3$ in \eqref{MetricCP3}.
\paragraph{Supergravity solution in AdS$_5\times\text{S}^5$} The classical equations of motion of type IIB supergravity afford a solution \cite{KimRomansNieuwenhuizen85} which consists of the AdS$_5\times\text{S}^5$ metric,
\begin{IEEEeqnarray}{c}
ds^2 = ds^2_{\text{AdS}_5} + \ell^2 d\Omega_5^2, \label{MetricAdS5xS5}
\end{IEEEeqnarray}
where $d\Omega_5$ is the line element of the unit 5-sphere. The solution \eqref{MetricAdS5xS5} is supported by $N_c$ units of self-dual 5-form Ramond-Ramond (RR) flux $\hat{F}_{(5)}$ through AdS$_5$ and S$^5$. The field strengths $\hat{F}_{(5)} = d\hat{C}_{(4)}$ (we set $\hat{C}_{(4)}$ for the corresponding 4-form potentials) through AdS$_5$ and S$^5$ read:
\begin{IEEEeqnarray}{c}
\hat{F}_{mnpqr} = \varepsilon_{mnpqr}, \qquad \hat{F}_{\mu\nu\rho\sigma\tau} = \varepsilon_{\mu\nu\rho\sigma\tau}. \qquad \label{RamondRamondPotentialAdS5xS5}
\end{IEEEeqnarray}
Note that we are using Latin indices ($m,n,p,q,r$) for the AdS$_5$ coordinates and Greek indices ($\mu, \nu, \rho, \sigma, \tau$) for the S$^5$ coordinates.
\paragraph{Supergravity solution in AdS$_4\times\CP^3$} The classical equations of motion of type IIA supergravity afford a solution \cite{NilssonPope84} which consists of the AdS$_4\times\CP^3$ metric,
\begin{IEEEeqnarray}{c}
ds^2 = ds^2_{\text{AdS}_4} + 4\ell^2 ds^2_{\CP^3}, \label{MetricAdS4xCP3}
\end{IEEEeqnarray}
where, for $\chi \in \left[0,\pi/2\right)$, $\theta_{1,2} \in \left[0,\pi\right]$, $\phi_{1,2} \in \left[0,2\pi\right)$ and $\psi \in \left[-2\pi,2\pi\right]$,
\begin{IEEEeqnarray}{ll}
ds_{\CP^3}^2 = &d\chi^2 + \cos^2\chi \, \sin^2\chi \left(d\psi + \frac{1}{2} \cos\theta_1 \, d\phi_1 - \frac{1}{2} \cos\theta_2 \, d\phi_2\right)^2 + \frac{1}{4} \cos^2\chi \Big(d\theta_1^2 + \sin^2 \theta_1 \, d\phi_1^2\Big) + \nonumber \\[6pt]
&+ \frac{1}{4}\sin^2\chi \left(d\theta_2^2 + \sin^2\theta_2 \, d\phi_2^2\right). \qquad \label{MetricCP3}
\end{IEEEeqnarray}
The solution \eqref{MetricAdS4xCP3}--\eqref{MetricCP3} is supported by $N_c$ units of 4-form RR flux $\hat{F}_{(4)}$ through AdS$_4$ and $k$ units of 2-form RR flux through $\CP^1\subset\CP^3$. The corresponding field strengths read:
\begin{IEEEeqnarray}{ll}
\hat{F}_{mnpq} = \frac{3\ell^3}{8}\varepsilon_{mnpq}, \quad F^{(2)}= k\Bigg(&-\cos\chi\sin\chi d\chi \wedge \left(2d\psi + \cos\theta_1 d\phi_1 - \cos\theta_2 d\phi_2\right) - \nonumber \\
& - \frac{1}{2} \cos^2\chi \sin\theta_1 d\theta_1 \wedge d\phi_1 - \frac{1}{2} \sin^2\chi \sin\theta_2 d\theta_2 \wedge d\phi_2\Bigg). \qquad
\end{IEEEeqnarray}
\paragraph{AdS propagators} The bulk-to-bulk propagator of a massive scalar field in Euclidean AdS$_{d+1}$ reads:
\begin{IEEEeqnarray}{c}
G_{\Delta}\left(x,z;y,w\right) = \frac{\Gamma\left(\Delta\right)\eta^{\Delta}}{2^{\Delta + 1}\pi^{d/2}\Gamma\left(\Delta - \frac{d}{2} + 1\right)}\cdot {_2}F_1\left(\frac{\Delta}{2},\frac{\Delta+1}{2},\Delta - \frac{d}{2} + 1,\eta^2\right), \qquad \label{PropagatorBulkToBulk1}
\end{IEEEeqnarray}
where $m$ is the mass and $\Delta$ is the scaling dimension of the scalar field, while
\begin{IEEEeqnarray}{c}
m^2\ell^2 = \Delta\left(\Delta - d\right), \qquad \eta \equiv \frac{2 z w}{z^2 + w^{2} + \left(x - y\right)^2}.
\end{IEEEeqnarray}
For small $w$ and integer $\Delta$, the propagator \eqref{PropagatorBulkToBulk1} can be expanded as follows:
\begin{IEEEeqnarray}{c}
G_{\Delta}\left(x,z;y,w\right) = \frac{\Gamma\left(\Delta\right)}{2\pi^{d/2}\Gamma\left(\Delta - \frac{d}{2} + 1\right)}\cdot\left\{1 + \frac{\Delta\Lambda_z w^2}{\left(\Delta - \frac{d}{2} + 1\right) K_z^2} + \OO\left(w^4\right)\right\}\cdot \left(\frac{z w}{K_z}\right)^{\Delta}, \qquad \label{PropagatorBulkToBulk2}
\end{IEEEeqnarray}
where $K_z \equiv z^2 + \left(x - y\right)^2$ and $\Lambda_z \equiv \left(\Delta + 1\right) z^2 - \left(\Delta - \frac{d}{2} + 1\right)K_z$. The asymptotic value of \eqref{PropagatorBulkToBulk1} near the boundary of AdS ($w = 0$) becomes
\begin{IEEEeqnarray}{c}
\G_{\Delta}\left(x,z;y\right) \equiv \lim_{w\rightarrow 0} \frac{1}{w^{\Delta}}\cdot G_{\Delta}\left(x,z;y,w\right) = \frac{\Gamma\left(\Delta\right)}{2\pi^{d/2}\Gamma\left(\Delta - \frac{d}{2} + 1\right)} \cdot \frac{z^{\Delta}}{K_z^{\Delta}} = \frac{\K_{\Delta}\left(x,z;y\right)}{2\Delta - d}, \qquad \label{PropagatorBulkToBoundary1}
\end{IEEEeqnarray}
where $\K_{\Delta}$ is the bulk-to-boundary propagator of the scalar field in AdS$_{d+1}$:
\begin{IEEEeqnarray}{c}
\K_{\Delta}\left(x,z;y\right) = \frac{c_{\Delta} \cdot z^{\Delta}}{\left[z^2 + \left(x - y\right)^2\right]^{\Delta}}, \qquad c_{\Delta} \equiv \frac{\Gamma\left(\Delta\right)}{\pi^{d/2}\Gamma\left(\Delta - \frac{d}{2}\right)}. \qquad \label{PropagatorBulkToBoundary2}
\end{IEEEeqnarray}
More information can be found e.g.\ in the book/lecture notes \cite{DHokerFreedman02, FreedmanVanProeyen12}, as well as the original set of references \cite{Fronsdal74a, BurgessLutken85, InamiOoguri85a, BurgesFreedmanGibbons86}.
\section[Chiral primary operators]{Chiral primary operators \label{Appendix:ChiralPrimaryOperators}}
The chiral primary operators (CPO's) of $SU(N_c)$, $\N = 4$ SYM theory are given by symmetrized single-trace products of the theory's six scalar fields $\varphi_i$ ($i=1,\ldots,6$):
\begin{IEEEeqnarray}{c}
\OO^{\text{CPO}}_I\left(x\right) = \frac{1}{\sqrt{L}} \left(\frac{8\pi^2}{\lambda}\right)^{\frac{L}{2}} \Psi^{\mu_1\ldots \mu_L}_I \text{tr}\left[\varphi_{\mu_1}\left(x\right)\ldots \varphi _{\mu_L}\left(x\right)\right], \label{ChiralPrimaryOperators}
\end{IEEEeqnarray}
where $x = \left(x_0,\ldots,x_3\right)$ and $\Psi^{\mu_1\ldots \mu_L}_I$ are traceless symmetric tensors of $SO(6)$. The tensors $\Psi^{\mu_1\ldots \mu_L}_I$ define the S$^5$ spherical harmonics:
\begin{IEEEeqnarray}{c}
Y_I\left(x_{\mu}\right) \equiv \Psi^{\mu_1\ldots \mu_L}_I x_{\mu_1}\ldots x_{\mu_L}, \quad \Psi^{\mu_1\ldots \mu_L}_I\Psi^{\mu_1\ldots \mu_L}_J = \delta_{IJ}, \quad \sum_{\mu=4}^{9} x_{\mu}^2 = 1, \qquad \label{SphericalHarmonicsSO6}
\end{IEEEeqnarray}
where $I,J$ are the corresponding quantum numbers. The overall factor in front of the CPO's \eqref{ChiralPrimaryOperators} normalizes their 2-point functions to unity \cite{LeeMinwallaRangamaniSeiberg98}:
\begin{IEEEeqnarray}{c}
\langle\OO^{\text{CPO}}_I\left(x_1\right) \OO^{\text{CPO}}_J\left(x_2\right)\rangle = \frac{\delta_{IJ}}{x_{12}^{2L}}.
\end{IEEEeqnarray}
Differentiating the definition of $Y_I$ in \eqref{SphericalHarmonicsSO6} we may also show,
\begin{IEEEeqnarray}{c}
\Box Y_I = -L\left(L + 4\right)Y_I. \label{SphericalHarmonicsSO6eigenvalues}
\end{IEEEeqnarray}
\paragraph{Supergravity fluctuations} The scalar supergravity modes $\phi_I(x_m)$ that are dual to the CPO's \eqref{ChiralPrimaryOperators} of $\N = 4$ SYM have been identified \cite{KimRomansNieuwenhuizen85, LeeMinwallaRangamaniSeiberg98}. They are linear combinations of the scalar modes of the AdS$_5$ metric and the RR potential \eqref{RamondRamondPotentialAdS5xS5}. Their mass is $m^2\ell^2 = L(L-4)$. The perturbation \eqref{FieldPerturbationIIB} can be expressed in terms of the modes $\phi(x_m, x_{\mu}) \equiv \phi_I(x_m) Y_I(x_{\mu})$ as follows:
\begin{IEEEeqnarray}{l}
\delta g_{mn} = \frac{2}{\N_L}\,\frac{1}{L+1} \cdot \left[2\ell^2\nabla_m \nabla_n - L(L - 1)\hat{g}_{mn}\right] \phi, \quad \delta g_{\mu\nu} = \frac{2L}{\N_L} \cdot \hat{g}_{\mu\nu} \, \phi \qquad \label{FluctuationsMetric1} \\[6pt]
\delta C_{mnpq} = \frac{\ell}{\N_L} \sqrt{\hat{g}_{\text{AdS}}} \, \varepsilon _{mnpqr} \nabla^r \phi, \quad \delta C_{\mu\nu\rho\sigma} = -\frac{\ell}{\N_L} \sqrt{\hat{g}_{\text{s}}} \, \varepsilon_{\mu\nu\rho\sigma\tau} \nabla^\tau \phi. \qquad \label{FluctuationsPotential1}
\end{IEEEeqnarray}
The normalization constant $\N_L$ is defined as
\begin{IEEEeqnarray}{l}
\N^2_{L} = \frac{N_c^2 L\left(L - 1\right)}{2^{L - 3} \pi^2(L + 1)^2}. \label{NormalizationFactor}
\end{IEEEeqnarray}
Therefore, in the case of the supergravity modes $\phi_I$ which are dual to the CPO's \eqref{ChiralPrimaryOperators}, we identify the vertex operators \eqref{VertexOperatorsIIB} as
\begin{IEEEeqnarray}{l}
V^I_{mn} = \frac{2}{\N_L}\,\frac{1}{L+1} Y_I\left[2\ell^2\nabla_m \nabla_n - L(L - 1)\hat{g}_{mn}\right], \quad V^I_{\mu\nu} = \frac{2L}{\N_L} Y_I \hat{g}_{\mu\nu} \qquad \label{VertexOperators1} \\[6pt]
v^I_{mnpq} = \frac{\ell}{\N_L} \sqrt{\hat{g}_{\text{AdS}}} \, \varepsilon _{mnpqr} \nabla^r Y_I, \quad v^I_{\mu\nu\rho\sigma} = -\frac{\ell}{\N_L} \sqrt{\hat{g}_{\text{s}}} \, \varepsilon_{\mu\nu\rho\sigma\tau} Y_I \nabla^\tau. \qquad \label{VertexOperators2}
\end{IEEEeqnarray}
Acting with the vertex operators \eqref{VertexOperators1}--\eqref{VertexOperators2} on the scalar AdS propagators \eqref{PropagatorBulkToBoundary1}--\eqref{PropagatorBulkToBoundary2} and assuming that the $Y_I$'s are constants (as in the case of the $SO(3)\times SO(3)$ or $SO(4)$ invariant spherical harmonics on S$^5$ in appendix \ref{Appendix:SphericalHarmonics}), we are led to the fluctuations
\begin{IEEEeqnarray}{l}
\frac{\delta g_{ij}^{\text{AdS}}}{\ell^2 Ls} = \frac{2}{\N_L}\left[-\frac{\delta_{ij}}{z^2} + \frac{8r_i r_j}{K_z^2}\right], \qquad\frac{\delta g_{iz}^{\text{AdS}}}{\ell^2 Ls} = \frac{8r_i}{\N_L}\left[\frac{2z}{K_z^2} - \frac{1}{zK_z}\right] \label{FluctuationsMetric2a} \\[6pt]
\frac{\delta g_{zz}^{\text{AdS}}}{\ell^2 Ls} = \frac{2}{\N_L}\left[\frac{1}{z^2} - \frac{8}{K_z} + \frac{8z^2}{K_z^2}\right], \qquad \frac{\delta g_{\mu\nu}^{\text{S}}}{Ls} = \frac{2}{\N_L} \cdot \hat{g}_{\mu\nu} \label{FluctuationsMetric2b} \\[6pt]
\frac{\delta C_{0123}}{\ell^4 Ls} = \frac{1}{\N_L}\cdot\frac{1}{z^3}\left[\frac{1}{z} - \frac{2z}{K_z}\right], \qquad \frac{\delta C_{012z}}{\ell^4 Ls} = -\frac{2}{\N_L} \cdot \frac{r_3}{z^3 K_z}, \label{FluctuationsPotential2}
\end{IEEEeqnarray}
where either $\phi \rightarrow Y_I \cdot \G_L\left(x,z;y\right)$ or $\phi \rightarrow Y_I \cdot \K_L\left(x,z;y\right)$, and we have set $r_i \equiv x_i - y_i$, $K_z \equiv r^2 + z^2$, for $i,j = 0,\ldots,3$.
\section[Spherical harmonics on S$^5$]{Spherical harmonics on S$^5$ \label{Appendix:SphericalHarmonics}}
\subsection{$SO(3)\times SO(3)$ invariant subset \label{Appendix:SphericalHarmonicsSO3}}
\noindent The definition of the S$^5$ spherical harmonics was given in \eqref{SphericalHarmonicsSO6}. To determine the subset of S$^5$ spherical harmonics that is invariant under the $SO(3)\times SO(3)$ subgroup of $SO(6)$, let us first write out the line element of the unit 5-sphere $d\Omega_5$ in a manifestly $SO(3)\times SO(3)$ invariant way:
\begin{IEEEeqnarray}{c}
d\Omega_5^2 = d\psi^2 + \cos^2\psi\left(d\theta^2 + \sin^2\theta d\chi^2\right) + \sin^2\psi\left(d\vartheta^2 + \sin^2\vartheta d\varrho^2\right), \qquad \label{MetricS5so3so3}
\end{IEEEeqnarray}
where $\psi \in \left[0, \pi/2\right]$. The corresponding Cartesian coordinates $x_4, \ldots, x_9$ read
\begin{IEEEeqnarray}{lll}
x_4 = \cos\psi \sin\theta \cos\chi, \quad & x_5 = \cos\psi \sin\theta \sin\chi, \quad & x_6 = \cos\psi \cos\theta, \label{CartesianCoordinatesS5a} \\
x_7 = \sin\psi \sin\vartheta \cos\varrho, \quad & x_8 = \sin\psi \sin\vartheta \sin\varrho, \quad & x_9 = \sin\psi \cos\vartheta. \label{CartesianCoordinatesS5b}
\end{IEEEeqnarray}
They obviously obey
\begin{IEEEeqnarray}{c}
\sum_{\mu=4}^{9} x_{\mu}^2 = 1, \qquad \sum_{\mu=4}^{6} x_{\mu}^2 = \cos^2\psi, \qquad \sum_{\mu=7}^{9} x_{\mu}^2 = \sin^2\psi, \label{CartesianCoordinatesS5c}
\end{IEEEeqnarray}
which subsequently also implies that the $SO(3)\times SO(3)$ invariant spherical harmonics on S$^5$ depend only on the angle $\psi$. Denoting these spherical harmonics by $Y\left(\psi\right)$, we can compute them from the eigenfunctions of the Laplace operator on the 5-sphere \eqref{MetricS5so3so3}:
\begin{IEEEeqnarray}{c}
\Box Y = \frac{1}{\sqrt{\hat{g}_{\text{s}}}}\,\partial_{\mu}\left[\sqrt{\hat{g}_{\text{s}}} \, \hat{g}^{\mu\nu}\partial_{\nu}Y\right] = \frac{1}{\cos^2\psi\sin^2\psi}\,\partial_{\psi}\left(\cos^2\psi\sin^2\psi\partial_{\psi}Y\left(\psi\right)\right). \qquad
\end{IEEEeqnarray}
Changing variables $z = \sin^2\psi$, the eigenvalue equation $\Box Y = -E Y$ is brought to the following form
\begin{IEEEeqnarray}{c}
z\left(1-z\right)\partial_z^2Y\left(z\right) + \left(\frac{3}{2} - 3z\right)\partial_zY\left(z\right) + \frac{E}{4} \, Y\left(z\right) = 0,
\end{IEEEeqnarray}
which is just the hypergeometric equation with solution
\begin{IEEEeqnarray}{l}
E = 2j\left(2j+4\right), \qquad Y_j\left(z\right) = \CC_{j} \cdot {}_2 F_1\big(-j,j+2,\frac{3}{2};z\big), \qquad j = 0,1,\ldots, \qquad \label{SphericalHarmonicsSO3}
\end{IEEEeqnarray}
and the normalization factor $\CC_j$ is determined from
\begin{IEEEeqnarray}{l}
\int_{\text{S}^5} \left|Y_j\right|^2 = \frac{1}{2^{2j-1}\left(2j+1\right)\left(2j+2\right)}\int_{\text{S}^5}1. \label{SphericalHarmonicsSO3normalization1}
\end{IEEEeqnarray}
We end up with the general formula,
\begin{IEEEeqnarray}{l}
Y_j\left(\psi\right) = \frac{(2j + 2)!}{2^{j + \frac{1}{2}}\sqrt{(2j+1)(2j+2)}} \cdot \sum_{p=0}^j \frac{(-1)^p \cos^{2p}\psi\sin^{2j-2p}\psi}{(2p+1)!(2j-2p+1)!},
\end{IEEEeqnarray}
which implies the following value for the normalization factor $\CC_j$:
\begin{IEEEeqnarray}{l}
\CC_j = \left(-\frac{1}{2}\right)^j\sqrt{\frac{j+1}{2j+1}}. \label{SphericalHarmonicsSO3normalization2}
\end{IEEEeqnarray}
Comparing the expression $E = 2j\left(2j+4\right)$ in \eqref{SphericalHarmonicsSO3} for the $SO(3)\times SO(3)$ eigenvalues with the $SO(6)$ eigenvalues \eqref{SphericalHarmonicsSO6eigenvalues}, we also obtain $L = 2j$.
\subsection{$SO(4)$ invariant subset}
\noindent In the present appendix we determine the subset of S$^5$ spherical harmonics (defined in \eqref{SphericalHarmonicsSO6} above) which is invariant under $SO(4) \subset SO(6)$. We start off with the cosine analog of the S$^5$ metric \eqref{MetricS5so6a}--\eqref{MetricS5so6b}:
\begin{IEEEeqnarray}{c}
ds^2 = d\theta^2 + \cos^2\theta \, d\Omega_4^2, \qquad \label{MetricS5so5}
\end{IEEEeqnarray}
where $\theta \in \left[-\pi/2,\pi/2\right]$. The corresponding Cartesian coordinates $x_4, \ldots, x_9$ are given by
\begin{IEEEeqnarray}{ll}
x_a = m_{(a-3)}\cos\theta, \qquad x_9 = \sin\theta, \qquad a = 4,\ldots,8, \qquad \sum_{a=1}^{5} m_{a}^2 = 1,
\end{IEEEeqnarray}
where the variables $m_a$ parametrize the unit 4-sphere as e.g.\ in \eqref{SineParametrization2}. Obviously, the corresponding spherical harmonics will depend only on the angle $\theta$. As such, they are eigenfunctions of the Laplace operator,
\begin{IEEEeqnarray}{c}
\Box Y = \frac{1}{\sqrt{\hat{g}_{\text{s}}}}\,\partial_{\mu}\left[\sqrt{\hat{g}_{\text{s}}} \, \hat{g}^{\mu\nu}\partial_{\nu}Y\right] = \sec^4\theta\,\partial_{\theta}\left(\sec^4\theta\,\partial_{\theta}Y\left(\theta\right)\right) = - E \, Y\left(\theta\right), \qquad
\end{IEEEeqnarray}
which reads, after changing variables $z = (1-\sin\theta)/2$:
\begin{IEEEeqnarray}{c}
z\left(1-z\right)\partial_z^2Y\left(z\right) + \left(\frac{5}{2} - 5z\right)\partial_zY\left(z\right) + E \, Y\left(z\right) = 0.
\end{IEEEeqnarray}
This is just the hypergeometric equation with solution
\begin{IEEEeqnarray}{l}
E = 2j\left(2j+4\right), \qquad Y_j\left(z\right) = \CC_{j} \cdot {}_2 F_1\big(-2j,2j+4,\frac{5}{2};z\big), \qquad j = 0,1,\ldots, \qquad \label{SphericalHarmonicsSO5}
\end{IEEEeqnarray}
where the normalization factor $\CC_j$ is determined from \eqref{SphericalHarmonicsSO3normalization1}
\begin{IEEEeqnarray}{l}
\int_{\text{S}^5} \left|Y_j\right|^2 = \frac{1}{2^{2j-1}\left(2j+1\right)\left(2j+2\right)}\int_{\text{S}^5}1.
\end{IEEEeqnarray}
The $SO(5)$ invariant spherical harmonics on S$^5$ are given by the expression
\begin{IEEEeqnarray}{l}
Y_j\left(\theta\right) = \frac{1}{2^j}\sqrt{\frac{(2j+2)(2j+3)}{6}} \cdot \sum_{p=0}^{2j} \frac{\Gamma\left(5/2\right)}{\Gamma\left(p + 5/2\right)} \frac{(2j + p +3)!(2j)!}{(2j-p)!(2j+3)!p!} \left(\frac{\sin\theta - 1}{2}\right)^p, \qquad
\end{IEEEeqnarray}
from which we may extract the normalization factor $\CC_j$:
\begin{IEEEeqnarray}{l}
\CC_j = \frac{1}{2^j}\sqrt{\frac{(2j+2)(2j+3)}{6}}.
\end{IEEEeqnarray}
By comparing the expression $E = 2j\left(2j+4\right)$ in \eqref{SphericalHarmonicsSO5} for the $SO(5)$ eigenvalues with the $SO(6)$ eigenvalues \eqref{SphericalHarmonicsSO6eigenvalues}, we get $L = 2j$.
\section[D-brane actions]{D-brane actions \label{Appendix:DbraneActions}}
\subsection{Probe D5-brane}
The (Abelian) action of the probe D5-brane is given by the sum of the Dirac-Born-Infeld (DBI) and Wess-Zumino (WZ) term:
\begin{IEEEeqnarray}{c}
S_{\text{D5}} = \frac{T_5}{g_s}\int \left[d^6\zeta\sqrt{\det\left(G_{ab} + 2\pi\alpha'F_{ab}\right)} + 2\pi\alpha' F\wedge C\right], \qquad \label{D5braneAction}
\end{IEEEeqnarray}
where $T_5 \equiv \left(2\pi\right)^{-5}\alpha'^{-3}$ is the D5-brane tension, $G_{ab}$ is the pullback of the IIB graviton field $g_{MN}$ (which is given by \eqref{FieldPerturbationIIB}) on the 5-brane, i.e.\
\begin{IEEEeqnarray}{c}
G_{ab} \equiv \partial_a \Y^M \partial_b \Y^N g_{MN}. \label{GravitonPullbackD5}
\end{IEEEeqnarray}
Moreover, $F_{ab}$ is the field strength of the worldvolume gauge field in \eqref{FieldStrengthD5a}--\eqref{FieldStrengthD5c} and $C$ is the 4-form (Ramond-Ramond) potential of type IIB supergravity (as given by \eqref{FieldPerturbationIIB}). We have also set
\begin{IEEEeqnarray}{c}
h_{ab} \equiv \partial_a \Y^M \partial_b \Y^N \hat{g}_{MN} + 2\pi\alpha'F_{ab}, \qquad h \equiv \det h_{ab}, \label{InducedMetricD5-1}
\end{IEEEeqnarray}
and use $(\zeta_0,\ldots,\zeta_5) = (y_0,y_1,y_2,w,\tilde{\theta},\tilde{\chi})$ for the worldvolume coordinates of the D5-brane.
\paragraph{D5-brane embedding} As we have explained in \S\ref{SubSubSection:D3D5intersection}, the probe D5-brane wraps an AdS$_4\times\text{S}^2$ geometry inside AdS$_5\times\text{S}^5$. The geometry is parametrized by \eqref{D5braneEmbedding1}, while it is also supported by $\kk$ units of (Abelian) magnetic flux through S$^2$ which obey \eqref{FieldStrengthD5a}. \\
\indent Plugging the (Euclidean) AdS$_5\times\text{S}^5$ metric \eqref{MetricAdS5xS5}, the D5-brane parametrization \eqref{D5braneEmbedding1} and the worldvolume flux \eqref{FieldStrengthD5a} into the induced metric \eqref{InducedMetricD5-1} we arrive at,
\begin{IEEEeqnarray}{c}
h_{ab} = \ell^2\left[\begin{array}{cccccc} w^{-2} &&&&& \\ & w^{-2} &&&& \\ && w^{-2} &&& \\ &&& \left(1 + \kappa^2\right)w^{-2} && \\ &&&&1&\kappa\sin\tilde{\theta} \\ &&&&-\kappa\sin\tilde{\theta}&\sin^2\tilde{\theta} \end{array}\right]. \label{InducedMetricD5-2}
\end{IEEEeqnarray}
The inverse metric reads,
\begin{IEEEeqnarray}{c}
h^{ab} = \ell^{-2}\left[\begin{array}{cccccc} w^2 &&&&& \\ & w^2 &&&& \\ && w^2 &&& \\ &&& w^2\left(1 + \kappa^2\right)^{-1} && \\ &&&&\left(1 + \kappa^2\right)^{-1}&-\kappa\csc\tilde{\theta}\left(1 + \kappa^2\right)^{-1} \\ &&&&\kappa\csc\tilde{\theta}\left(1 + \kappa^2\right)^{-1}& \csc^2\tilde{\theta}\left(1 + \kappa^2\right)^{-1} \end{array}\right], \label{InducedMetricD5-3}
\end{IEEEeqnarray}
while the corresponding determinant becomes:
\begin{IEEEeqnarray}{c}
h = \frac{\ell^{12}}{w^8}\left(1 + \kappa^2\right)^2\sin^2\tilde{\theta}.
\end{IEEEeqnarray}
\subsection{Probe D7-brane}
For the probe D7-branes that we consider in the present article, we need the non-Abelian generalization of the DBI + WZ action \cite{Tseytlin97c}:
\begin{IEEEeqnarray}{c}
S_{\text{D7}} = \frac{T_7}{g_s} \cdot \text{STr}\int \left[d^8\zeta\sqrt{\det\left(G_{ab} + 2\pi\alpha'F_{ab}\right)} + 2\pi\alpha' F\wedge F\wedge \hat{C}\right], \qquad \label{D7braneAction}
\end{IEEEeqnarray}
where $T_7 \equiv \left(2\pi\right)^{-7}\alpha'^{-4}$ is the tension of the D7-brane and STr stands for the symmetrized (color) trace.\footnote{The symmetrized trace is computed as $\text{STr}\left[A_1A_2\ldots A_n\right] \equiv \frac{1}{n!}\cdot\text{tr}\left[A_1A_2\ldots A_n + \text{permutations}\right]$.} The determinant which shows up in the action \eqref{D7braneAction} is computed with respect to the D-brane worldvolume indices $a,b$. As before, $G_{ab}$ is the pullback of the graviton field $g_{MN}$ on the 7-brane,
\begin{IEEEeqnarray}{c}
G_{ab} \equiv \partial_a \Y^M \partial_b \Y^N g_{MN}, \qquad G \equiv \det G_{ab}, \label{GravitonPullbackD7}
\end{IEEEeqnarray}
while $F_{ab}$ is the field strength of the worldvolume gauge field and $\hat{C}$ is the 4-form (Ramond-Ramond) potential. Also, $(\zeta_0,\ldots,\zeta_7) = (y_0,y_1,y_2,w,\tilde{\theta},\tilde{\chi},\tilde{\vartheta},\tilde{\varrho})$ denote the worldvolume coordinates of the D7-brane. \\
\indent A practical form of the probe D7-brane action in the non-Abelian case can be obtained by expanding the determinant and the square root of the DBI component in \eqref{D7braneAction}. Following \cite{Tseytlin97c},
\begin{IEEEeqnarray}{c}
\text{STr} \sqrt{\det\left(G_{ab} + 2\pi\alpha'F_{ab}\right)} = \sqrt{G} \cdot \text{STr} \sqrt{\det\left(\delta^{a}_{b} + 2\pi\alpha'F^{a}_{b}\right)} = \sqrt{G} \cdot \text{STr}\Bigg\{I + (\pi\alpha')^2 F_{ab}F^{ab} + \nonumber \\[6pt]
+ 2(\pi\alpha')^4 \left[\frac{1}{4}(F_{ab}F^{ab})^2 - F_{ab}F^{bc}F_{cd}F^{da}\right] + \ldots \Bigg\} = \sqrt{G} \cdot \text{Tr}\Bigg\{I + (\pi\alpha')^2 F_{ab}F^{ab} + \nonumber \\[6pt]
+ \frac{4(\pi\alpha')^4}{3} \left[\frac{1}{4} F_{ab}F^{ab}F_{cd}F^{cd} + \frac{1}{8} F_{ab}F^{cd}F_{ab}F^{cd} - F_{ab}F^{cb}F_{ad}F^{cd} - \frac{1}{2}F_{ab}F^{cb}F_{cd}F^{ad}\right] + \ldots \Bigg\}. \qquad
\end{IEEEeqnarray}
\paragraph{Abelian D7-brane embedding} In the $SU(2)\times SU(2)$ symmetric case, the probe D7-brane wraps an AdS$_4\times\text{S}^2\times\text{S}^2$ geometry inside AdS$_5\times\text{S}^5$ which is parametrized by \eqref{D7braneEmbedding1}. The supporting $\kk_{1,2}$ units of (Abelian) magnetic flux through each S$^2$ obey \eqref{FieldStrengthD7a}--\eqref{D7braneConstraint}. \\
\indent By using the expressions for the AdS$_5$ metric \eqref{MetricAdS5xS5}, the D7-brane parametrization \eqref{D7braneEmbedding1} and the Abelian flux \eqref{FieldStrengthD7a} we are led to the following D7-brane induced metric \eqref{InducedMetricD5-1}:
\begin{IEEEeqnarray}{c}
h_{ab} = \ell^2\left[\begin{array}{cccccccc} w^{-2} &&&&&&& \\ & w^{-2} &&&&&& \\ && w^{-2} &&&&& \\ &&& \left(1 + \Lambda^2\right)w^{-2} &&&& \\ &&&&1&\kappa_1\sin\tilde{\theta} && \\ &&&&-\kappa_1\sin\tilde{\vartheta}&\sin^2\tilde{\vartheta} && \\ &&&&&&1&\kappa_2\sin\tilde{\vartheta} \\ &&&&&&-\kappa_2\sin\tilde{\vartheta}&\sin^2\tilde{\vartheta} \end{array}\right]. \qquad \label{InducedMetricD7}
\end{IEEEeqnarray}
The corresponding determinant reads,
\begin{IEEEeqnarray}{c}
h = \frac{\ell^{16}}{w^8}\left(1 + \Lambda^2\right)\left(1 + \kappa_1^2\right)\left(1 + \kappa_2^2\right)\sin^2\tilde{\theta}\sin^2\tilde{\vartheta}.
\end{IEEEeqnarray}
\subsection{Probe D4-brane}
The (Abelian) DBI + WZ action for a probe D4-brane reads:
\begin{IEEEeqnarray}{c}
S_{\text{D4}} = \frac{T_4}{g_s}\int \left[d^5\zeta\sqrt{\det\left(G_{ab} + 2\pi\alpha'F_{ab}\right)} + 2\pi\alpha' F\wedge C\right], \qquad \label{D4braneAction}
\end{IEEEeqnarray}
where $T_4 \equiv \left(2\pi\right)^{-4}\alpha'^{-5/2}$ is the brane tension, $G_{ab}$ is the pullback of the IIA graviton field $g_{MN}$ on the 4-brane (as defined by \eqref{GravitonPullbackD5}), while the definition of the induced metric $h_{ab}$ can once more be found in \eqref{InducedMetricD5-1}. $F_{ab}$ is the field strength of the worldvolume gauge field and $C$ is the 3-form (Ramond-Ramond) potential of type IIA supergravity. The worldvolume coordinates of the D4-brane are denoted by $(\zeta_0,\ldots,\zeta_5) = (y_0,y_1,w,\tilde{\theta}_1,\tilde{\phi}_1)$.
\section[Integrals for defect correlators]{Integrals for defect correlators \label{Appendix:Integrals}}
\paragraph{$\I$ integrals} The integrals $\I_{a,b}$ were defined in \eqref{IntegralsI} as follows:
\begin{IEEEeqnarray}{l}
\I_{a,b}\left(\kappa\right) \equiv \int\displaylimits_{0}^{\infty} du \, \frac{u^a}{\left[u^2 + (1 - \kappa u)^2\right]^b}, \qquad b > \frac{1}{2}.
\end{IEEEeqnarray}
Following \cite{NagasakiYamaguchi12, KristjansenSemenoffYoung12b}, we change variables $\tan\theta = \left(\kappa^2 + 1\right)u - \kappa$,
\begin{IEEEeqnarray}{ll}
\I_{a,b}\left(\kappa\right) &= \kappa^{2b-a-2}\left[1+\frac{1}{\kappa^2}\right]^{b-a-1} \int_{-\arctan\kappa}^{\pi/2} \left(\cos\theta\right)^{2b-2} \left[1 + \frac{1}{\kappa}\tan\theta\right]^a d\theta = \qquad \nonumber \\[6pt]
& \hspace{-.3cm}= \sum_{n=0}^{a}\binom{a}{n}\kappa^{2b-a-n-2}\left[1+\frac{1}{\kappa^2}\right]^{b-a-1} \int_{-\arctan\kappa}^{\pi/2} \left(\cos\theta\right)^{2b-2} \tan^n\theta \, d\theta, \qquad
\end{IEEEeqnarray}
and perform the integrals which gives
\begin{IEEEeqnarray}{ll}
\I_{a,b}\left(\kappa\right) = &\kappa^{2b-a-2}\left[1+\frac{1}{\kappa^2}\right]^{b-a-1} \sum_{n=0}^{a}\binom{a}{n} \bigg\{\frac{1+(-1)^n}{2\kappa^n} B\Big(b-\frac{n+1}{2},\frac{n+1}{2}\Big) + \nonumber \\[6pt]
& + \frac{(-1)^n\left(1+\frac{1}{\kappa^2}\right)^{-b}}{\left(2b-n-1\right)\kappa^{2b-1}} \cdot {_2}F_1\left(1,b,b-\frac{n-1}{2},\frac{1}{1+\kappa^2}\right)\bigg\}. \qquad
\end{IEEEeqnarray}
Expanding around $\kappa = \infty$ we obtain, for $b > 3/2$ (in order for the contribution of the hypergeometric term to be subleading),
\begin{IEEEeqnarray}{ll}
\I_{a,b} = &\kappa^{2b-a-2}\left(1 + \left(b-a-1\right)\frac{1}{\kappa^2} + \ldots\right) \cdot \bigg\{B\Big(b-\frac{1}{2},\frac{1}{2}\Big) + \nonumber \\[6pt]
& + \frac{a\left(a-1\right)}{2\kappa^2}\,B\Big(b-\frac{3}{2},\frac{3}{2}\Big) + \ldots \bigg\}. \qquad
\end{IEEEeqnarray}
In particular we find (assuming $j>1/2$):
\begin{IEEEeqnarray}{l}
\I_{2j-2,2j+\frac{1}{2}} = \kappa^{2j+1}B\big(2j,\frac{1}{2}\big) \cdot \left\{1 + \left[\frac{3}{2} + \frac{\left(2j-3\right)\left(j-1\right)}{2\left(2j-1\right)}\right] \frac{1}{\kappa^2} + \ldots\right\} \qquad \label{IntegralI1} \\[6pt]
\I_{2j,2j+\frac{3}{2}} = \kappa^{2j+1}B\big(2j+1,\frac{1}{2}\big) \cdot \left\{1 + \frac{(2j+1)}{4\kappa^2} + \ldots\right\} \qquad \label{IntegralI2} \\[6pt]
\I_{2j-1,2j+\frac{3}{2}} = \kappa^{2j+2}B\big(2j+1,\frac{1}{2}\big) \cdot \left\{1 + \left[\frac{3}{2} + \frac{\left(2j-1\right)\left(j-1\right)}{4j}\right] \frac{1}{\kappa^2} + \ldots\right\} \qquad \label{IntegralI3} \\[6pt]
\I_{2j-2,2j+\frac{3}{2}} = \kappa^{2j+3}B\big(2j+1,\frac{1}{2}\big) \cdot \left\{1 + \left[\frac{5}{2} + \frac{\left(2j-3\right)\left(j-1\right)}{4j}\right] \frac{1}{\kappa^2} + \ldots\right\} \qquad \label{IntegralI4} \\[6pt]
\I_{2j+2,2j+\frac{5}{2}} = \kappa^{2j+1}B\big(2j+2,\frac{1}{2}\big) \cdot \left\{1 + \frac{j}{2\kappa^2} + \ldots\right\} \qquad \label{IntegralI5} \\[6pt]
\I_{2j+1,2j+\frac{5}{2}} = \kappa^{2j+2}B\big(2j+2,\frac{1}{2}\big) \cdot \left\{1 + \frac{(j+1)}{2\kappa^2} + \ldots\right\} \qquad \label{IntegralI6} \\[6pt]
\I_{2j,2j+\frac{5}{2}} = \kappa^{2j+3}B\big(2j+2,\frac{1}{2}\big) \cdot \left\{1 + \left[\frac{3}{2} + \frac{j\left(2j-1\right)}{2\left(2j+1\right)}\right] \frac{1}{\kappa^2} + \ldots\right\}. \qquad \label{IntegralI7}
\end{IEEEeqnarray}
\paragraph{$\J$ integrals} For $\uu \equiv 1/\sqrt{\xi}$, the integrals $\J_{a,b}$ are defined as
\begin{IEEEeqnarray}{l}
\J_{a,b} \equiv \int\displaylimits_{-\infty}^{+\infty} \frac{\text{sech}^{2a+2}s\cdot ds}{\left(\uu + \tanh s\right)^b} = 2^a \, \Gamma(a+1) \cdot \sum_{n = 0}^{a} \left(-1\right)^n \uu^{a-n} \left[\begin{array}{c} a \\ n\end{array}\right]_b \J_{0,b-a-n}, \qquad \label{IntegralsJ}
\end{IEEEeqnarray}
where the bracket coefficients are defined recursively for $n = 0, \ldots, a$ by
\begin{IEEEeqnarray}{l}
\left[\begin{array}{c} a \\ n\end{array}\right]_b = \frac{1}{b-a-(n-1)}\left[\begin{array}{c} a-1 \\ n-1 \end{array}\right]_b + \frac{1}{b-a-n}\left[\begin{array}{c} a-1 \\ n\end{array}\right]_b,
\end{IEEEeqnarray}
and the initial conditions
\begin{IEEEeqnarray}{l}
\left[\begin{array}{c} 1 \\ 0\end{array}\right]_b = \left[\begin{array}{c} 1 \\ 1 \end{array}\right]_b = \frac{1}{b-1}. \qquad
\end{IEEEeqnarray}
For $j,n = 0,1,2,\ldots$ we find,
\begin{IEEEeqnarray}{ll}
\J_{j+n,2j+2n+2} &= \frac{\Gamma\left(\frac{1}{2}\right)}{\Gamma\left(j+n+\frac{3}{2}\right)} \sum_{m=j+n+1}^{\infty} \frac{\Gamma(m)}{\Gamma(m-j-n)}\,\frac{1}{\uu^{2m}} = \nonumber \\[6pt]
& = \frac{B\big(j+n+1,\frac{1}{2}\big)}{\uu^{2j+2n+2}} \cdot \Bigg\{1 + \left(j+n+1\right)\frac{1}{\uu^2} + \ldots \Bigg\}, \quad j+n = 0,1,\ldots \qquad \ \label{IntegralJ1} \\[6pt]
\J_{j+n-1,2j+2n+1} \, &= \frac{\Gamma\left(\frac{1}{2}\right)}{\left(j+n\right)\Gamma\left(j+n+\frac{1}{2}\right)} \sum_{m=j+n+1}^{\infty} \frac{\Gamma(m)}{\Gamma(m-j-n)}\,\frac{1}{\uu^{2m-1}} = \nonumber \\[6pt]
& = \frac{B\big(j+n,\frac{1}{2}\big)}{\uu^{2j+2n+1}}\cdot\Bigg\{1 + \left(j+n+1\right)\frac{1}{\uu^2} + \ldots \Bigg\}, \quad j+n = 1,2,\ldots \qquad \label{IntegralJ2} \\[6pt]
\J_{j+n-1,2j+2n+2} \, &= \frac{\Gamma\left(\frac{3}{2}\right)}{\left(j+n\right)\Gamma\left(j+n+\frac{3}{2}\right)} \sum_{m=j+n+1}^{\infty} \left(2m-1\right)\frac{\Gamma(m)}{\Gamma(m-j-n)}\,\frac{1}{\uu^{2m}} = \nonumber \\[6pt]
& = \frac{B\big(j+n,\frac{1}{2}\big)}{\uu^{2j+2n+2}}\cdot\Bigg\{1 + \frac{(j+n+1)(2j+2n+3)}{(2j+2n+1)}\,\frac{1}{\uu^2} + \ldots \Bigg\}, \quad j+n = 1,2,\ldots \qquad \quad \label{IntegralJ3}
\end{IEEEeqnarray}

\newpage\bibliographystyle{style}
\bibliography{Bibliography_Physics, Bibliography_Math}

\end{document}